\documentclass[12pt]{article}
\makeatletter \@addtoreset{equation}{section} \makeatother
\renewcommand{\theequation}{\thesection.\arabic{equation}}
\newcommand{\ba}{\begin{array}}
\newcommand{\ea}{\end{array}}
\newcommand{\beq}{\begin{equation}}
\newcommand{\eeq}{\end{equation}}
\newcommand{\bea}{\begin{eqnarray}}
\newcommand{\eea}{\end{eqnarray}}

\def\bce{\begin{center}}
\def\ece{\end{center}}

\def\nonu{\nonumber}

\def\pa{\partial}

\def\be{\beta}

\def\la{\lambda}

\def\eps6{{\displaystyle \mathop{\epsilon}^{6}}{}}
\def\g6{{\displaystyle \mathop{g}^{6}}{}}

\def\nab6{{\displaystyle \mathop{\nabla}^{6}}{}}

\def\0{{\sst{(0)}}}
\def\1{{\sst{(1)}}}
\def\2{{\sst{(2)}}}
\def\3{{\sst{(3)}}}
\def\4{{\sst{(4)}}}
\def\5{{\sst{(5)}}}
\def\6{{\sst{(6)}}}
\def\7{{\sst{(7)}}}
\def\8{{\sst{(8)}}}

\def\ba{\begin{array}}
\def\ea{\end{array}}
\def\beq{\begin{equation}}
\def\eeq{\end{equation}}
\def\be{\begin{equation}}
\def\ee{\end{equation}}

\def\la{\lambda}
\def\eps{\epsilon}

\def\ba{\begin{array}}
\def\ea{\end{array}}
\def\beq{\begin{equation}}
\def\eeq{\end{equation}}
\def\be{\begin{equation}}
\def\ee{\end{equation}}

\def\la{\lambda}
\def\eps{\epsilon}

\def\eps6{{\displaystyle \mathop{\epsilon}^{6}}{}}

\def\nab6{{\displaystyle \mathop{\nabla}^{6}}{}}

\newcommand{\bean}{\begin{eqnarray*}}
\newcommand{\eean}{\end{eqnarray*}}

\begin{document}
\thispagestyle{empty} \addtocounter{page}{-1}
   \begin{flushright}
PUPT-2395 \\
\end{flushright}

\vspace*{1.3cm}
  
\centerline{ \Large \bf   
The Coset Spin-4 Casimir Operator and }
\centerline{ \Large \bf 
Its Three-Point Functions with Scalars}
\vspace*{1.5cm}
\centerline{{\bf Changhyun Ahn 
}} 
\vspace*{1.0cm} 
\centerline{\it 
Department of Physics, Kyungpook National University, Taegu
702-701, Korea} 
\centerline{\it 
Department of Physics, Princeton University, Jadwin Hall, 
Princeton, NJ 08544, USA}
\vspace*{0.8cm} 
\centerline{\tt ahn@knu.ac.kr 
} 
\vskip2cm

\centerline{\bf Abstract}
\vspace*{0.5cm}

We find the GKO coset construction of the dimension $4$ Casimir
operator  
that contains the quartic WZW currents contracted with completely symmetric 
$SU(N)$ invariant tensors of ranks $4, 3$, and $2$. 
The requirements, that the operator product expansion with the diagonal
current is regular and it should be primary 
under the coset Virasoro generator of dimension $2$, fix all the
coefficients
in spin-$4$ current,
up to two unknown coefficients.
The operator product expansion of coset primary spin-$3$ field with itself 
fixes them completely.
We compute the three-point functions 
with scalars  for all values of the 't Hooft
coupling in the large $N$ limit.
At fixed 't Hooft coupling,  these three-point functions are dual to that 
found by Chang and Yin recently
in the undeformed $AdS_3$ bulk theory (higher spin gravity with matter). 

\baselineskip=18pt
\newpage
\renewcommand{\theequation}
{\arabic{section}\mbox{.}\arabic{equation}}

\section{Introduction}

It was conjectured in \cite{GG} that the large $N$ limit  of the 
$W_N$ (WZW coset) minimal model \cite{GKO85,GKO86,BBSS,BBSS1} is dual to 
a particular $AdS_3$ higher spin theory of Vasiliev  \cite{PV,PV1,Vasiliev}. 
The bulk theory has an infinite tower of massless fields with spins
$s=2, 3, \cdots $  coupled to two complex scalars. 
The higher spin Lie algebra, where a free parameter $\lambda$ is present, 
describes interactions between the higher spin fields and the scalars.
The scalars  have equal mass  determined  by the algebra, $M^2 =
-(1-\lambda^2)$. They  are quantized
with opposite boundary conditions 
and their conformal dimensions are $h_{+}= \frac{1}{2}(1+\lambda)$ and 
$h_{-}=\frac{1}{2}(1-\lambda)$.
The boundary theory is an $A_{N-1}$ (WZW coset) minimal model  
which has a higher spin $W_N$ symmetry generated by currents of 
spins $2, 3, \cdots, N$ \cite{FL}. See also \cite{BS} 
for a detailed description of $W_N$-symmetry in conformal field theory.
The theory is labeled by two positive integers $N$ and $k$
where $k$ is the level of the WZW current algebra.  
The above bulk mass parameter is identified with the boundary 
't Hooft coupling $\lambda=\frac{N}{N+k}$ which is fixed in the large
$N$ limit.

Recently, in \cite{GGHR}, the partition
function of the $W_N$ minimal model was obtained. 
Due to the fact that certain states become
null and decouple from correlation functions (and therefore have to be
removed from the spectrum), the careful limiting procedure shows that the resulting
states that survive exactly match the gravity prediction for all
values of the 't Hooft coupling. 
In this
computation, they have considered the `strict' infinite $N$ limit where the
sum of the number of boxes and antiboxes in the Young tableau 
has maximum value in the
conformal field theory partition function.
Furthermore, the 
zero mode eigenvalues (extension of the conformal dimensions) 
of the primary generators for spins $s=2,3,4$ are computed by using the 
classical Miura transformation (nonprimary basis) and transforming to
primary basis, based on the work of \cite{DIZ}. 
Similarly, the eigenvalues in bulk higher spin algebra 
by using the Drinfeld-Sokolov reduction \cite{DS}   
are also computed with an appropriate limit. 
The degenerate representations of the two algebras match.

In \cite{CY}, the three point functions with scalars at tree level in
the undeformed bulk theory (with an appropriate normalization) 
are computed and then this result is
compared to the three-point functions in the $W_N$ minimal model, in the
large $N$ limit, at 't Hooft coupling $\lambda=\frac{1}{2}$. In
particular, they have checked for spin-$3$ current and made
predictions for the three-point functions of spin $s \geq 4$, at
fixed 't Hooft coupling $\la=\frac{1}{2}$, 
in the $W_N$ coset conformal field theory. 

In this paper, we would like to compute the three-point functions for
spin-$4$ current in
the $W_N$ minimal model, in the large $N$ limit, for all values of 't
Hooft coupling.    
So, at first, one need to 
find spin-$4$ (or dimension-4) Casimir operator in the $W_N$ minimal model.
So far,  
there are no known coset generators of spin greater than $3$, although
some partial attempts to this direction can be seen from the work of 
\cite{Watts,BS}.
The main obstacles to construct the coset generators of spin greater
than $3$ are that it is not straightforward to 
find the right candidate for the coset generator. For spin-$3$
generator, there are only four independent terms that can be written
in terms of two kinds of WZW currents and then they are cubic in these currents. 
As a spin $s$ increases, the possible terms are increased very fast.
Once a candidate for the higher spin generator is constructed, then 
one should discover all the coefficient functions, by using 
both 

1) the operator product expansion of diagonal spin-$1$ current and
higher spin generator should not contain any singular terms and 

2)
the higher spin generator should transform as a primary field of
dimension-$4$ under the coset spin-$2$ Virasoro
generator.   

For the spin-$3$ case described in \cite{BBSS1,GH}, the only overall coefficient is not
determined from the above two requirements. For the spin-$4$ case in
this paper, in general, there exist two unknown coefficients which depend
on the levels and $N$. From the higher spin representations, one can
find a linear equation between these two coefficients.    
Moreover, the operator product expansion of spin-$3$ with itself   
restricts to these two coefficients. It turns out that 
all the coefficients are fixed completely 
and we are left with the spin-$4$ coset
Casimir operator described in \cite{BBSS1} (its explicit expression is
not known so far and we present them for the first time in this paper).

Although there is a mathematica package for the operator product
expansion by Thielemans \cite{Thielemans} some time ago, one should perform all the
computations by hand because one is interested in the case where the
$N$ is arbitrary. Of course, for example, one can compute the operator
product expansion for $N=4$ with $SU(4)$ group 
using his package and check whether the
calculations by hand are correct or not.
For all the computations on the operator product expansions 
in this paper, his package for $N=4$ has been used, as a
consistency check \footnote{Of course, the computations for $N=3$ are
simpler than those for $N=4$, but the former does not tell us any
nontrivial checks for the quartic Casimir operator because the $d$
symbol of rank $4$ is zero identically. For the computations that do
not involve this $d$ symbol, it is O.K. to take $N=3$ for consistency check.}. 

In section 2, after reviewing the basic contents on the two
dimensional conformal field theory in the context of WZW model,
we construct spin-$4$ Casimir operator using a completely symmetric
traceless tensor $d$ symbol of rank $4$. A generalization of
\cite{BBSS,BBSS1} to higher spin-$4$ current is described. 

In section 3, after reviewing the Goddard, Kent and Olive (GKO)
construction \cite{GKO85,GKO86}, we find coset spin-$4$ Casimir 
operator. 
 There are twenty three
independent terms.
Along the line of \cite{GH}, we describe the large $N$ limit
of this coset spin-$4$ Casimir operator.
We compute the three-point functions with scalars and compare to the
recent findings by Chang and Yin \cite{CY}.

In section 4, we summarize what we have found in this paper and
describe the relevant future works.

In the Appendices, we present all the details that are necessary to the
sections $2$, and $3$.

There exist some relevant papers \cite{Camp}-\cite{DMR} 
which are related to the work of \cite{GG} and
see also \cite{Ozer} where the quantum Miura transformation is
introduced and the spins-$3$ and $4$ primary fields are constructed
from free massless scalar fields. Of course, these fields are quantum
version of \cite{DIZ,GGHR}, but it is not clear, at the moment, 
how the eigenvalues of
a Casimir operator of order $4$ have two arbitrary constants. It would
be interesting to analyze the eigenvalues of the primary fields and to
find the precise relations with the work of \cite{GGHR} further.        

\section{The fourth-order Casimir operator of $SU(N)$}

\subsection{Review}
\label{third}

The conformal field $J(z)$ takes values in the Lie algebra $SU(N)$ and
its components $J^a(z)$($a=1, 2, \cdots, N^2-1$) 
with respect to an antihermitian basis  $T^a$
where $\mbox{Tr} (T^a T^b) = -\delta^{ab}$
satisfy the operator product expansion defined as \cite{BS}
\bea
J^a (z) J^b (w) = -\frac{1}{(z-w)^2} k \delta^{ab} +\frac{1}{(z-w)}
f^{abc} J^c(w) + \cdots,
\label{JJ}
\eea
where the positive integer $k$ is the level and $f^{abc}$
are the structure constants of $SU(N)$. 
The dots stand for the regular terms in the limit $z \rightarrow w$.

The Sugawara stress energy tensor corresponding to the second order
Casimir operator, in terms of currents $J^a(z)$,
is given by
\bea
T(z) = -\frac{1}{2(k+N)} (J^a J^a) (z),
\label{T}
\eea
where the normal ordered field product denoted by brackets \cite{BS} is assumed.
By using the operator product expansion (\ref{JJ}), one can compute 
the operator product expansion between $T(z)$ and $T(w)$ and it turns
out the well-known result 
\footnote{Notice that the singular term $\frac{1}{(z-w)^4}$ in (\ref{TT}) comes from the 
series expansions of $\frac{1}{(z-x)}$ and $\frac{1}{(z-x)^2}$ with
respect to $x$ around $w$ where $x$ is
an intermediate contour integral variable \cite{BBSS}.
This operator product expansion (\ref{TT}) can be written in terms of a
commutator relation for the modes $L_m$ where $T(z) =\sum_{m \in {\bf
    Z}} 
\frac{L_m}{z^{m+2}}$ as follows:
$[ L_m, L_n] = \oint_{C_0} \frac{d w}{2\pi i} w^{n+1} \oint_{C_w}
\frac{dz}{2\pi i} z^{m+1} T(z) T(w)=(m-n) L_{m+n} +\frac{c}{12}
m(m^2-1) \delta_{m+n,0}$ \cite{BS}.}
\bea
T(z) T(w) =\frac{1}{(z-w)^4} \frac{c}{2} +\frac{1}{(z-w)^2} 2 T(w) +
\frac{1}{(z-w)} \pa T(w) +\cdots. 
\label{TT}
\eea
The central charge $c$ is read off and it is found to be
\bea
c =\frac{k}{k+N} (N^2-1).
\label{centralcharge}
\eea

Furthermore, the current $J^a(z)$ is primary field of dimension $1$
with respect to (\ref{T}) as follows:
\bea
T(z) J^a(w) =\frac{1}{(z-w)^2} J^a(w) +\frac{1}{(z-w)} \pa J^a(w) +
\cdots.
\label{TJ}
\eea
The standard way to obtain this operator product expansion (\ref{TJ}) 
is to compute the operator product
expansion between $J^a(z)$ and $T(w)$ first. Then one reverses the
arguments $z$ and $w$ and uses the series expansion around $w$.
Note $ J^a(z) T(w) = \frac{1}{(z-w)^2} J^a(w) +\cdots$. The stress
energy tensor (\ref{T}) has dimension $2$.

The third-order Casimir operator for $SU(N)$ is defined as \cite{BBSS}
\bea
Q(z) \equiv d^{abc} (J^a (J^b J^c)) (z),
\label{Q}
\eea
where $d^{abc}$ are the $d$-symbols of $SU(N)$ and they are completely
symmetric traceless $SU(N)$-invariant tensor of rank $3$ \footnote{
From now on, we denote this normal ordered product without any
brackets, for simplicity. That is, $(J^a(J^b J^c))(z) = J^a J^b J^c(z)$.
We follow the convention for the fully normal ordered product given in \cite{BBSS}.}.
This is unique choice for the dimension 3 operator because the other 
choice $d^{abc} ((J^a J^b) J^c)(z)$ reduces to the one (\ref{Q}) above
by using the property (\ref{fd}) in this paper and the relation $(A.15)$ of \cite{BBSS}.

Let us introduce the dimension 2 operator 
\bea
Q^a (z) \equiv d^{abc} J^b J^c (z).
\label{spin2Q}
\eea
For the computation of the operator product expansion 
$Q(z) Q(w)$, it is convenient to use this
dimension 2 operator as well.
The dimension 2 operator
$f^{abc} J^b J^c(z)$ is nothing but $N \pa J^a$ which can be obtained 
by using the identities (\ref{ff}) of this paper and $(A.10)$ of \cite{BBSS}.
Then it is straightforward to 
compute the following operator product expansions from the relations 
(\ref{Q}) and (\ref{spin2Q}) \cite{BBSS}
\bea
J^a(z) Q^b(w)  & = &  -\frac{1}{(z-w)^2} (2k+N) d^{abc} J^c(w)
+\frac{1}{(z-w)} f^{abc} Q^c(w) +\cdots,
\label{JaQb}
\\
J^a (z) Q(w) & = & -\frac{1}{(z-w)^2} 3(k+N) Q^a(w) +\cdots.
\label{JQ}
\eea
We have used the properties (\ref{dff}) and (\ref{fdJacobi}) in order to 
obtain the operator product expansion (\ref{JaQb}) 
and similarly, the relations (\ref{ff}) and (\ref{fd}) are used 
for the operator product expansion (\ref{JQ}). 
One can compute the commutator, $[J_0^a, Q^b(w)]=\oint_w 
\frac{dz}{2\pi i} J^a(z) Q^b(w)=f^{abc} Q^c(w)$, where
$J_0^a$
is a Fourier component $J_0^a =\oint \frac{dz}{2\pi i} J^a(z)$ and the
operator product expansion (\ref{JaQb}) is used.
This implies that the field $Q^a(z)$ transforms under the adjoint representation
of the horizontal finite dimensional Lie algebra $SU(N)$.
Similarly, from the commutator, $[J_0^a, Q(w)]=0$, in which we used
the fact that  there is no 
$\frac{1}{(z-w)}$ term in the operator product expansion 
(\ref{JQ}), the field $Q(z)$ transforms under the
singlet of the horizontal finite Lie algebra $SU(N)$ \footnote{ 
Also the stress energy tensor 
$T(z)$ is a singlet under the horizontal (finite-dimensional)
subalgebra $SU(N)$ of $\widehat{SU}(N)$ because the operator product
expansion between $J^a(z)$ and $T(w)$ does not contain
$\frac{1}{(z-w)}$ term as before and the commutator $[J_0^a, T(w)]$ vanishes.}.

Furthermore, the fields $Q^a(z)$  and $Q(z)$ 
are primary fields of dimension 2 and 3, with respect
to the stress energy tensor (\ref{T}), respectively
\bea
T(z) Q^a(w) & = & \frac{1}{(z-w)^2} 2 Q^a(w) +\frac{1}{(z-w)} \pa Q^a(w) +
\cdots,
\label{TQa}
\\
T(z) Q(w)  & = & \frac{1}{(z-w)^2} 3 Q(w) +\frac{1}{(z-w)} \pa Q(w) +
\cdots,
\label{TQ}
\eea
where we have used the properties (\ref{d}) and (\ref{fd}) in order to
simplify the terms 
$\frac{1}{(z-w)^4}$ and $\frac{1}{(z-w)^3}$
in (\ref{TQa}) respectively.
For the computation of the operator product expansion 
(\ref{TQ}), we have used the fact that there is no singular
term in the operator product expansion  
$J^a(z) Q^a(w)$ from the relation (\ref{JaQb}) and
this is the reason why there are no higher order singular terms in the
equation (\ref{TQ}).

The singlet algebra (or Casimir algebra) 
of primary fields of dimension $3$, $Q(z)$, and dimension
$2$, $T(z)$, together with
the operator product expansions (\ref{TT}) and (\ref{TQ}),
can be obtained by computing the operator product expansion 
$Q(z) Q(w)$ explicitly \cite{BBSS}. In general, the new primary field
of dimension $4$ arises in this operator product expansion. However,
for $N=3, k=1$, the new spin $4$ primary field vanishes identically.  

\subsection{Fourth order Casimir }
\label{fourth}

How one can determine the fourth-order Casimir of $SU(N)$ with $N > 3$?
The completely symmetric traceless $d$ symbols are introduced by
Schoutens in \cite{Schoutens} (See also \cite{DP})
as follows:
\bea
d^{abcd} = d^{abe} d^{ecd} +d^{ace} d^{ebd} +d^{ade} d^{ebc}
-\frac{4(N^2-4)}{N(N^2+1)} \left( \delta^{ab} \delta^{cd} +\delta^{ac}
\delta^{bd} +\delta^{ad} \delta^{bc} \right).
\label{dabcd}
\eea
The $N$-dependent coefficient above can be fixed by requiring that 
this $d$ tensor of rank $4$ should be traceless. 
When the indices $a,b$ are equal, $b=a$, the first term of (\ref{dabcd}) vanishes due to the
property (\ref{d}), the second and third terms of (\ref{dabcd}) contribute to
$\frac{4}{N}(N^2-4)$ via the identity (\ref{dd}), the fourth term contains 
$(N^2-1)$ and the fifth and sixth terms of (\ref{dabcd}) contribute to $1$. 
Therefore, the requirement $d^{aacd}=0$ implies that the 
$N$ dependent coefficient should be $-\frac{4(N^2-4)}{N(N^2+1)}$.
The symmetric property between the indices of $d^{abcd}$ can be seen from
the relation (\ref{dabcd})
by using  the symmetry in the symmetric tensors $d^{abc}$ and $\delta^{ab}$.
The higher order $d$ tensors of rank greater than $4$ are also introduced  
in \cite{Macfarlaneetal} \footnote{
Let us comment on the notations for the $d$ symbol and $f$ symbol we
are using in the paper.
We follow the convention of \cite{BBSS}.
There are some  differences between the notations in \cite{BBSS,BBSS1} and
\cite{BS}. For example, the minus sign in the third equation of $(7.23)$
of \cite{BS} and some minus signs in the equation $(7.22)$ of \cite{BS}.
The normalization of $\sqrt{2}$ between 
\cite{Macfarlaneetal} and \cite{BBSS,BBSS1} appears.
We insert the Jacobi identity which was missed in 
\cite{BBSS,BBSS1} into the
identity
(\ref{ffJacobi}) of the Appendix $A$.
We have checked that for $N=3$, the quantity on the right hand side of
(\ref{dabcd})
is zero identically, by using the explicit $f$ tensor and $d$ tensor
elements with a factor of $\sqrt{2}$ in \cite{Macfarlaneetal}.
For the nontrivial $d$ tensor of rank $4$, the $N$ should be greater
than $3$.}.

There were some attempts, by Watts in \cite{Watts} and by Bouwknegt and
Schoutens in the preprint version of \cite{BS}, 
to the generalization for the coset spin-3
generator. It originates from the equation $(2.12)$ of \cite{TM}, but 
this identity is not valid.
For the spin-3 generator, the $N$-dependence of $\frac{1}{(z-w)^2}$
term in the operator product expansion (\ref{JaQb}) was obtained by
using the relation (\ref{dff}).
The contraction with one $d$ symbol and two $f$ symbols 
leads to other $d$ symbol, but this is not true for higher $d$ symbol. 
For example, 
the triple product $d^{abcd} f^{fga} f^{fhb}$ does not produce the
$d^{ghcd}$ tensor of rank $4$ exactly,
but there exist some extra structures.
This can be checked explicitly by taking the 4-th order $d$ symbol 
in (\ref{dabcd}) with two $f$ symbols.
So far, there are no known coset generators of spin greater than $3$. 

It is convenient to introduce the following dimension $1, 2, 3$ and
$4$ 
operators, as in spin-3 generator, along the line of \cite{BS}:
\bea
T_{abc}^{(1,3)} (z) & \equiv & d^{abcd} J^d (z),
\label{13}
\\
T_{ab}^{(2,2)} (z) & \equiv &  d^{abcd} J^c J^d (z), 
\label{22}
\\
T_a^{(3,1)} (z) & \equiv &  d^{abcd} J^b J^c J^d (z),  
\label{31}
\\
T^{(4,0)} (z) & \equiv &  d^{abcd} J^a J^b J^c J^d (z).  
\label{40}
\eea
Here $d$ symbol is given by (\ref{dabcd}).
It turns out that they are primary fields of dimension $1, 2, 3$ and
$4$
with respect to the stress energy tensor (\ref{T}):
\bea
T(z) T_{abc}^{(1,3)} (w) & = & \frac{1}{(z-w)^2} T_{abc}^{(1,3)}(w) 
+\frac{1}{(z-w)} \pa T_{abc}^{(1,3)}(w) +
\cdots,
\label{T13}
\\
T(z) T_{ab}^{(2,2)} (w) & = & \frac{1}{(z-w)^2} 2 T_{ab}^{(2,2)}(w) 
+\frac{1}{(z-w)} \pa T_{ab}^{(2,2)}(w) +
\cdots,
\label{T22}
\\
T(z) T_{a}^{(3,1)} (w) & = & \frac{1}{(z-w)^2} 3 T_{a}^{(3,1)}(w) 
+\frac{1}{(z-w)} \pa T_{a}^{(3,1)}(w) +
\cdots,
\label{T31}
\\
T(z) T^{(4,0)} (w)  & = & \frac{1}{(z-w)^2} 4 T^{(4,0)}(w) 
+\frac{1}{(z-w)} \pa T^{(4,0)}(w) +
\cdots.
\label{T40}
\eea
It is easy to see the operator product expansion 
(\ref{T13}) by using the defining equation
(\ref{TJ}).
The higher order singular terms in the operator product expansion 
(\ref{T22}) vanish due to the fact that 
the products of $d^{abcd}$ and $\delta^{cd}$ or $f^{cde}$ contribute to zero.
Also the higher order singular terms in the operator product expansion
(\ref{T31}) do not occur 
because the operator product expansion of $J^b(z) T_{ab}^{(2,2)}(w)$
does not contain any singular terms (note that
the $d^{abcd}$ symbol is symmetric and traceless).
Similarly, there are no higher order singular terms in the operator
product expansion (\ref{T40})
because the operator product expansion between $J^a(z)$ and
$T_a^{(3,1)}(w)$ does not produce any singular terms.

Now it is ready to compute the operator product expansions of
$J^a(z)$ and four primary fields (\ref{13}), (\ref{22}), (\ref{31})
and (\ref{40}),
as we did in the subsection \ref{third}.
Let us present them here as follows:
\bea
J^a (z) T_{bcd}^{(1,3)} (w)  & = & -\frac{1}{(z-w)^2} k d^{abcd} +
\frac{1}{(z-w)} f^{aef} d^{bcde} J^f(w) + \cdots,
\label{j13}
\\
J^a (z) T_{bc}^{(2,2)} (w)  & = & \frac{1}{(z-w)^2}  \left[-2k d^{abcd}
  J^d(w) + f^{adf} f^{feg} d^{bcde} J^g(w) \right]
\nonu \\
&+& \frac{1}{(z-w)}  \left[ f^{adf} d^{bcde} J^f J^e(w) +
f^{aef} d^{bcde} J^d J^f(w) \right] +\cdots,
\label{j22}
\\
J^a (z) T_{b}^{(3,1)} (w)  & = &  \frac{1}{(z-w)^3}  f^{acf} d^{bcde}
f^{geh} f^{fdg} J^h(w) 
 + \frac{1}{(z-w)^2} \left[ - 3k d^{abcd} J^c J^d(w)
\right. \nonu \\ 
& + & \left.
f^{acf} f^{fdg} d^{bcde} J^g J^e(w) +
f^{acf} f^{feg} d^{bcde} J^d J^g(w) + 
f^{adf} f^{feg} d^{bcde} J^c J^g(w) \right]
\nonu \\
&+&  \frac{1}{(z-w)} \left[ f^{acf} d^{bcde} J^f J^d J^e(w) + 
f^{adf} d^{bcde} J^c J^f J^e(w) +
f^{aef} d^{bcde} J^c J^d J^f(w) \right] \nonu \\
& + &
\cdots, 
\label{JandJJJ} 
\\
J^a (z) T^{(4,0)} (w)  & = & \frac{1}{(z-w)^4} f^{abf} f^{fci} d^{bcde} f^{geh} f^{idg} 
J^h(w) \nonu \\
& + & \frac{1}{(z-w)^3} \left[ f^{abf} d^{bcde} ( f^{hdg} f^{fch} J^g
  J^e(w) + 
f^{heg} f^{fch} J^d J^g(w) +
  f^{heg} f^{fdh} J^c J^g(w) ) \right. \nonu \\
&+ & \left. f^{acf} d^{bcde} f^{geh} f^{fdg} J^b J^h(w) \right] \nonu \\
&+&  \frac{1}{(z-w)^2} \left[-4k d^{abcd} J^b J^c J^d(w) +
f^{abf} f^{fcg} d^{bcde} J^g J^d J^e(w) \right. \nonu \\
& + &  f^{abf} f^{fdg} d^{bcde} J^c J^g J^e(w) +
f^{abf} f^{feg} d^{bcde} J^c J^d J^g(w) + f^{acf} f^{fdg} d^{bcde} J^b
J^g J^e(w) 
\nonu \\
& +& \left. f^{acf} f^{feg} d^{bcde} J^b J^d J^g(w) + 
f^{adf} f^{feg} d^{bcde} J^b J^c J^g(w) \right]
+\cdots. 
\label{JandJJJJ}
\eea
It is easy to see the operator product expansion 
(\ref{j13}) from the defining equation (\ref{JJ}).
There is no $\frac{1}{(z-w)^3}$ term in the operator product expansion 
(\ref{j22}) because the product of
$d^{bcde}$ and $f^{ade}$ vanishes.

In the operator product expansion 
(\ref{JandJJJ}), the result of the operator product expansion (\ref{j22}) is used and
the $k$-dependent term in the singular term of $\frac{1}{(z-w)^3}$  
vanishes due to the fact that 
the product of $d^{fbcd}$ and $f^{acf}$ becomes zero. 
One need to simplify the singular term of 
$\frac{1}{(z-w)^3}$ further in the operator product expansion (\ref{JandJJJ}). 
In order to simplify this, one should use the $d$
symbol in (\ref{dabcd}). By plugging the first term of (\ref{dabcd})
into the above singular term, $fdff$ term, one realizes that $d f f$ term can be
reexpressed in terms of a single $d$ term via the identity (\ref{dff}). Then this
reduces to the product of $d d f $ and this leads to a single $f$ term
from the relation (\ref{ddf}). Finally, one obtains $-N d^{bha}$
coming from the first term of (\ref{dabcd}) \footnote{
Here we have ignored the index structures of $d$ and $f$ symbols for simplicity.}.  
Now let us consider the second term in (\ref{dabcd}). In the above
singular term, this provides $d d f f $ term  which can be further
reduced to the relation (\ref{ffdd}). With the remaining $f$ in
$\frac{1}{(z-w)^3}$ term, it leads to a single
$f$ term and $d d f$ terms. These $d d f $ terms can be simplified
further by using the relation (\ref{ddf}). It turns out that 
the final expression coming
from the second term of (\ref{dabcd}) is
equal to zero. For the third term of (\ref{dabcd}), the singular term
can be written similarly. It gives $-(N^2-4) f^{abh}$. Finally, the
$\delta$ terms in (\ref{dabcd}) can be reduced to $f f f $ terms and
these can be written in terms of a single $f$ terms by the identity (\ref{fff}).

By collecting all the contributions we have obtained so far, the
$\frac{1}{(z-w)^3}$ term in the operator product expansion 
(\ref{JandJJJ}) can be summarized by $-\frac{2(N^2-4)(N^2-9)}{(N^2+1)}
f^{abc} J^c(w)$. The other lower singular terms are simplified further and the
details for these computations are presented in the Appendix $B$.
The complete operator product expansion (\ref{JandJJJ}) 
is presented in (\ref{JT31detail}).

Moreover, in the operator product expansion 
(\ref{JandJJJJ}), the result of the operator product expansion 
(\ref{JandJJJ}) is also used
in the $\frac{1}{(z-w)^4}$ and $\frac{1}{(z-w)^3}$ terms.
Let us consider the highest singular term, $\frac{1}{(z-w)^4}$ term. 
By substituting the $d$
symbol (\ref{dabcd}) into there, one has $f f d d f f $ terms.
One uses the identity (\ref{dff}) and obtains
$2N(N^2-4) \delta^{ah}$ coming from the first term of (\ref{dabcd}). 
The second term of (\ref{dabcd}) with other
four $f$ symbols can be simplified via the identity (\ref{ffdd}) with 
the relation (\ref{ddf}). It turns out that 
there is no contribution. The third term of 
(\ref{dabcd}) gives a nonzero contribution. We again use the identity
(\ref{ffdd})
with (\ref{ddf}) and (\ref{ff}) and the final expression is
$2N(N^2-4)\delta^{ah}$.
The remaining terms of (\ref{dabcd}) contribute to
$-\frac{40N(N^2-4)}{(N^2+1)}$
where the identities (\ref{ff}) and (\ref{fff}) are used.  
Then, one obtains  $\frac{4N(N^2-4)(N^2-9)}{(N^2+1)} J^a(w)$ in
$\frac{1}{(z-w)^4}$ term.

Let us consider the next higher order singular term,
$\frac{1}{(z-w)^3}$. 
The first term of
$d$ symbol in (\ref{dabcd}) contributes to $(N^2-4) f^{eag}$ via the
identities (\ref{dff}) and (\ref{ddf}).
There is no contribution from the second term of $d$ symbol where the
identities (\ref{ffdd}) and (\ref{ddf}) are used. The third term of (\ref{dabcd})
gives 
$(N^2-4)f^{age}$ via the relations 
(\ref{ffdd}) and (\ref{ddf}). The delta terms of 
$d$ symbol can be simplified, through the identities (\ref{fff}) and (\ref{ff}),  
as $5N f^{age}$ with prefactor $-\frac{4(N^2-4)}{N(N^2+1)}$.  
By summing over all the contributions, one obtains
$\frac{2N(N^2-4)(N^2-9)}{(N^2+1)} \pa J^a(w)$ in $\frac{1}{(z-w)^3}$.
The next term can be arranged as the sum of $f d f f $ term above 
and  $f d f f f $ term by moving the current
$J^g$ to the left. This last term can be
further simplified as $-\frac{4N(N^2-4)(N^2-9)}{(N^2+1)} \pa J^a(w)$.  
The $J^c J^g$ term is the same as the $J^d J^g$ term above by
interchanging
between the indices $c$ and $d$ that are dummy indices.
The $J^b J^h$ term in $\frac{1}{(z-w)^3}$ can be written as 
the sum of above first contribution and 
$-\frac{4N(N^2-4)(N^2-9)}{(N^2+1)} \pa J^a(w)$ term.  

Therefore, one has the final
term as $-\frac{4N(N^2-4)(N^2-9)}{(N^2+1)} \pa J^a(w)$ in the  $\frac{1}{(z-w)^3}$.
The other terms in lower singular terms are simplified further and the
details for these computations are also presented in the Appendix $B$.
The full operator product expansion (\ref{JandJJJJ}) is 
presented in (\ref{JT40detail}).
There are no $\frac{1}{(z-w)}$ terms in the operator product expansion
(\ref{JandJJJJ}) for $N=4$. We expect this is also true for any $N >5$.
The singlet field $T^{(4,0)}(z)$ has a vanishing commutator 
$[J_0^a, T^{(4,0)}(w)]=0$ under $SU(N)$ because there is no
$\frac{1}{(z-w)}$ term in the operator product expansion (\ref{JandJJJJ}).

The higher order singular terms, $\frac{1}{(z-w)^4}$ term and
$\frac{1}{(z-w)^3}$ term, 
in the operator product expansions 
(\ref{JandJJJ}) and (\ref{JandJJJJ}) 
allow us to consider the extra terms for the coset spin $4$ current in the
coset construction which will be discussed in next section. 
We also present the operator product expansions between $J^a(z)$ and
other spin $4$ fields in the Appendix $C$. 

We have seen that the field $T^{(4,0)}$ is primary field of dimension
$4$ from the operator product expansion (\ref{T40}). However,
for general spin $4$ field, the condition for primary field under the 
stress energy tensor (\ref{T}) is not straightforward. 
Sometimes the higher order singular terms arise in the operator
product expansion with the stress energy tensor $T(z)$.
For example, one can think of spin $4$ field $Q^a Q^a(z)$ which 
is quadratic in dimension $2$ field $Q^a(z)$ (\ref{spin2Q}).
In order to compute  the operator product expansion $T(z) Q^a Q^a(w)$ explicitly, 
one should obtain the operator product expansion $Q^a(z) Q^a(w)$ and
more generally, one has the following operator product expansion with
different indices:
\bea
Q^a (z) Q^b (w) & = & \frac{1}{(z-w)^4} \frac{2}{N}(N^2-4) k (N+2k) \delta^{ab}
- \frac{1}{(z-w)^3} \frac{2}{N}(N^2-4)(N+2k) f^{abc} J^c(w)
\nonu \\
&-& \frac{1}{(z-w)^2} \left[ N d^{abc} Q^c(w) +2(N+2k) d^{ace} d^{bde}
  J^c J^d(w) 
\right] \nonu \\
& + & \frac{1}{(z-w)} 
\left[ -(N+2k) d^{ace} d^{bde} \pa J^c J^d(w) + f^{ace} d^{bcd}
  Q^e J^d(w) + f^{ade} d^{bcd} J^c Q^e(w) \right. \nonu \\
& - & \left.  (N+2k) d^{ace} d^{bde} J^d \pa J^c(w) \right] 
+ \cdots. 
\label{QQ}
\eea
Here one uses the identity $(A.8)$ of \cite{BBSS} with the operator product
expansion (\ref{JaQb}) and obtains the operator product expansion 
$Q^a(z) J^c(w)$. In order to have the complete expression for the 
higher order singular terms, one uses the relations
(\ref{dd}) and (\ref{ddf}). 

It is ready to write down the operator product expansion 
$T(z)$ with $Q^aQ^a(w)$ and it turns out to be
\bea
T(z) Q^a Q^a (w) & = & \frac{1}{(z-w)^6} \frac{12k}{N}(N^2-4)(N^2-1)(N+2k)
\nonu \\
&+ &
\frac{1}{(z-w)^4} \frac{32}{N}(N^2-4)(N+2k)(N+k) T(w)\nonu \\
& + &
\frac{1}{(z-w)^3} \frac{12}{N}(N^2-4)(N+2k)(N+k) \pa T(w)
\nonu \\
& + & \frac{1}{(z-w)^2} 4 Q^a Q^a(w) +
\frac{1}{(z-w)} \pa (Q^a Q^a)(w) + \cdots.
\label{TQQ}
\eea
The higher order singular terms in the operator product expansion 
(\ref{TQQ}) are determined by
the operator product expansion (\ref{QQ}) and 
the identity (\ref{dd}).
These higher order terms will play an important role in next section.
We should find the correct candidate for the coset spin $4$ field
which transforms as a primary field of dimension $4$ under the coset
Virasoro field. In other words, the higher order singular terms
$\frac{1}{(z-w)^n}$ 
with $n > 2$ in the
operator product expansion between the coset Virasoro field and the
coset spin $4$ field should vanish. 
We also present the operator product expansions between $T(z)$ and
the other spin $4$ fields in the Appendix $C$. 

The singlet algebra (or Casimir algebra) 
of primary fields $T^{(4,0)}(z), Q(z)$, and $T(z)$,
can be obtained, in principle, by computing the operator product expansions 
$Q(z) T^{(4,0)}(w)$  and $T^{(4,0)}(z) T^{(4,0)}(w)$ explicitly. 
In general, we expect that the new primary fields will
arise in these operator product expansions. 
For particular values of $N,k$, these new primary fields will vanish
and the algebra reduces to \cite{BFKNRV,KW}, along the line of \cite{BBSS}. 
Of course, it would be interesting to perform all the operator product
expansions to see how the algebra is different from the algebra in 
\cite{BFKNRV,KW}, but we are interested in a minimal model conformal
field theory. 
In next section, we want to construct a new primary field of dimension
$4$, based on the operator product expansions we have described 
in this section. 

Therefore, we have seen the fourth-order Casimir operator 
$d^{abcd} J^a J^b J^c J^d(z)$ which will play an important role in
next section.

\section{The GKO coset construction}

\subsection{Review}

Let us consider the diagonal coset WZW model given by
\bea
\frac{\widehat{SU}(N)_k \oplus \widehat{SU}(N)_1}{\widehat{SU}(N)_{k+1}}.
\label{coset}
\eea
Denoting the spin 1 current fields as $K^a(z)$ and $J^a(z)$ of level
$k$ and $1$ respectively and the spin 1 current field as 
$J'^a(z)$ of level $k+1$,
the operator product expansion between
$J^a(z)$ and $J^b(w)$ is
\bea
J^a (z) J^b (w) = -\frac{1}{(z-w)^2} k_1 \delta^{ab} +\frac{1}{(z-w)}
f^{abc} J^c(w) + \cdots,
\label{JJcoset}
\eea
where the level is characterized by $k_1(\equiv 1)$
and the other operator product expansion between $K^a(z)$ and $K^b(w)$
is
\bea
K^a (z) K^b (w) = -\frac{1}{(z-w)^2} k_2 \delta^{ab} +\frac{1}{(z-w)}
f^{abc} K^c(w) + \cdots,
\label{KKcoset}
\eea
where the level is given by $k_2(\equiv k)$. 
We follow the convention of \cite{GH}.
These two currents are independent
in the sense that the operator product expansion 
$J^a(z) K^b(w)=0$. That is, there are no singular terms.
The diagonal current is given by
\bea
J'^a (z) = J^a (z) +K^a (z).
\label{diag}
\eea
Then it is easy to check, by adding the operator product expansions (\ref{JJcoset}) 
and (\ref{KKcoset}), that 
\bea
J'^a (z) J'^b (w) = -\frac{1}{(z-w)^2} k' \delta^{ab} +\frac{1}{(z-w)}
f^{abc} J'^c(w) + \cdots, \qquad k' \equiv k_1 +k_2 \equiv k+1.
\label{opej'j'}
\eea

The stress energy tensor for the coset model (\ref{coset}) is given by 
\bea
\widetilde{T} (z) = T_{(1)}(z) +T_{(2)}(z) - T'(z),
\label{Ttilde}
\eea
where the stress energy tensors in terms of the spin 1 currents are as follows: 
\bea
T_{(1)}(z) & = &  -\frac{1}{2(k_1+N)} (J^a J^a) (z),
\nonu \\
T_{(2)}(z) & = &  -\frac{1}{2(k_2+N)} (K^a K^a) (z),
\nonu \\
T'(z) & = &   -\frac{1}{2(k_1 + k_2 +N)} (J'^a J'^a) (z).
\label{stressetal}
\eea
It is easy to check that 
the operator product expansion 
between $J'^a(z)$ and $\widetilde{T}(w)$ has no singular term
because the $\frac{1}{(z-w)^2}$ term is given by $(J^a(w) + K^a(w) -J'^a(w))$
which is identically zero from (\ref{diag}). 
One computes the operator product expansion  
$\widetilde{T}(z) \widetilde{T}(w)$ that is equivalent to 
the operator product expansion 
$\widetilde{T}(z) (T_{(1)} +T_{(2)})(w)$ because there is no singular
term in the operator product expansion 
between $\widetilde{T}(z)$ and $T'(w)$, by construction. Then by substituting
(\ref{Ttilde}) into the operator product expansion $\widetilde{T}(z) (T_{(1)} +T_{(2)})(w)$, 
one has $T_{(1)}(z) T_{(1)}(w) +T_{(2)}(z) T_{(2)}(w)-T'(z)T'(w)$.
Here we used the fact that 
the operator product expansion of $T'(z) (T_{(1)} +T_{(2)})(w)$
can be rewritten as $T'(z)(\widetilde{T}(w) +T'(w))$ and this becomes
$T'(z) T'(w)$. 

Therefore, one obtains \footnote{ The
stress energy tensor $T_{(1)}(z)$ has a central charge
$c_{(1)}=(N^2-1)\frac{k_1}{N+k_1}$ (\ref{centralcharge}),
the stress energy tensor $T_{(2)}(z)$ has a central charge
$c_{(2)}=(N^2-1)\frac{k_2}{N+k_2}$, and 
the stress energy tensor $T'(z)$ has a central charge
$c'=(N^2-1)\frac{k_1+k_2}{N+k_1+k_2}$ with (\ref{opej'j'}).}
\bea 
\widetilde{T}(z) \widetilde{T}(w) & = &
T_{(1)}(z) T_{(1)}(w) +T_{(2)}(z) T_{(2)}(w)-T'(z)T'(w),
\nonu \\
& = & \frac{1}{(z-w)^4} \frac{\widetilde{c}}{2} +\frac{1}{(z-w)^2} 2 \widetilde{T}(w) +
\frac{1}{(z-w)} \pa \widetilde{T}(w) +\cdots.
\label{cosettt}
\eea 
Finally, after reading off the $\frac{1}{(z-w)^4}$ terms in (\ref{cosettt}),
one has the following coset central charge as follows:
\bea
\widetilde{c} & = & c_{(1)} +c_{(2)} -c' =
(N^2-1) \left( \frac{k_1}{k_1+N} +\frac{k_2}{k_2+N} -
\frac{k_1+k_2}{k_1 +k_2 +N} \right) 
\nonu \\
& \equiv & (N-1) \left[
1-\frac{N(N+1)}{(N+k)(N+k+1)}
\right],
\label{ctilde}
\eea
where we put $k_1=1, k_2=k$.

It is known that the coset spin $3$ primary field
$\widetilde{T}^{(3)}(z)$ 
is found in \cite{BBSS1}. 
Together with (\ref{cosettt}), 
the operator product expansion of
$\widetilde{T}(z)\widetilde{T}^{(3)}(z)$
satisfies the standard one for the primary field of dimension $3$ and 
the operator product expansion of $\widetilde{T}^{(3)}(z) \widetilde{T}^{(3)}(w)$ 
provides the full structure of the operator algebra. The 
spin $3$ field $\widetilde{T}^{(3)}(z)$ consists of four terms which
can be written as the spin $1$ currents $J^a(z), K^a(z)$. The
requirement that  the field $\widetilde{T}^{(3)}(z)$ is a primary
field of dimension $3$ under the $\widetilde{T}(z)$ fixes
$\widetilde{T}^{(3)}(z)$
up to a normalization factor which depends on $N,k_1$ and $k_2$.

\subsection{Coset primary spin-4 current $\widetilde{W}(z)$}
\label{spin4section}

Along the line of coset spin 3 current \cite{BBSS1}, one 
can think of the following fields as a candidate for the coset spin 4 current 
\bea
d^{abcd} \left( c_1  J^a J^b J^c J^d +
c_2  J^a J^b J^c K^d +
c_3  J^a J^b K^c K^d +
c_4  J^a K^b K^c K^d+
c_5 K^a K^b K^c K^d \right),
\label{five}
\eea
where the coefficient functions $c_i(i=1, 2, \cdots 5)$ depend on $N,
k_1$ and $k_2$.
We simply generalize the field $T^{(4,0)}(z)$ to 
have both spin $1$ currents, $J^a(z)$ and $K^a(z)$. 
How does one determine the coset spin 4 primary field?
As we described in the introduction,
first of all, the coset spin 4 primary field should commmute with the 
diagonal current (\ref{diag}) as follows:
\bea
J'^a(z) \widetilde{W}(w) = \mbox{regular}.
\label{regular}
\eea
Secondly, the coset spin 4 primary field should transform
as dimension 4 under the stress energy tensor (\ref{Ttilde}) 
as follows:
\bea
\widetilde{T}(z) \widetilde{W}(w) = \frac{1}{(z-w)^2} 4
\widetilde{W}(w) +\frac{1}{(z-w)} \pa \widetilde{W}(w)
+\cdots.
\label{tprimary}
\eea

One can easily see that if there exist only five candidate terms as in
(\ref{five}), then there are no consistent coefficient functions
satisfying the conditions (\ref{regular}) by checking the operator
product expansions given in the Appendix (\ref{JJJJ})-(\ref{KKKK}). 
This is one indication why the construction in \cite{Watts} where the
field contents are given by (\ref{five}) is not correct.
Therefore, one should consider the extra terms in order to satisfy
(\ref{regular})
by requiring that they contribute to 
the vanishing of any singular terms.
Since the $d$ symbol of rank $4$ is defined as (\ref{dabcd}),
one also considers the following candidates for the coset spin 4
primary field by looking at the first term of (\ref{dabcd})
\bea
d^{abe} d^{cde}( c_6 J^a J^b J^c J^d+  
c_7   J^a J^b J^c K^d+
c_8  J^a J^b K^c K^d + c_9  J^a K^b K^c K^d + 
c_{10}  K^a K^b K^c K^d ).
\label{anotherfive}
\eea
For given the first term in (\ref{anotherfive}), we 
simply consider the possible terms with the current $K^a(z)$ as we did
in (\ref{five}).
Then it is ready to check whether 
the sum of (\ref{five}) and (\ref{anotherfive})
is consistent with the condition (\ref{regular}) or not. 
It turns out that they do not satisfy the regularity condition (\ref{regular}).
This implies that we need to have  
more terms and we will have a chance to
remove all the singular terms.
The delta term in (\ref{dabcd}) gives us to allow the following terms
\bea
c_{11} J^a J^a J^b J^b +
c_{12} J^a J^a J^b K^b + 
c_{13} J^a J^a K^b K^b  +
c_{14} J^a K^a K^b K^b +
c_{15} K^a K^a K^b K^b.
\label{Five}
\eea
For given first term in (\ref{Five}), we 
also consider the possible terms by realizing the other current
$K^a(z)$.
Still the contractions in the indices are fixed.
Moreover, one should also consider the following terms
\bea
&& c_{16} \pa^2 J^a J^a+
c_{17} \pa^2 J^a K^a+
c_{18} \pa^2 K^a K^a+
c_{19} \pa J^a \pa J^a + c_{20} \pa J^a \pa K^a +
c_{21} \pa K^a \pa K^a \nonu \\
&& +
c_{22} J^a \pa^2 K^a.
\label{anotherFive}
\eea
These terms appear in the second derivative of coset stress energy
tensor (\ref{stressetal}).

By adding the terms (\ref{five}), (\ref{anotherfive}), (\ref{Five}), 
and (\ref{anotherFive}), one writes down the following coset spin 4 current 
\bea
\widetilde{W}(z) & = &  d^{abcd} \left( c_1  J^a J^b J^c J^d +
c_2  J^a J^b J^c K^d +
c_3  J^a J^b K^c K^d
+
c_4  J^a K^b K^c K^d
+ 
c_5  K^a K^b K^c K^d \right) \nonu \\
& + &   d^{abe} d^{cde} \left( c_6 J^a J^b J^c J^d+  
c_7   J^a J^b J^c K^d+
c_8   J^a J^b K^c K^d
+   c_9  J^a K^b K^c K^d + 
c_{10}  K^a K^b K^c K^d \right)
\nonu \\
& +&  c_{11} J^a J^a J^b J^b +
c_{12} J^a J^a J^b K^b 
+
c_{13} J^a J^a K^b K^b  +
c_{14} J^a K^a K^b K^b +
c_{15} K^a K^a K^b K^b 
\nonu \\
&+ &  c_{16} \pa^2 J^a J^a+
c_{17} \pa^2 J^a K^a 
+
c_{18} \pa^2 K^a K^a+
c_{19} \pa J^a \pa J^a + c_{20} \pa J^a \pa K^a \nonu \\
& + &
c_{21} \pa K^a \pa K^a + 
c_{22} J^a \pa^2 K^a.
\label{cosetspin4}
\eea
We also checked whether the other contractions with different indices
provide other independent terms we are missing or not in the Appendix $G$. 
In particular, the equations (\ref{independent}), (\ref{indepen1}),
(\ref{g3}) 
and (\ref{g4}).
It turns out that there are no other independent terms except the
field $J^a J^b K^a K^b(z)$. We will discuss on the presence of this
field later.
The field contents in (\ref{cosetspin4}) are complete except the field
$J^a J^b K^a K^b(z)$ in this sense.
Due to the fact that
there exists an identity
\bea
&& d^{abcd} J^a J^b J^c J^d(z)   =  3 d^{abe} d^{cde} J^a J^b J^c J^d(z)
-\frac{12(N^2-4)}{N(N^2+1)}
J^a J^a J^b J^b(z) \nonu \\
&& -  \frac{3(N^2-4)(N^2-3)}{N^2+1}  
\pa J^a \pa J^a(z)
+\frac{2(N^2-4)(N^2-3)}{N^2+1} \pa^2 J^a J^a(z),
\label{dddd}
\eea
and the terms appearing in the 
coefficients $c_1, c_6, c_{11}, c_{16}$ and $c_{19}$ (and similarly
the coefficients $c_5, c_{10}, c_{15}, c_{18}$ and $c_{21}$) are not
independent. For example, this allows us to put $c_{19}=0=c_{21}$ from
the beginning
in order to keep the quartic term on the left hand side of (\ref{dddd}).

Let us apply the condition (\ref{regular}) to the coset spin 4 current 
(\ref{cosetspin4}). In Appendix D, we present the twenty two operator
product expansions (\ref{JJJJ})-(\ref{JJJJend}). 
Next we consider the property of (\ref{tprimary}) on the coset spin 4
current.
Since we require the condition (\ref{regular}) and the stress energy
tensor $T'(z)$ is given by (\ref{stressetal}), one has 
no singular terms in the operator product expansion $T'(z)$ and $\widetilde{W}(w)$.  
This implies that we introduce 
\bea
T_{(1)}(z) + T_{(2)}(z) \equiv \hat{T}(z),
\label{That}
\eea
and consider the operator product expansion with $\widetilde{W}(w)$.
We collect the operator product expansions in the Appendix $E$:
(\ref{cosetJJJJfin})-(\ref{finfin}).
Then the twenty three linear equations from the vanishing of singular
terms
in the operator product expansions $J'^a(z) \widetilde{W}(w)$
are summarized in the equations 
(\ref{rel1}), (\ref{rel2}) and (\ref{rel3}) of Appendix $F$. 
Similarly, the eight linear equations from the vanishing of higher
order singular terms (i.e., $\frac{1}{(z-w)^n}$ term with $n > 2$) 
in the operator product expansions $\hat{T}(z) 
\widetilde{W}(w)$
are given in the equations (\ref{rel4}), (\ref{rel5}) and
(\ref{relfin}) of the Appendix $F$.
By solving these equations, the coefficients appearing in the coset
spin 4 current (\ref{cosetspin4}) are determined  except the $c_1$
coefficient
and they are in the
equations (\ref{coeff})
and (\ref{dcoeff}) of the Appendix $F$.  
The large $N$ limit for these coefficients is given in (\ref{limitc})
that will be used in next subsection. 

\subsection{ Coset primary spin-$4$ current 
$\widetilde{W}(z)$ in the large $N$ 't Hooft limit}
\label{firstlimit}

The large $N$ 't Hooft limit \cite{GG} is defined as 
\bea
N, k \rightarrow \infty, \qquad \lambda \equiv \frac{N}{N+k}
\qquad \mbox{fixed}.
\label{limit}
\eea
Let us describe the large $N$ 't Hooft limit for the 
coset spin 4 current (\ref{cosetspin4}) with the coefficients
(\ref{coeff}) and (\ref{dcoeff}).
In this limit, 
the term proportional to $ d^{abcd} K^a K^b K^c K^d(z)$, $c_5$, behaves as
$\frac{1}{N}$ and goes to zero.
Similarly, the term with $d^{abe} d^{cde} K^a K^b K^c K^d(z)$, $c_{10}$, 
behaves as $\frac{1}{N}$ also and vanishes.  
The terms with $c_{11}, c_{12} $ vanish, the terms  with $c_{13}, c_{14}$ have 
$\frac{1}{N}$ dependence and the term with $c_{15}$
has $\frac{1}{N^2}$ dependence. See the relations (\ref{limitc}) for details.
Then, all of these vanish in the large $N$ limit.
One obtains the coset spin 4 current, in the large $N$ 't Hooft limit,
together with the unknown coefficient $c_1$, 
as follows:
\bea
\widetilde{W}(z) &= & \left( d^{abcd} \left[ J^a J^b J^c J^d(z) +
\frac{4\lambda(5\lambda+4)}{3(\lambda-1)(\lambda+2)} 
 J^a J^b J^c K^d(z) +
\frac{2\lambda^2(5\lambda+7)}{(\lambda-1)^2(\lambda+2)} 
 J^a J^b K^c K^d(z) \right. \right.
\nonu \\
&+& \left. 
\frac{20\lambda^2(\lambda+1)}{3(\lambda-3)(\lambda-1)(\lambda+2)} 
 J^a K^b K^c K^d(z) \right]
+ d^{abe} d^{cde} \left[ \frac{2\lambda}{\lambda+2}  J^a J^b J^c J^d(z) 
\right. \nonu \\
&+&   
\frac{8\lambda}{(\lambda-1)(\lambda+2)}   J^a J^b J^c
K^d(z)
 +
\frac{12\lambda^2(\lambda+3)}{(\lambda-2)(\lambda-1)^2(\lambda+2)} 
  J^a J^b K^c K^d(z) 
\nonu \\ 
&+& \left.
\frac{40\lambda^2(\lambda+1)}{(\lambda-3)(\lambda-2)(\lambda-1)(\lambda+2)} 
 J^a K^b K^c K^d(z)  \right] 
-  N^2 \left[
\frac{4\lambda}{3(\lambda+2)} \pa^2 J^a K^a(z)
\right.
\nonu \\
&- & \left. \left.
\frac{4\lambda (\lambda+1)}{(\lambda-1)(\lambda+2)} \pa J^a \pa K^a(z) 
 +   
 \frac{4\lambda(\lambda+1)}{3(\lambda-1)(\lambda+2)} 
J^a \pa^2 K^a(z) \right] \right) c_1.
\label{WWWW}
\eea
At first sight, the last three terms become very large compared to the
rest but this is not true.
We will see that the $N$ dependence of each term behaves equally.   
Let us focus on the zero mode of spin 4 current. 
By realizing the equation (\ref{dddd}), one can reexpress the zero
mode as follows:
\bea
d^{abcd} J_0^a J_0^b J_0^c J_0^d & \rightarrow &  
3 d^{abe} d^{cde} J_0^a J_0^b J_0^c J_0^d
-  3N^2 (\pa J^a)_0 (\pa J^a)_0
+2N^2 (\pa^2 J^a)_0 (J^a)_0
\nonu \\
& = & 3 d^{abe} d^{cde} J_0^a J_0^b J_0^c J_0^d + N^2 J_0^a J_0^a,
\label{conditi1}
\eea
where the Laurent mode expansion is used \footnote{
That is, $J^a(z)=\sum_{m \in \bf{Z}} 
\frac{J_m^a}{z^{m+1}}$, $\pa J^a(z) =- \sum_{m \in \bf{Z}} 
(m+1) \frac{J_m^a}{z^{m+2}}$ and $\pa^2 J^a(z) = \sum_{m \in \bf{Z}} 
(m+1)(m+2) \frac{J_m^a}{z^{m+3}}$.}.
The contribution from $J^a_0 J^a_0 J^b_0 J^b_0$
is removed under (\ref{limit}) because the behavior of this term
in (\ref{dddd}) depends on $N=\frac{1}{N} \times N^2$. 
The zero mode of (\ref{WWWW}) can be written as 
\bea
\widetilde{W}_0 & = &  \left[ -\frac{(18+15\lambda+111\lambda^2-
7\lambda^3 +7 \lambda^4)}{3(\lambda-3)(\lambda-1)^2(\lambda+2)} 
d^{abcd}  J_0^a J_0^b J_0^c J_0^d \right.
\nonu \\
&+ &
 \frac{2\la(30-95\lambda+41\lambda^2-25\lambda^3+\lambda^4)}
{(\lambda-3)(\lambda-2)(\lambda-1)^2(\lambda+2)} 
d^{abe} d^{cde}  J_0^a J_0^b J_0^c J_0^d 
\nonu \\
& + & \left. N^2 \frac{4\lambda(\lambda-3)}{3(\lambda-1)(\lambda+2)} J^a_0
J^a_0 \right] c_1, 
\label{conditi2}
\eea
where the singlet condition $K_0^a =-J_0^a$ is used.
Note that, in \cite{GG}, the second element of highest weight representations 
corresponds to the diagonal current $J'^a(z) =
J^a(z) + K^a(z)$ and the first element corresponds
to the current $K^a(z)$. We are looking for the eigenvalue equation
of (\ref{conditi2}) acting on the representation $(f; 0)\otimes (f;
0)$ where $J_0^a +K_0^a =0$. 
The first term in (\ref{conditi2}) 
is the sum of the first four terms in (\ref{WWWW}) with alternative signs.
Similarly, the second term in (\ref{conditi2}) 
is the sum of the next four terms in (\ref{WWWW}) with alternative
signs
and finally, the third term in (\ref{conditi2}) 
is the sum of the last three terms in (\ref{WWWW}) with appropriate
signs and multiplicities.

One should compute the spin $4$ zero mode (\ref{conditi2}) 
on the fundamental representation. 
The product of $SU(N)$ generators we are using in this paper 
has the following decompositions 
with $\delta, d$ and $f$ symbols
\bea
T^a T^b = -\frac{1}{N} \delta^{ab} -\frac{i}{2} d^{abc} T^c
+\frac{1}{2} f^{abc} T^c.
\label{TT1}
\eea
From the relation (\ref{TT1}), one obtains the quartic product with two 
$d$ symbols. 
Among nine terms, the nonzero five terms (due to the traceless
condition for antihermitian basis) can be obtained and using the 
property (\ref{fd}), one finds 
\bea
d^{abe} d^{cde} \mbox{Tr} (T^a T^b T^c T^d) & = & 
\frac{1}{4} d^{abe} d^{cde} d^{abf} d^{cdf} 
=\frac{1}{4} \frac{2}{N} (N^2-4) \frac{2}{N} (N^2-4) \delta^{aa}
\nonu \\
&= & \frac{1}{N^2}(N^2-4)^2(N^2-1),
\label{trace}
\eea
where the property (\ref{dd}) is used.
Since we want to obtain the eigenvalue not the trace as in
(\ref{trace}),
we should divide this by $N$.
Then, under (\ref{limit}), one arrives at the zero mode that appears
in the second term in (\ref{conditi2}) acting on the fundamental
representation, 
\bea
d^{abe} d^{cde} J_0^a J_0^b J_0^c J_0^d | f > = N^3 |f >.
\label{conditi3}
\eea 
Similarly, the zero mode that appears
in the last term in (\ref{conditi2}) acts on the fundamental
representation, 
\bea
\delta^{ab} \mbox{Tr} (T^a T^b) =
-\delta^{aa} = -(N^2-1) \rightarrow J_0^a J_0^a | f > = -N | f >.
\label{conditi4}
\eea
One also has the same expressions (\ref{conditi3}) and
(\ref{conditi4})
on the antifundamental representation.

By using the equations (\ref{conditi1}), (\ref{conditi3}) and
(\ref{conditi4}),
one simplifies the right hand side of (\ref{conditi2})
and surprisingly, it turns out that it is very simple factorized form 
\bea
&&  N^3 \left[  -\frac{(18+15\lambda+111\lambda^2-
7\lambda^3 +7 \lambda^4)}{3(\lambda-3)(\lambda-1)^2(\lambda+2)} (3-1)
+ \frac{2\la(30-95\lambda+41\lambda^2-25\lambda^3+\lambda^4)}
{(\lambda-3)(\lambda-2)(\lambda-1)^2(\lambda+2)} \right. \nonu \\
&& -\left.
\frac{4\lambda(\lambda-3)}{3(\lambda-1)(\lambda+2)} \right] c_1
= N^3 \left[ 
\frac{4(1+\la)^2(3+\la)}{(3-\lambda)(2-\lambda)(1-\lambda)}  
\right] c_1.
\label{above}
\eea
For the primary $ ( f ; 0) \otimes ( f ; 0)$ where 
$J_0^a + K_0^a =0$, from the result of 
(\ref{above}), one can
compute 
\bea
\widetilde{W}_0 | {\cal{O}}_{+} > & = &  N^3 \left[ 
\frac{4(1+\la)^2(3+\la)}{(3-\lambda)(2-\lambda)(1-\lambda)}  
\right] c_1 | {\cal{O}}_{+} >, \qquad {\cal{O}}_{+} \equiv 
( f ; 0) \otimes ( f ; 0)
\label{eigen1}
\eea
and for the primary $( 0; f) \otimes (0; f)$ where $K_0^a=0$,
the spin $4$ current acts as
\bea
\widetilde{W}_0 | {\cal{O}}_{-} > & = &   N^3 \left[
\frac{4(1+\lambda)}{(2+\lambda)} \right]
c_1 | {\cal{O}}_{-} >, \qquad {\cal{O}}_{-} \equiv 
( 0; f ) \otimes ( 0; f ),
\label{eigen2}
\eea
where one uses $\left[3-1 + \frac{2\la}{\la+2} \right]N^3=  N^3 \left[
\frac{4(1+\lambda)}{(2+\lambda)} \right]
c_1$ that corresponds to the equation (\ref{conditi2})
by choosing the first and fifth terms in (\ref{WWWW}).
For the other primaries $  \overline{{\cal{O}}}_{+} \equiv
( \overline{f} ; 0) \otimes ( \overline{f} ; 0)$ and  
 $ \overline{{\cal{O}}}_{-} \equiv
( 0; \overline{f}) \otimes (0; \overline{f})$, one has similar
 relations.
The generators in the antifundamental representation 
has an extra minus sign, compared to the generators in the fundamental
representation, but this does not affect the relations (\ref{trace})
and 
(\ref{conditi4}) because the number of the power of $T^a$ is even.
Then we have same equations with (\ref{eigen1}) and (\ref{eigen2}) by
replacing $ |{\cal{O}}_{\pm}>$ with $|\overline{ {\cal{O}}}_{\pm}>$ respectively.   
There are no sign changes in the eigenvalues, contrary to the spin $3$
case \cite{GH,GGHR} where the number of the power of $T^a$ is odd.
According to the normalization of \cite{GGHR}, 
the eigenvalues of the spin $4$ zero mode acting on the above
primaries
appear as follows: 
\bea
U_0^4 | {\cal{O}}_{+} > & = & (1+\lambda)(2+\lambda)(3+\lambda) |
{\cal{O}}_{+} >,  \qquad {\cal{O}}_{+} \equiv 
( f ; 0) \otimes ( f ; 0),
\nonu \\
U_0^4  | {\cal{O}}_{-} > & = &
(1-\lambda)(2-\lambda)(
3-\lambda) | {\cal{O}}_{-} >, \qquad  {\cal{O}}_{-} \equiv 
( 0; f ) \otimes ( 0; f ).
\label{Eigen}
\eea
The different normalization in \cite{GH} is used. 
For given relations (\ref{eigen1}) and (\ref{eigen2}), it is 
possible to determine the unknown coefficient function $c_1(N,k)$ 
in order to satisfy the condition (\ref{Eigen}) as follows:
\bea
c_1(N,\la) =\frac{(1-\la)(2-\la)(3-\la)}{\left[N^3 \frac{4(1+\la)}{(2+\la)} \right]}.
\label{c1condition}
\eea
This relation (\ref{c1condition}) is a generalization of 
$(4.24)$ of \cite{GH} for the higher spin greater than $3$. 
According to (\ref{conditi3}), the factor $N^3$ comes from the quartic
Casimir for the fundamental representation.
For the spin $4$ case in our paper, the eigenvalue equation has other
factor from (\ref{eigen2}). In other words, 
the $\widetilde{W}(z)$ in the large $N$ limit has $\la$ dependent term
in $c_6$ coefficient:$\frac{2\la}{\la+2}$ from (\ref{limitc}).
Combined with the constant piece, the overall factor depends on the 't
Hooft coupling constant $\la$ explicitly.
Therefore,  we have 
\bea
\widetilde{W}_0 | {\cal{O}}_{+} > & = &   
(1+\lambda)(2+\lambda)(3+\lambda)
| {\cal{O}}_{+} >, \qquad {\cal{O}}_{+} \equiv 
( f ; 0) \otimes ( f ; 0), \nonu \\
\widetilde{W}_0 | {\cal{O}}_{-} > & = &   
(1-\lambda)(2-\lambda)(3-\lambda)
| {\cal{O}}_{-} >, \qquad  {\cal{O}}_{-} \equiv 
( 0; f ) \otimes ( 0; f ).
\label{EEigeneigen}
\eea
In next subsection, 
we would like to look for more general coset spin
$4$ field which satisfies the above condition (\ref{Eigen}).

Following the analysis by Chang and Yin \cite{CY}, from the action of 
$\widetilde{W}_0$ on the primary states, (\ref{EEigeneigen}), 
the three-point function with scalars is summarized as 
\bea
<\overline{\cal{O}}_{+} {\cal{O}}_{+} \widetilde{W}> & = & 
(1+\lambda)(2+\lambda)(3+\lambda), \qquad  \overline{{\cal{O}}}_{+} \equiv 
( \overline{f} ; 0) \otimes ( \overline{f} ; 0), \nonu \\
<\overline{\cal{O}}_{-} {\cal{O}}_{-} \widetilde{W}> & = & 
(1-\lambda)(2-\lambda)(3-\lambda), \qquad
 \overline{{\cal{O}}}_{-} \equiv 
( 0; \overline{f}) \otimes ( 0; \overline{f}).
\label{three}
\eea
Note that in \cite{CY}, the normalization for the spin $3$ current is
the same as that in \cite{GH} \footnote{
In the notation of \cite{GGHR}, the three-point function for the spin
$3$ current is $<\overline{\cal{O}}_{+} {\cal{O}}_{+} \widetilde{T}^{(3)}>  =  
-(1+\lambda)(2+\lambda)$ and $
<\overline{\cal{O}}_{-} {\cal{O}}_{-} \widetilde{T}^{(3)}>  =  
(1-\lambda)(2-\lambda)$ corresponding to the equation $(5.14)$ of \cite{CY}.}.
As a spin increases, the extra $\la$ dependent factors occur in the
three-point functions.

How does one compare the three-point function (\ref{three}) to 
that from the bulk computation in \cite{CY}?
With the normalization $<J^{(s)}(z) J^{(s)}(w)>=1$, 
the three-point function at $\lambda=\frac{1}{2}$ is given by 
\bea
<\overline{\cal{O}}_{+} {\cal{O}}_{+} J^{(s)}> = N^{-\frac{1}{2}}
\Gamma(s) \sqrt{\frac{2s-1}{\Gamma(2s-1)}}, \qquad
<\overline{\cal{O}}_{-} {\cal{O}}_{-} J^{(s)}> = (-1)^s N^{-\frac{1}{2}}
 \frac{\Gamma(s)}{\sqrt{\Gamma(2s)}}.
\label{threeCY}
\eea
In our case, we did not compute the operator product expansion
$\widetilde{W}(z) \widetilde{W}(w)$ explicitly where
$\widetilde{W}(z)$ is given by (\ref{WWWW}).
We will have 
$
<\widetilde{W}(z) \widetilde{W}(w) > =
A(N,k) + (\frac{1}{N} \, \mbox{correction})$ by computing 
the highest singular terms $\frac{1}{(z-w)^8}$.
Let us divide the two relations (\ref{three}). Then we do not have to
worry about the normalization for $\widetilde{W}(z)$.
\bea
\frac{<\overline{\cal{O}}_{+} {\cal{O}}_{+} \widetilde{W}>}{
<\overline{\cal{O}}_{-} {\cal{O}}_{-} \widetilde{W}>}
=\frac{(1+\lambda)(2+\lambda)
(3+\lambda)}{(1-\lambda)(2-\lambda)(3-\lambda)}
\rightarrow \left(\frac{3+\lambda}{1-\lambda} \right)|_{\lambda=\frac{1}{2}} = 7,
\label{7}
\eea
where we put $\lambda=\frac{1}{2}$ at the final stage.
Note that the factor $(1+\lambda)(2+\lambda)$ in the numerator cancels the 
factor $(2-\lambda)(3-\lambda)$ in the denominator at $\lambda=\frac{1}{2}$.
On the other hand, by taking the ratio in (\ref{threeCY}),
one obtains
\bea
\frac{<\overline{\cal{O}}_{+} {\cal{O}}_{+}
  J^{(s)}>}{<\overline{\cal{O}}_{-} 
{\cal{O}}_{-} J^{(s)}>} = (-1)^s(2s-1). 
\label{77}
\eea
For $s=4$, this relation (\ref{77}) is exactly the same as the relation (\ref{7})
\footnote{ This feature occurs for $s=2, 3$ case. For $s=3$, 
the ratio $\frac{<\overline{\cal{O}}_{+} {\cal{O}}_{+} \widetilde{T}^{(3)}>}{
<\overline{\cal{O}}_{-} {\cal{O}}_{-} \widetilde{T}^{(3)}>}
=-\frac{(1+\lambda)(2+\lambda)}
{(1-\lambda)(2-\lambda)} \rightarrow -\left(\frac{2+\lambda}{1-\lambda}
  \right)|_{\lambda=\frac{1}{2}}=-5$ which is the same as (\ref{77})
  for $s=3$. Similarly, for $s=2$, we have $
\frac{<\overline{\cal{O}}_{+} {\cal{O}}_{+} \widetilde{T}^{(2)}>}{
<\overline{\cal{O}}_{-} {\cal{O}}_{-} \widetilde{T}^{(2)}>}
=\frac{\frac{1}{2}(1+\lambda)}
{\frac{1}{2}(1-\lambda)} \rightarrow \left(\frac{1+\lambda}{1-\lambda}
  \right)|_{\lambda=\frac{1}{2}}=3$ which coincides with (\ref{77})
  for $s=2$.}. 
One can continue to analyze for higher spin greater than $4$.
From
$<\overline{\cal{O}}_{+} {\cal{O}}_{+} \widetilde{W}^{(s)}>  =  
(-1)^s (1+\lambda)(2+\lambda)(3+\lambda) \cdots (s-1+\lambda)$
and $
<\overline{\cal{O}}_{-} {\cal{O}}_{-} \widetilde{W}^{(s)}>  =  
(1-\lambda)(2-\lambda)(3-\lambda) \cdots (s-1-\lambda)$,
one takes the ratio and realizes that the factors in the numerator
cancel the factors in the denominator  at $\lambda=\frac{1}{2}$,
\bea
\frac{<\overline{\cal{O}}_{+} {\cal{O}}_{+} \widetilde{W}^{(s)}>}{
<\overline{\cal{O}}_{-} {\cal{O}}_{-} \widetilde{W}^{(s)}>}
& = & \frac{(-1)^s (1+\lambda)(2+\lambda)
(3+\lambda) \cdots (s-1+\lambda)}{(1-\lambda)(2-\lambda)(3-\lambda) \cdots (s-1-\lambda)}
\nonu \\
& \rightarrow & (-1)^s
\left(\frac{s-1+\lambda}{1-\lambda} \right)|_{\lambda=\frac{1}{2}} =
(-1)^s (2s-1).
\label{eqeq}
\eea
This relation (\ref{eqeq}) is the same as the relation (\ref{77}).
It would be interesting to construct the three-point functions in the
deformed $AdS_3$ bulk theory, as mentioned in \cite{CY}, 
that generalize to the equations (\ref{threeCY})  
and compare to the three-point functions (\ref{three}) for all values
of 't Hooft coupling constant in the $W_N$
coset conformal field theory in the large $N$ 't Hooft limit.

\subsection{The general coset primary 
spin-4 current $\widetilde{W}^{(4)}(z)$ and its
  large  $N$ 't Hooft limit}
\label{other}

More generally, one can consider the coset spin $4$ field that 
contains the field $\widetilde{T}^2(z)$ given by (\ref{tt}).
Or one can understand that according to the second equation of (\ref{indepen1}),
if we write down the field $d^{abcd} J^a J^b K^c K^d(z)$ using the
identity (\ref{dabcd}), then there exists an independent field 
$J^a J^b K^a K^b(z)$ appears naturally.
The only new term from the relation (\ref{tt}), compared to the
relation (\ref{cosetspin4}), is 
$J^a J^b K^a K^b(z)$
and let us add this term with the coefficient function $c_{23}$ to the
previous coset spin $4$ field (\ref{cosetspin4}):  
\bea
\widetilde{W}^{(4)}(z)  = \widetilde{W}(z) + c_{23} J^a J^b K^a K^b(z).
\label{newW}
\eea
Then one should have the above two properties (\ref{regular}) and
(\ref{tprimary})
for the new field (\ref{newW}).
That is, $J'^a(z) \widetilde{W}^{(4)}(w) = \mbox{regular}$
and $\widetilde{W}^{(4)}(z)$ is a primary field of dimension $4$ under
the stress energy tensor (\ref{Ttilde}) with (\ref{stressetal}).
From the relations (\ref{jprimeeq}) and (\ref{tprimeeq}),
one should include $c_{23}$ dependent terms into (\ref{rel1})-(\ref{relfin}).
It turns out that the more general spin $4$ coset field 
is
\bea
\widetilde{W}^{(4)}(z) & = &  d^{abcd} \left[ c_1  J^a J^b J^c J^d +
(c_2+b_2)  J^a J^b J^c K^d +
(c_3+b_3)  J^a J^b K^c K^d
+
(c_4+b_4)  J^a K^b K^c K^d
\right.
\nonu \\
& +&  \left. 
(c_5+b_5)  K^a K^b K^c K^d \right] 
 +    d^{abe} d^{cde} \left[ (c_6+b_6) J^a J^b J^c J^d+  
(c_7+b_7)   J^a J^b J^c K^d \right. 
\nonu \\
& + & \left.
(c_8+b_8)   J^a J^b K^c K^d
+   (c_9+b_9)  J^a K^b K^c K^d + 
( c_{10} +b_{10})  K^a K^b K^c K^d \right]
\nonu \\
& + &  (c_{11}+b_{11}) J^a J^a J^b J^b +
( c_{12}+b_{12}) J^a J^a J^b K^b 
 + 
(c_{13}+b_{13}) J^a J^a K^b K^b  \nonu \\
& + &
(c_{14}+b_{14})  J^a K^a K^b K^b 
+
(c_{15}+ b_{15}) K^a K^a K^b K^b 
+
(c_{17}+b_{17}) \pa^2 J^a K^a \nonu \\
 & + & (c_{20}+b_{20}) \pa J^a \pa K^a   
 +  
(c_{22} +b_{22}) J^a \pa^2 K^a +
c_{23} J^a J^b K^a K^b,
\label{generalW}
\eea
where the coefficients $c_i$ are given by the equations (\ref{coeff}) 
and the coefficients $b_i$ are given by 
the equations (\ref{b}).
In order to see how the modification arises, we keep the old
coefficient functions $c_i$ and present the extra  terms characterized
by the coefficients $b_i$.
The two unknown coefficient functions $c_1$ and $c_{23}$ cannot be
fixed by the above requirements.

In Appendix $I$, we have found the coset spin $4$ field discussed in
\cite{BBSS1} explicitly.
From the beginning, the coset spin $3$ field, $\widetilde{T}^{(3)}$, is
completely determined by the regularity condition (\ref{regular}) and
the condition for the primary field corresponding to 
(\ref{tprimary}) except the overall
factor which is fixed by the highest singular term, in the operator product expansion
$\widetilde{T}^{(3)}(z) \widetilde{T}^{(3)}(w)$, that behaves as
$\frac{1}{3} \widetilde{c}$ with the relation (\ref{ctilde}).
By focusing on the $\frac{1}{(z-w)^2}$ terms in the above operator
product expansion, one obtains a new primary coset spin $4$ field
explicitly (this is not known so far) and 
it is given by the equations (\ref{r4}), (\ref{dcoeff1}),
$\widetilde{\Lambda}(w)$ and $\pa^2 \widetilde{T}(w)$.
Alternatively, it is easy to see that this new primary field
$\widetilde{R}^{(4)}(z)$
in the Appendix $I$ is nothing but the $\widetilde{W}^{(4)}(z)$ in
(\ref{generalW})
with fixed $c_1$ and $c_{23}$ that are given on (\ref{c1c23}) and (\ref{ddefinition}).
This is summarized in the relation (\ref{r4w4}).
Note that 
the field $J^a J^b K^a K^b(z)$ arises in the equation (\ref{r4}) very naturally.
That is, it comes from the $d_{23}$ term and $\widetilde{T}^2(w)$ term.

In the large $N$ limit (\ref{limit}), 
the term proportional to $ d^{abcd} K^a K^b K^c K^d(z)$ behaves as
$\frac{1}{N}$ and goes to zero.
Similarly, the term with $d^{abe} d^{cde} K^a K^b K^c K^d(z)$ 
behaves as $\frac{1}{N}$ also and vanishes.  
The terms    with $b_{13}, b_{14}$ have 
$\frac{1}{N}$ dependence and the term with $b_{15}$
has $\frac{1}{N^2}$ dependence. See the equations (\ref{limitb}) for details.
Then, all of these vanish in the large $N$ limit.
The large $N$ 't Hooft limit of (\ref{WWWW}) is generalized to the
following expression,  from (\ref{limitb}) and large $N$ limits 
of $\widetilde{\Lambda}(z)$ and $\pa^2 \widetilde{T}(z)$,
as follows:
\bea
\widetilde{W}^{(4)}(z)  & = &   \widetilde{W}(z) +
\left( d^{abcd} \left[ -
\frac{(-1+\lambda)(1+\lambda)}{2\lambda(2+\lambda)}   
J^a J^b J^c K^d(z)  
-\frac{3(1+\lambda)}{8(2+\lambda)}   J^a J^b K^c K^d(z)
\right. \right.
\nonu \\
&- & \left.   \frac{5(-1+\lambda)(1+\lambda)}{12(-3+\lambda)(2+\lambda)} 
  J^a K^b K^c K^d(z)  \right] 
+    d^{abe} d^{cde} \left[  
- \frac{(-1+\lambda)^2(1+\lambda)}
{4\lambda^2(2+\lambda)}  J^a J^b J^c J^d(z)
\right.  \nonu \\
&+ &    
\frac{(-1+\lambda)(1+\lambda)}{2\lambda(2+\lambda)}    J^a J^b J^c K^d(z)  
- \frac{(1+\lambda)(1+2\lambda)}{4(-2+\lambda)(2+\lambda)}    J^a J^b K^c K^d(z)
\nonu \\
&-& \left.  
\frac{5(-1+\lambda)(1+\lambda)}{2(-3+\lambda)
(-2+\lambda)(2+\lambda)}   J^a K^b K^c K^d(z)  \right] 
+ \frac{(-1+\la)^2}{4\la^2} J^a J^a J^b J^b(z)
\nonu \\
& + & \frac{(-1+\la)}{\la} J^a J^a J^b K^b(z)
 +  J^a J^b K^a K^b(z)
+ N^2 \left[ 
   \frac{(-1+\lambda)(1+\lambda)}
{4\lambda(2+\lambda)} 
 \pa^2 J^a K^a(z)    
\right.  \nonu \\
&- & \left. \left.
 \frac{(-1+\lambda)(1+\lambda)}{4\lambda(2+\lambda)}   \pa J^a \pa K^a(z)   
 +  
 \frac{(-1+\lambda)(1+\lambda)}{12\lambda(2+\lambda)} 
  J^a \pa^2 K^a(z) \right]  \right) c_{23}.
\label{limitW}
\eea
One can compute the contribution from $J^a J^b K^a K^b(z)$ by
realizing that the trace $\mbox{Tr} (T^a T^b T^a T^b)$ with (\ref{TT1}) 
can be calculated. There is no $N^2$ term and the next leading term
behaves as $\frac{1}{N}$.  
In other words, the field $J^a J^b K^a K^b(z)$ itself does not
contribute to the final answer, but its presence does affect all the
other coefficient functions. 
The terms with $c_{23}$ in the relation (\ref{limitW}) 
also appear in the relation (\ref{WWWW}).
Therefore, in (\ref{limitW}), the independent terms are the same as
the one in (\ref{WWWW}) with different coefficient functions.   

By taking the same analysis in the previous subsection
\ref{firstlimit} 
(i.e., (\ref{conditi3}) and 
(\ref{conditi4})),
the eigenvalue (\ref{eigen1}) acting on $(f;0) \otimes (f;0)$ is
generalized to  
\bea
\widetilde{W}_0^{(4)} | {\cal{O}}_{+} > & = &  
N^3 \frac{(1+\la)^2(3+\la)}{(1-\la)(2-\la)(3-\la)}
\left[4 c_1 -\frac{(1-\la)^2}{4\la^2} c_{23} 
\right]| {\cal{O}}_{+} >.
\label{Eeigen1}
\eea
Similarly, one has the following generalized expression on $(0;f)\otimes (0;f)$
\bea
\widetilde{W}_0^{(4)} | {\cal{O}}_{-} > & = & 
N^3 \frac{(1+\la)}{(2+\la)} \left[ 4  c_1 -\frac{(1-\la)^2}{4\la^2} c_{23}
\right]
| {\cal{O}}_{-} >.
\label{Eeigen2}
\eea
By choosing the two undetermined coefficients $c_1(N,k)$ and
$c_{23}(N,k)$,
\bea
c_1(N,\la)   =   0,
\qquad
c_{23}(N,\la)   =   -\frac{(1-\la)(2-\la)(3-\la)}{
\left[ N^3 \frac{(1-\lambda)^2(1+\lambda)}
{4\lambda^2(2+\lambda)} \right]}, 
\label{123}
\eea
and plugging (\ref{123}) into (\ref{Eeigen1}) and (\ref{Eeigen2}) 
one has the following relations,
\bea
\widetilde{W}_0^{(4)} | {\cal{O}}_{+} > & = &   
(1+\lambda)(2+\lambda)(3+\lambda)
| {\cal{O}}_{+} >, \nonu \\
\widetilde{W}_0^{(4)} | {\cal{O}}_{-} > & = &   
(1-\lambda)(2-\lambda)(3-\lambda)
| {\cal{O}}_{-} >,
\label{EEigen}
\eea
which are exactly the same as the equations (\ref{Eigen}).
In this case, there is no 
$d^{abcd} J^a J^b J^c J^d(z)$ term because $c_1(N,\la)=0$. 
The similar relations hold for $| \overline{{\cal{O}}}_{\pm} >$ as we
explained before.

More generally, as long as the following relation,  
\bea
4  c_1(N,\la)  -\frac{(1-\la)^2}{4\la^2} c_{23}(N,\la)
=\frac{(1-\la)(2-\la)(3-\la)}{\left[N^3 \frac{(1+\la)}{(2+\la)}
  \right]},
\label{c1c23sol}
\eea
holds, the above equation (\ref{EEigen}) is satisfied.
The case with (\ref{c1condition}) and the case with (\ref{123}) are the
particular solutions of the relation (\ref{c1c23sol}).
Recall that the denominator in (\ref{c1c23sol}) is exactly the same as
the overall factor on the right hand side of (\ref{Eeigen2}) that
comes from the $d^{abe} d^{cde} J^a J^b J^c J^d(z)$ and other terms in
$\widetilde{W}(z)$ we have discussed before. 
It seems that this behavior can be generalized to the higher spin
greater than $4$ and the right hand side of (\ref{Eeigen2})  for
$\widetilde{W}^{(s)}_0$ 
should behave
as $\prod_{i=1}^{s-1}(i-\la)$ where $s$ is a spin. Then the right hand
side of (\ref{Eeigen1}) for $\widetilde{W}^{(s)}$ can be fixed automatically.  

For the field $\widetilde{W}^{(4)}(z)$, the only
requirements we used so far are given in $1)$ and $2)$ in the introduction and it
turns out that there are two undetermined coefficients with the
constraint (\ref{c1c23sol}). 
This implies that these requirements are not enough to fix them completely.
Are there any ways to determine these two undetermined coefficients?
Recall that the field contents of $W_N$ minimal model are given by the
spin-$2$ field, the spin-$3$ field, the spin-$4$ field and so on
 \footnote{ 
The vacuum character, the equation $(7.18)$ of \cite{BS} or the
equation $(2.18)$
of \cite{GG},  is given by
$
\frac{1}{\prod_{s=2}^{N} \prod_{n=s}^{\infty} (1-q^n)} =
1+q^2+2 q^3+4 q^4+6 q^5+12 q^6+ {\cal O }(q^7).
$
The $4$ in front of $q^4$  indicates that there
are four spin-$4$ fields, $\widetilde{T}^2(z), \pa^2 \widetilde{T}(z),
\pa \widetilde{T}^{(3)}(z)$ and $\widetilde{W}^{(4)}(z)$.
Among these four fields, the only primary field is
$\widetilde{W}^{(4)}(z)$ up to normalization. 
We thank the referee for pointing out this observation.}.
The operator product expansion of primary spin-$3$ field with itself
should provide further constraint on the above spin-$4$ field.

From the result of the Appendix $I$,
one can think of the following spin $4$ primary field  
\bea
\alpha(N,k) \, \widetilde{R}^{(4)}(z),
\label{newnewW}
\eea
where $\widetilde{R}^{(4)}(z)$ is the spin-$4$ primary field
described in \cite{BBSS1} or in the Appendix $I$ and
$\alpha(N,k)$ is an arbitrary constant which depends on $N,k$.
One takes the large $N$ 't Hooft limit for (\ref{c1c23}) and
(\ref{ddefinition}) (or (\ref{c1c23limit}))
and then using the equation (\ref{limitW}) or 
the equations (\ref{Eeigen1}) and (\ref{Eeigen2}),
one obtains
\bea
\widetilde{R}_0^{(4)} | {\cal{O}}_{+} > & = &   
\frac{2(1+\lambda)(3+\lambda)}{5(2-\lambda)}
| {\cal{O}}_{+} >, \nonu \\
\widetilde{R}_0^{(4)} | {\cal{O}}_{-} > & = &   
\frac{2(1-\lambda)(3-\lambda)}{5(2+\lambda)}
| {\cal{O}}_{-} >.
\label{Reigen}
\eea
In this case, the eigenvalues are very symmetric in the 't Hooft
coupling $\la$. The first equation in (\ref{Reigen}) 
goes to the second equation in (\ref{Reigen}) by taking $\la
\rightarrow -\la$ and vice versa.
Then it is straightforward to compute the three-point functions from
the relations (\ref{Eeigen1}) and (\ref{Eeigen2}) or  the equation 
(\ref{Reigen}).
For $
\alpha(N,k) = \frac{5}{2} (2-\la)(2+\la)$, one has 
the same equations as (\ref{EEigen}) for the zero mode of
(\ref{newnewW}). 
One can easily see that the $c_1$ and $c_{23}$ in (\ref{c1c23limit})
multiplied by the $\alpha(N,k)$ satisfy the above relation
(\ref{c1c23sol}).
This is an expected result from the identity (\ref{r4w4}). 
Note that the field $\widetilde{R}^{(4)}(z)$ is obtained from the operator
product expansion of two spin-$3$ fields and all the coefficient
functions are known. 
Once the highest singular term
$\frac{1}{(z-w)^8}$ in the operator product expansion of spin-$4$
field with itself, in the large $N$, 
is computed, then one can normalize such that 
the two-point function is equal to $1$ and this will fix the overall constant
completely.

\section{Conclusions and outlook }

We have found the coset primary spin-$4$ field
$\widetilde{W}^{(4)}(z)$ 
in (\ref{generalW})
together with  the coefficient functions (\ref{coeff}), (\ref{dcoeff})
and (\ref{b})
where  the two coefficient functions 
$c_1$ and $c_{23}$ are given by the equations (\ref{c1c23}).
The large $N$ limit of (\ref{generalW}) 
satisfies the zero mode acting on the primary states (\ref{EEigen}).
Furthermore, the three-point functions for $\widetilde{W}^{(4)}(z)$ with
the scalars take the form (\ref{three}) and they are dual to the
three-point functions in \cite{CY}.  

$\bullet$ How does one perform the operator algebra consisting of spins-$2, 3$
and $4$? Among the possible six operator product expansions, one of
them is given by the operator product expansion (\ref{cosettt}). There
are also two operator product expansions, (\ref{tprimary}) and similar
operator product expansion for the spin $3$ field
$\widetilde{T}^{(3)}(z)$ (\ref{t3tilde}).  
The operator product expansion  $\widetilde{T}^{(3)}(z)
\widetilde{T}^{(3)}(w)$ is known in \cite{BBSS1}.
Then there are two unknown operator product expansions 
$\widetilde{T}^{(3)}(z) \widetilde{W}^{(4)}(w)$ and 
$\widetilde{W}^{(4)}(z) \widetilde{W}^{(4)}(w)$. 
It would be interesting to find these and to see how they differ from
the algebras found in \cite{BFKNRV,KW} although this work will be tedious.
Furthermore, $\frac{1}{(z-w)^3}$ terms of the operator product
expansion $\widetilde{W}^{(4)}(z) \widetilde{W}^{(4)}(w)$ or 
$\frac{1}{(z-w)^2}$ terms of the operator product
expansion $\widetilde{T}^{(3)}(z) \widetilde{W}^{(4)}(w)$ will provide
the structure of spin-$5$ field.
See also the works in \cite{Bouw,HT,Zhang} for the operator product algebra between
spins-$2, 4$.

$\bullet$ It would be interesting to find 
the zero modes on all other representations.
The question is how one can write down the generators for these higher
order representations?
For example, one can act the zero mode on 
the representation $(\mbox{adj}; 0)$ that appears in the
fusion of $(f; 0)$ and $(\overline{f}; 0)$. In this case, the element of
the generator $T^a$ in the adjoint representation for 
$SU(N)$ can be written as $(T^a)_{bc}=f^{abc}$. Then the
trace $\mbox{Tr} (T^a T^b T^c T^d)$ is nothing but the quartic in $f$
symbols.
Using the identity (\ref{quarticf}) in the Appendix $A$, one can simplify (\ref{trace})
as $2N^3$ in the large $N$ limit 
where one should use the identity (\ref{cubicd}). Similarly, 
the equation (\ref{conditi4}) becomes $-2N$. 
This reflects an overall extra factor $2$ in this representation.
This implies that 
one has
$\widetilde{W}_0^{(4)} | (\mbox{adj};0) \otimes (\mbox{adj}; 0)  >  =    
2(1+\lambda)(2+\lambda)(3+\lambda)
|(\mbox{adj}; 0) \otimes (\mbox{adj}; 0) >$ and 
$\widetilde{W}_0^{(4)} | (0; \mbox{adj}) \otimes (0; \mbox{adj})  >  =    
2(1-\lambda)(2-\lambda)(3-\lambda)
|(0; \mbox{adj}) \otimes (0; \mbox{adj}) >$.  
These can be seen from the equations (\ref{EEigen}) and their
eigenvalue equations for $ | \overline{{\cal{O}}}_{\pm} >$.
Since they have same eigenvalues (as we mentioned before), 
this is compatible with the above 
observation \footnote{ 
This feature looks similar to the eigenvalue equations of spin-$3$
zero mode on the higher order representations
$(\frac{N(N-1)}{2}; 0)$ or $(0; \frac{N(N-1)}{2})$.
Recall that in spin-$3$ case, the zero mode eigenvalues of
these states  $(\mbox{adj}; 0)$  or 
 $(0; \mbox{adj})$ vanish due to the fact that the $d$ symbol of rank $3$
contracted with three $f$'s is equal to zero according to the identity
(\ref{fff}).}.

$\bullet$ The coset primary spin-$4$ field is written in terms of 
WZW currents with finite $N, k$-dependent coefficient functions  
in the $W_N$ minimal model. How does this behavior appear in the
higher spin gravity with matter in $AdS_3$ theory?

$\bullet$ It would be interesting to find the supersymmetric theory
of \cite{GG}
where there exist a field of dimension $\frac{3}{2}$ \cite{ASS}, a field 
of dimension $\frac{5}{2}$ \cite{ASS,Ahn1106} and 
probably a field of dimension $\frac{7}{2}$.
The first one and the stress energy tensor consists of the usual
${\cal N}=1$ super stress energy tensor, the second one and the dimension $3$
primary field consists of ${\cal N}=1$ super field of dimension
$\frac{5}{2}$
and the third one and the dimension $4$ primary field 
gives another ${\cal N}=1$
super field of dimension $\frac{7}{2}$. 
One way to see the field contents of dimension $\frac{7}{2}$ is to compute the
operator product expansion of the field of dimension $\frac{5}{2}$ and
a field of dimension $3$ by focusing on $\frac{1}{(z-w)^2}$ terms.

$\bullet$ Can one generalize to the higher coset Casimir operator of spin
greater than $4$ and more generally of arbitrary spin $N$? 
As we described above, we should construct 
the completely symmetric traceless $d$ symbol
of rank $5$ which generalizes the $d$ symbol (\ref{dabcd}).
The possible terms for $d$ symbol of rank $5$ are  $d^{abcf} d^{fde}$
or
$d^{abc} \delta^{de}$ \cite{Macfarlaneetal}.

$\bullet$ 
It would be interesting to find the coset spin-$4$ primary field in
the other types of minimal models \cite{Ahn1106,GV}.
In doing this, one should understand higher order invariant symmetric
polynomials for $D_{\frac{N}{2}}, B_{\frac{N-1}{2}}$ that can be
written in terms of the second lowest symmetric invariant polynomial
for  $D_{\frac{N}{2}}, B_{\frac{N-1}{2}}$ \cite{Macfarlaneetal}. 
It will be useful to consider the coset model for fixed $N$
\cite{Ahn1992} in which there is a half integer spin as well as an
integer spin.

$\bullet$
As described in the introduction, the eigenvalues of the zero modes of
spins $2,3,4$, and $5$ on the highest weight states of Fock space of
$SU(N)$ Lie algebras  were
constructed in \cite{Ozer} using the quantum Miura transformation with
Feigin-Fuchs type of free massless scalar fields. It would be
interesting to find any relations between the findings in \cite{Ozer}
and the results of \cite{GGHR} where the classical Miura
transformation was used. The strategy for this classical limit \cite{BW}
is as
follows. At the classical level, any composite field term (product of
$n$ fields) on the right
hand side of operator product expansion should have its denominator
proportional to $(n-1)$-th power of $c$. Therefore, as we take $c
\rightarrow \infty$ in quantized operator product expansion,  
only those terms survive and any composite fields that do not satisfy
this $c$-dependence will disappear in the classical limit.
 
\vspace{.7cm}

\centerline{\bf Acknowledgments}

We would like to thank the following people for 
correspondence on the following topics: C.-M. Chang on three point function, 
T. Hartman on the zero modes of
degenerate representations, K. Thielemans on his mathematica package
for operator product expansions, and G. Watts
on higher spin generalization. 
This work was supported by the Mid-career Researcher Program through
the National Research Foundation of Korea (NRF) grant 
funded by the Korean government (MEST) (No. 2009-0084601).
CA acknowledges warm hospitality and partial support from 
the Department of Physics, Princeton University (I.R. Klebanov).

\newpage

\appendix

\renewcommand{\thesection}{\large \bf \mbox{Appendix~}\Alph{section}}
\renewcommand{\theequation}{\Alph{section}\mbox{.}\arabic{equation}}

\section{Some properties of $f$ and $d$ tensors of $SU(N)$ }

We list some properties between $d^{abc}$ and $f^{abc}$ symbols as follows:
\bea
d^{aab} & = & 0,
\label{d}
\\
f^{abc} f^{dbc} & = & 2N \delta^{ad}, 
\label{ff}
\\
f^{abc} d^{dbc} & = & 0,
\label{fd}
\\
d^{abc} d^{dbc} & = & \frac{2}{N}(N^2-4) \delta^{ad}, 
\label{dd}
\\
f^{adb} f^{bec} f^{cfa} & = & - N f^{def}, 
\label{fff}
\\ 
d^{adb} f^{bec} f^{cfa} & = & - N d^{def},
\label{dff}
\\
d^{adb} d^{bec} f^{cfa} & = & \frac{(N^2-4)}{N} f^{def},
\label{ddf}
\\
d^{adb} d^{bec} d^{cfa} & = & \frac{(N^2-12)}{N} d^{def},
\label{cubicd}
\\
f^{ade} f^{ebc} + f^{bde} f^{eca} + f^{cde} f^{eab}  &= & 0,
\\
f^{ade} d^{ebc} + f^{bde} d^{eca} + f^{cde} d^{eab}  &= & 0,
\label{fdJacobi}
\\
f^{abc} f^{cde} & = & -\frac{4}{N} (\delta^{ae} \delta^{bd}
-\delta^{ad} \delta^{be}) 
- (d^{bdc} d^{cae} -d^{adc} d^{cbe}), 
\label{ffJacobi}
\\
f^{hae} f^{ebf} f^{fcg} f^{gdh} & = & 4 \delta^{ab} \delta^{cd} +4
\delta^{ad} \delta^{bc} \nonu \\
&+ & \frac{N}{2}( d^{abe} d^{cde} + d^{ade}
d^{ebc} -
d^{ace} d^{bde}),
\label{quarticf}
\\
f^{hae} f^{ebf} f^{fcg} d^{gdh} & = & -\frac{N}{2} d^{abe}
f^{cde}-\frac{N}{2} f^{abe} d^{cde},
\\
f^{hae} f^{ebf} d^{fcg} d^{gdh} & = & \frac{4(4-N^2)}{N^2}
(\delta^{ab} \delta^{cd}-
\delta^{ac} \delta^{bd}) \nonu \\
& + & \frac{(8-N^2)}{2N} (d^{abe} d^{cde} -d^{ace} d^{bde})
-\frac{N}{2} d^{ade} d^{bce},
\label{ffdd}
\\ 
f^{hae} d^{ebf} f^{fcg} d^{gdh} & = & \frac{N}{2} ( d^{ace} d^{bde} - d^{ade} d^{bce}
)-\frac{N}{2} d^{abe} d^{cde},
\\ 
f^{hae} d^{ebf} d^{fcg} d^{gdh} & = & \frac{(N^2-12)}{2N} f^{abe}
d^{cde} +\frac{N}{2} d^{abe} f^{cde} \nonu \\
& + & \frac{2}{N} (f^{ade} d^{bce} - f^{ace} d^{bde} ),
\\
d^{hae} d^{ebf} d^{fcg} d^{gdh} & = & 
\frac{4(N^2-4)}{N^2} (\delta^{ab} \delta^{cd} +\delta^{ad}
\delta^{bc})-\frac{N}{2}
d^{ace} d^{bde} \nonu \\
&+& \frac{(N^2-16)}{2N} ( d^{abe} d^{cde} + d^{ade} d^{bce}). 
\label{quarticd}
\eea
Two and three products of these and the Jacobi identities (\ref{d})-(\ref{fdJacobi}) 
are given in \cite{BBSS}.
The last Jacobi identity which was missing in \cite{BBSS} is included in (\ref{ffJacobi}). 
The four products between $d$ and $f$ symbols (\ref{quarticf})-(\ref{quarticd})
are given in \cite{Macfarlaneetal} with an 
appropriate normalization.
In this paper, the identities (\ref{d})-(\ref{ddf}), (\ref{fdJacobi}),
(\ref{ffJacobi}) and (\ref{ffdd}) are used mainly
and the other identities can be used for 
higher order representations of $SU(N)$. 

\section{The operator product expansions between the 
spin 1 current and
primary spin 3 and 4 currents 
in subsection \ref{fourth} }

Let us continue to simplify the $\frac{1}{(z-w)^2}$ terms in
the operator product expansion (\ref{JandJJJ}).
First of all, by using the $d$ symbol (\ref{dabcd}) one can write down
$k$-dependent term as follows:
\bea
d^{abcd} J^c J^d(w) & = & d^{abc} Q^c(w) -\frac{(N^2-4)(N^2-3)}{N(N^2+1)}
f^{abc} \pa J^c(w) + 2 d^{ace} d^{bde} J^c J^d(w) \nonu \\
& - & \frac{8(N^2-4)}{N(N^2+1)} J^a J^b(w)
 -  \frac{4(N^2-4)}{N(N^2+1)} \delta^{ab} J^c J^c(w). 
\label{dJJ}
\eea
Now let us consider the next term. The $d f f$ terms can be simplified
by the identity (\ref{ffdd}).
\bea
d^{bcde} f^{fdg} f^{acf} J^g J^e(w) & = & 
\frac{4(N^2-4)}{(N^2+1)} f^{abc} \pa J^c(w) -\frac{4(2N^2-3)}{N(N^2+1)} d^{abc} Q^c(w)
\nonu \\
&+ & \frac{4(N^2-4)(N^2+3)}{N^2(N^2+1)} J^a J^b(w) +
\frac{4(N^2-4)(N^2-3)}{N^2(N^2+1)} \delta^{ab} J^c J^c(w)
\nonu \\
& - & \frac{2(N^4-3N^2+6)}{N(N^2+1)} d^{ace} d^{bde} J^c J^d(w).  
\label{dffJJ}
\eea
We also used the identity (\ref{ffJacobi}) for the $f f$ terms.
For the next term, one writes down 
\bea
d^{bcde} f^{feg} f^{acf} J^d J^g(w)
=d^{bcde} f^{fdg} f^{acf} J^g J^e(w) + d^{bcde} f^{fdg} f^{acf} f^{egh} \pa J^h(w),
\label{dffJJ1}
\eea
by interchanging an index $d$ and an index $e$ that are dummy indices.
Moreover, the second term can be further simplified as before and this
becomes $\frac{2(N^2-4)(N^2-9)}{(N^2+1)} f^{abh} \pa J^h(w)$. Of course the first
term of the equation (\ref{dffJJ1}) is given by the relation (\ref{dffJJ}).
Similarly, one obtains the final term 
\bea
d^{bcde} f^{feg} f^{adf} J^c J^g(w)
=d^{bcde} f^{fdg} f^{acf} J^g J^e(w) +
\frac{2(N^2-4)(N^2-9)}{(N^2+1)} f^{abc}
\pa J^c(w).
\label{dffJJ2}
\eea
Therefore, by collecting the four terms characterized by 
(\ref{dJJ}), (\ref{dffJJ}),
(\ref{dffJJ1}) 
and (\ref{dffJJ2}), one arrives at the final simplified expression for
the $\frac{1}{(z-w)^2}$ terms 
in the operator product expansion $J^a(z) T_b^{(3,1)}(w)$.
See the operator product expansion (\ref{JT31detail}) for the complete
expression we will present below.

Let us move on the $\frac{1}{(z-w)}$ terms.
By multiplying $ f^{acf} J^f$ into the relation (\ref{dJJ}) to the left,
one has the following expression
\bea
&& f^{acf} d^{bcde} J^f J^d J^e(w)   =  f^{acf} d^{bcd} J^f Q^d(w) -
\frac{(N^2-4)(N^2-3)}{
N(N^2+1)} f^{acf} f^{bcd} J^f \pa J^d(w)
\nonu \\
&& -  \frac{4(N^2-4)}{N(N^2+1)} f^{abc} J^c J^d J^d(w)
+   2 f^{acf} d^{bdg} d^{ceg} J^f J^d J^e(w) +
\frac{8(N^2-4)}{(N^2+1)} J^b \pa J^a(w) \nonu \\
&& -
\frac{8(N^2-4)}{N(N^2+1)} f^{acf} f^{fbg}
\pa J^g J^c(w).
\label{dfJJJ}
\eea
Here we also rewrite $f^{acf} J^f J^b J^c$ term in terms of $J^b \pa
J^a$ term and $
f^{acf} f^{fbg} \pa J^g J^c$ term.
By moving the current $J^f$ to the left and interchanging between the
index $d$ and the index $c$ that are dummy indices, one has
\bea
d^{bcde} f^{adf} J^c J^f J^e(w) = f^{acf} d^{bcde} J^f J^d J^e(w) +
d^{bcde} f^{adf} f^{cfg} \pa J^g J^e(w).
\label{rel}
\eea
The first term of the equation (\ref{rel}) is equal to the expression 
(\ref{dfJJJ}).
How does one  determine the second term above?
With $4$ index $d$ symbol (\ref{dabcd}), 
one can use the identities (\ref{ffdd}) and (\ref{dff}) 
and it leads to
\bea
&& d^{bcde} f^{adf} f^{cfg} \pa J^g J^e(w)  = 
\frac{8(N^2-4)}{N^2(N^2+1)} \pa J^a J^b(w)
+\frac{4(4-N^2)}{N^2} \delta^{ab} \pa J^c J^c(w)  
\nonu \\
&& +  \frac{2(N^2-4)}{N}
d^{ace} d^{bde} \pa J^c J^d(w)
 +  \frac{4}{N} d^{abc} d^{cde} \pa J^d J^e(w)   
+\frac{4(4-N^2)}{N^2} \pa J^b J^a(w)
\label{dffjj}
\\
&& +  \frac{4}{N} d^{ace} d^{bde} \pa J^d J^c(w)+
\frac{4(N^2-4)}{N(N^2+1)} f^{ace} f^{bde} \pa J^d J^c(w)
- 
\frac{4(N^2-4)}{N(N^2+1)} f^{abe} f^{cde} \pa J^c J^d(w).
\nonu
\eea
One also has the following term by changing the index $d$
and the index $f$ that are dummy indices
\bea
d^{bcde} f^{adf}   J^c J^f J^e(w) & = & d^{bcfe} f^{afd}   J^c J^d
J^e(w) =d^{bcde} f^{adf}   J^c J^f
J^e(w),
\label{dfJJJ1}
\eea
where we changed the index $d$ and $f$ again and 
this is equal to the relation (\ref{dfJJJ}).
Let us consider the final term by moving the current $J^f$ to the
left.
\bea
d^{bcde} f^{aef}   J^c J^d J^f(w) & = & d^{bcde} f^{adf}   J^c J^f
J^e(w)
+ d^{bcde} f^{adf}  f^{efg} J^c \pa J^g(w),
\label{dfJJJ2}
\eea 
where the first term is exactly the same as the relation 
(\ref{dfJJJ}) via the relation (\ref{dfJJJ1}) and one
should simplify the second term.
By moving the current $\pa J^g$ to the left, one gets that the second
term is given by both  the equation (\ref{dffjj}) and  $-\frac{(N^2-4)(N^2-9)}{(N^2+1)}
f^{abc} \pa^2 J^c(w)$. 
By combining the expressions given in (\ref{dfJJJ}), (\ref{rel}), 
and (\ref{dfJJJ2}) altogether with correct multiplicities, one obtains
the final expression which will be present in the operator product
expansion (\ref{JT31detail}).

Then finally, one arrives at the following operator product expansion
from the result in the previous subsection \ref{fourth} and the
results 
in this Appendix
\footnote{Sometimes we do not specify the argument of $w$ for the
  field on the right
  hand side of any operator product expansion as in (\ref{JT31detail})
  for simplicity. }
\bea
&& J^a(z) T_b^{(3,1)}(w) = -\frac{1}{(z-w)^3}
\frac{2(N^2-4)(N^2-9)}{(N^2+1)} f^{abc} J^c
\nonu \\
& & + \frac{1}{(z-w)^2} \left[
( -3k -\frac{12(2N^2-3)}{N(N^2+1)}) d^{abc} Q^{c} +
\frac{(N^2-4)}{(N^2+1)} ( \frac{3k}{N}(N^2-3) +4 (N^2-6))
f^{abc}\pa J^c \right.
\nonu \\
&& + ( -6k -\frac{6}{N(N^2+1)} (N^4-3N^2+6) ) d^{ace} d^{bde} J^c J^d 
\nonu \\
&& + \left. \frac{12(N^2-4)}{N(N^2+1)} ( 2k + \frac{(N^2+3)}{N} )
J^a J^b 
+ \frac{12(N^2-4)}{N(N^2+1)} ( k + \frac{(N^2-3)}{N} )
\delta^{ab} J^c J^c \right]
\nonu \\
&  & + \frac{1}{(z-w)} \left[
-3 f^{ace} d^{bde} J^c Q^d-\frac{3(N^2-4)(N^2-3)}{N(N^2+1)} f^{ace} f^{bde} J^c
\pa J^d + 6 f^{acf} d^{bdg} d^{ceg} J^f J^d J^e
\right.
\nonu \\
&& +\frac{24(N^2-4)}{(N^2+1)} J^b \pa J^a +\frac{12(N^2-6)}{N(N^2+1)}
f^{ace} f^{bde} \pa J^d J^c -\frac{12(N^2-4)}{N(N^2+1)} f^{abc}J^c
J^d J^d
\nonu \\
&& +\frac{24(N^2-4)}{N^2(N^2+1)} \pa J^a J^b -\frac{12(N^2-4)}{N^2}
\delta^{ab}
\pa J^c J^c + \frac{6(N^2-4)}{N} d^{ace} d^{bde} \pa J^c J^d
\nonu \\
&& +\frac{12}{N} d^{abc} d^{cde} \pa J^d J^e -\frac{12(N^2-4)}{N^2}
\pa J^b J^a + \frac{12}{N} d^{ace} d^{bde} \pa J^d J^c
\nonu \\
&& \left. -\frac{12(N^2-4)}{N(N^2+1)} f^{abc} f^{cde} \pa J^d J^e -
\frac{(N^2-4)(N^2-9)}{(N^2+1)} f^{abc} \pa^2 J^c
\right] +\cdots.
\label{JT31detail}
\eea

Let us consider the $\frac{1}{(z-w)^2}$ term in (\ref{JandJJJJ}).
Let us first note 
\bea
d^{abcd} J^b J^c J^d(w)  & = & 3 d^{abc} J^b Q^c(w)
 -\frac{12(N^2-4)}{N(N^2+1)} J^a J^b J^b(w) +
 \frac{2(N^2-4)(N^2-3)}{N(N^2+1)}
f^{abc} \pa J^b J^c(w) \nonu \\
&- &\frac{(N^2-4)(N^2-3)}{N(N^2+1)} f^{abc} J^b \pa J^c(w).
\label{tripleJJJ}
\eea
This can be seen from the relation (\ref{dJJ}) by multiplying one more current
$J^b$ where we use the identity (\ref{ddf}) and rearrange the currents
in order to simplify.
The first term of $d$ symbol in (\ref{dabcd}) contributes to $-N d^{abc} J^b Q^c(w)$
via the relation (\ref{dff}). The second and third terms of $d$ symbol
in (\ref{dabcd}) 
can be used by the identity (\ref{ffdd}). 
The $J^g J^d J^e(w)$ term in the $\frac{1}{(z-w)^2}$ terms 
can be summarized by
\bea
f^{abf} f^{fcg} d^{bcde}  J^g J^d J^e(w)  & = &
-2N d^{abc} J^b Q^c(w) -\frac{12(N^2-4)}{N(N^2+1)} J^a J^b J^b(w) + 
\frac{4(N^2-4)}{N(N^2+1)} f^{abc} J^b \pa J^c(w) \nonu \\
&- &
\frac{8(N^2-4)}{N(N^2+1)} 
f^{abc} \pa J^b J^c(w).
\label{ffdJJJ}
\eea
The $J^c J^g J^e(w)$ term consists of the relation (\ref{ffdJJJ})
and $-\frac{2(N^2-4)(N^2-9)}{(N^2+1)} f^{abc} \pa J^b J^c(w)$.
The $J^c J^d J^g(w)$ term can be written in terms of 
$J^c J^g J^e(w)$ term plus $\frac{2(N^2-4)(N^2-9)}{(N^2+1)}f^{abc} (J^b
\pa J^c(w)- \pa J^b J^c(w))$. Moreover the $J^b J^g J^e(w)$ term is the same as 
the $J^c J^g J^e(w)$ term, the $J^b J^d J^g(w)$ term becomes the $J^c J^d J^g(w)$
term and similarly,  the $J^b J^c J^g(w)$ term becomes the $J^c J^d J^g(w)$
term. One obtains 
six $J^g J^d J^e(w)$ terms and $\frac{2(N^2-4)(N^2-9)}{(N^2+1)} 
f^{abc} ( 3 J^b \pa J^c(w) -5 \pa J^b J^c(w))$ by combining all these
terms. 
See the
operator product expansion (\ref{JT40detail}) for complete expression below.   

Finally, one arrives at the following operator product expansion
from the result in the previous subsection \ref{fourth} and the results in this Appendix
as follows:
\bea
J^a(z) T^{(4,0)}(w)  & = &  \frac{1}{(z-w)^4}
\frac{4N(N^2-4)(N^2-9)}{(N^2+1)} J^a - \frac{1}{(z-w)^3}
\frac{4N(N^2-4)(N^2-9)}{(N^2+1)} \pa J^a
\nonu \\
& + &  \frac{1}{(z-w)^2}\left[
-12 (N+k) d^{abc} J^b Q^c + \frac{48(N^2-4)(N+k)}{N(N^2+1)} J^a
J^b J^b \right. \nonu \\
& + &  
\frac{2(N^2-4)}{N(N^2+1)} ( 3N(N^2-5) +2k(N^2-3) ) f^{abc}
J^b \pa J^c
\nonu \\
&-& \left.  
\frac{2(N^2-4)}{N(N^2+1)} ( N(5N^2-21)+ 4k(N^2-3) )
f^{abc} \pa J^b J^c \right] +\cdots.
\label{JT40detail}
\eea

There exist $\frac{1}{(z-w)}$ terms in the operator product expansion 
(\ref{JT40detail}).
They are 
\bea
d^{bcde} \left[ f^{abf} J^f J^c J^d J^e(w) + f^{acf} J^b J^f J^d J^e(w) + f^{adf}
J^b J^c J^f J^e(w) +
f^{aef} J^b J^c J^d J^f(w) \right].
\label{4spin} 
\eea
Let us look at the first term of the relation 
(\ref{4spin}). This can be written, by
multiplying the current $J^f$ into the relation (\ref{tripleJJJ}) with
$f^{abf}$, as follows:
\bea
&& f^{abf} d^{bcde} J^f J^c J^d J^e(w)  =  -3 f^{abc} d^{cde} J^b J^d Q^e(w) +
\frac{12(N^2-4)}{(N^2+1)} \pa J^a J^b J^b(w)
\label{fdJJJJnew}
\\
&& +  \frac{(N^2-4)(N^2-3)}{N(N^2+1)}
f^{abc} f^{cde} J^b J^d \pa J^e(w)  -  
\frac{2(N^2-4)(N^2-3)}{N(N^2+1)} f^{abc} f^{cde}
J^b \pa J^d J^e(w).
\nonu
\eea
In order to simplify the second term of (\ref{4spin}), we move the
current $J^f$ to the left and then the first term of (\ref{fdJJJJnew})
occurs and the extra piece has the following form 
\bea
&& f^{abf} f^{cfg} d^{bcde} \pa J^g J^d J^e(w)  = 
\frac{2(N^2-4)(N^2-3)}{N(N^2+1)} d^{abc} \pa J^b Q^c(w) + 
\frac{4(N^2-4)}{(N^2+1)} f^{abc} \pa J^b \pa J^c(w)
\nonu \\
&& -  \frac{24(N^2-4)}{N^2(N^2+1)} \pa J^a J^b J^b(w) -
\frac{8(N^2-4)(N^2-3)}{N^2(N^2+1)} \pa J^b J^b J^a(w) 
\nonu \\
&& +  \frac{8(2N^2-3)}{N(N^2+1)} d^{abc} d^{cde} \pa J^d J^e J^b(w).
\label{ffdJJJnew}
\eea
In the relation (\ref{dJJ}), we have simplified expression and we multiply $\pa
J^g$ with $f$ symbols. Using the identities (\ref{dff}), (\ref{fff}),
(\ref{ffdd}) and (\ref{ffJacobi}) we obtain the right hand side of 
(\ref{ffdJJJnew}). For the third term of (\ref{4spin}), we move the
current $J^f$ to the left. Then we have the second term of (\ref{4spin}), 
the extra piece (\ref{ffdJJJnew}) and $\frac{(N^2-4)(N^2-9)}{(N^2+1)}
f^{abc} \pa^2 J^b J^c(w)$.
The fourth term of (\ref{4spin}) can be written as the third term of
(\ref{4spin})
and the extra piece (\ref{ffdJJJnew}) and $\frac{(N^2-4)(N^2-9)}{(N^2+1)}
f^{abc} (\pa^2 J^b J^c(w)- J^b \pa J^c(w))$.

It turns out that the final expression consists of 
four of (\ref{fdJJJJnew}), six of (\ref{ffdJJJnew}) and 
\bea
\frac{(N^2-4)(N^2-9)}{(N^2+1)}  f^{abc} \left[ 3 \pa^2 J^b J^c(w) - J^b \pa^2
J^c(w) \right].
\label{inter}
\eea
So combining the relations (\ref{fdJJJJnew}), (\ref{ffdJJJnew}) and (\ref{inter}),
it turns out to be
\bea
&& -12 f^{abc} d^{cde} J^b J^d Q^e(w) + 6(N^2-4) f^{abc} 
J^b \pa^2 J^c(w) + 12 N d^{abc} d^{cde} \pa J^d J^e J^b(w) \nonu \\
&& -
2N(N^2-4) \pa^3 J^a(w).
\label{ffinal}
\eea
For $N=4$, we have checked that the equation (\ref{ffinal}) vanishes.

The fundamental results of this Appendix will be used in next
Appendices.

\section{The operator product expansions between the 
spins $1, 2$ currents and
various spin 4 currents in subsection  \ref{fourth}  }

Using the identity $(A.15)$ of \cite{BBSS}, one writes
the following relation 
\bea
(J^a J^a) (J^b J^b)(z) = - 2 (k+N) \pa^2 J^a J^a(z) + J^a J^a J^b J^b(z), 
\label{JJandJJ}
\eea
and from the defining equation (\ref{T}), one has 
\bea
\pa^2 T(z) = -\frac{1}{(k+N)} \left[\pa J^a \pa J^a(z) + \pa^2 J^a
  J^a(z) \right],
\label{ddt}
\eea
and also the following relation holds 
\bea
Q^a Q^a(z) = d^{abc} d^{ade} J^b J^c J^d J^e(z) -\frac{2}{N} (N^2-4) (2k+N)
\pa^2 J^a J^a(z).  
\label{QQJJJJ}
\eea
We present some operator product expansions between the primary spin 1
current and other spin 4 currents as follows: 
\bea
J^a(z) Q^b Q^b(w) & = &
\frac{1}{(z-w)^4} \frac{2}{N} (N^2-4)(N+2k)(5N+6k)J^a
\nonu \\
&+ & \frac{1}{(z-w)^3} 2(N^2-4)(N+2k) \pa J^a
\nonu \\
& + & 
\frac{1}{(z-w)^2} \left [ -2(N+k) d^{abc} (J^b Q^c + Q^b J^c)
\right. \nonu \\
& + & \left. 
\frac{(N^2-4)(N+2k)}{N} f^{abc}( \pa J^b J^c - J^b \pa J^c)
\right] +\cdots,
\label{jqq}
\\
J^a(z) T^2(w) & = & \frac{1}{(z-w)^4} 3J^a +\frac{1}{(z-w)^2}\left[ J^a
  T + T J^a \right] +\cdots,
\label{jtt}
\\
J^a(z) \pa^2 T(w) & = & \frac{1}{(z-w)^4} 6 J^a +\frac{1}{(z-w)^3} 4 \pa
J^a +
\frac{1}{(z-w)^2} \pa^2 J^a +\cdots,
\label{jd2t}
\\
J^a(z) \pa^2 J^b J^b(w) & = & -\frac{1}{(z-w)^4} 2(2N+3k)J^a -
\frac{1}{(z-w)^3} 2N \pa J^a \nonu \\
& + & 
\frac{1}{(z-w)^2} \left[-2 f^{abc} \pa J^b J^c
  -k \pa^2 J^a \right]+\cdots.
\label{Jjjexp}
\eea
Then one obtains the operator product expansions 
between $J^a(z)$ and the currents $J^b J^b J^c J^c(w)$, $\pa J^b \pa J^b(w)$
and $d^{bcf} d^{def} J^b J^c J^d J^e(w)$ through the relations (\ref{jqq}),
(\ref{jtt}),
(\ref{jd2t}), and (\ref{Jjjexp}).
We also consider 
some operator product expansions between the stress energy
tensor and other spin 4 currents as follows: 
\bea
T(z)  T^2(w) & = & \frac{1}{(z-w)^6} \frac{3k(N^2-1)}{N+k} 
+\frac{1}{(z-w)^4} \left[ 8 +\frac{k}{N+k}(N^2-1)\right] T \nonu
\label{Tsquare}
\\
& + & \frac{1}{(z-w)^3} 3 \pa T +\frac{1}{(z-w)^2} 4 T^2
+\frac{1}{(z-w)} \pa T^2 +\cdots,
\\
T(z) \pa^2 T(w) & = & \frac{1}{(z-w)^6} \frac{10k}{N+k}(N^2-1) +
\frac{1}{(z-w)^4} 12 T +\frac{1}{(z-w)^3} 10 \pa T
\nonu 
\\
& +& \frac{1}{(z-w)^2} 4 \pa^2 T +\frac{1}{(z-w)} \pa^3 T +\cdots,
\label{tddt}
\\ 
T(z) \pa^2 J^b J^b(w) & = & -\frac{1}{(z-w)^6} 6k(N^2-1)
+\frac{1}{(z-w)^4} 6 J^a J^a  +  \frac{1}{(z-w)^3} 6 \pa J^a J^a 
\nonu \\
& + & \frac{1}{(z-w)^2} 4 \pa^2 J^a J^a
+ \frac{1}{(z-w)} \pa (\pa^2 J^a J^a)+\cdots.
\label{derJJ}
\eea
The operator product expansion $T(z) Q^a Q^a(w)$ is given in (\ref{TQQ}).
The operator product expansions 
between $T(z)$ and the currents $J^b J^b J^c J^c(w)$, $\pa J^b \pa J^b(w)$
and $d^{bcf} d^{def} J^b J^c J^d J^e(w)$ can be obtained similarly
from the above relations (\ref{Tsquare}), (\ref{tddt}) and (\ref{derJJ}).
One should have the operator product expansions between the coset
stress energy tensor (\ref{Ttilde}) with (\ref{stressetal}) and other
currents in next Appendices. Then it is better to write down the
relations (\ref{Tsquare}), (\ref{tddt}) and (\ref{derJJ}) in terms of
the coset central charge (\ref{ctilde}).  
For example, the operator product expansion $\widetilde{T}(z)
\widetilde{T}^2(w)$
can be read off from the equation (\ref{Tsquare}) by taking 
the central charge $c$ (\ref{centralcharge}) 
as the coset central charge 
$\widetilde{c}$ (\ref{ctilde}) and the stress energy tensor 
$T(w)$ as the coset stress energy tensor $\widetilde{T}(w)$ on
the right hand side of (\ref{Tsquare}). It is not right to replace $k$
with $k_1+k_2$
and $T(w)$ as $\widetilde{T}(w)$.
That is, the $\frac{1}{(z-w)^4}$ term of the 
operator product expansion $\widetilde{T}(z)
\widetilde{T}^2(w)$ is $(8+\widetilde{c})\widetilde{T}(w)$ that is not
equal to $[8+\frac{k_1+k_2}{N+k_1+k_2}(N^2-1)]\widetilde{T}(w)$.

\section{ 
The operator product expansions between the diagonal spin 1 current and
various spin 4 currents in subsection \ref{spin4section} }

We summarize the operator product expansions between the diagonal
primary spin 1 current (\ref{diag}) and various spin 4 currents as follows:
\bea
&& J'^a(z) d^{bcde} J^b J^c J^d J^e(w)  = 
\frac{1}{(z-w)^4} \frac{4N(N^2-4)(N^2-9)}{(N^2+1)}
J^a \nonu \\
&& -  \frac{1}{(z-w)^3}  \frac{4N(N^2-4)(N^2-9)}{(N^2+1)}
\pa J^a 
+  \frac{1}{(z-w)^2} \left[
 -12 (N+k_1) d^{abc} J^b Q^c \right. 
\nonu \\
&& +   \frac{48(N^2-4)(N+k_1)}{N(N^2+1)} J^a
J^b J^b  
+ 
\frac{2(N^2-4)}{N(N^2+1)} ( 3N(N^2-5) +2k_1(N^2-3) ) f^{abc}
J^b \pa J^c
\nonu \\
&&- \left. \frac{2(N^2-4)}{N(N^2+1)} ( N(5N^2-21)+ 4k_1(N^2-3) )
f^{abc} \pa J^b J^c \right] +\cdots,
\label{JJJJ}
\\
&& J'^a(z) d^{bcde} J^b J^c J^d K^e(w)  = 
\frac{1}{(z-w)^3} \frac{2(N^2-4)(N^2-9)}{(N^2+1)} f^{abc} J^b K^c
\nonu \\
&& +  
\frac{1}{(z-w)^2} \left[(-3 k_1 -\frac{12(2N^2-3)}{N(N^2+1)} 
) d^{abc} Q^b K^c 
-
 \frac{(N^2-4)}{(N^2+1)}(\frac{3k_1(N^2-3)}{N}+4(N^2-6))f^{abc}
  \pa J^b K^c  \right. \nonu \\
&&-   
6(k_1 +\frac{N^4-3N^2+6}{N(N^2+1)}) d^{ace} d^{bde} J^c J^d K^b   +   
\frac{12(N^2-4)}{N(N^2+1)} (2k_1 +\frac{N^2+3}{N}) J^a J^b K^b
\nonu \\
&& +   \frac{12(N^2-4)}{N(N^2+1)} (k_1+ \frac{N^2-3}{N}) J^b J^b K^a
-3k_2 d^{abc} J^b Q^c \nonu \\
&& +   \left. \frac{12k_2(N^2-4)}{N(N^2+1)} J^a J^b J^b  
 -  
\frac{k_2(N^2-4)(N^2-3)}{N(N^2+1)} f^{abc} ( 2 \pa J^b J^c - J^b \pa J^c) 
\right] +\cdots,
\label{JJJK}
\\
&& J'^a(z) d^{bcde} J^b J^c K^d K^e(w)  = 
\frac{1}{(z-w)^2} \left[-(4k_1 + \frac{8(2N^2-3)}{N(N^2+1)}) 
d^{ace} d^{bde} J^b K^c K^d  \right. \nonu \\
&&-   
2(k_1 +\frac{(N^2-4)(N^2-3)}{N(N^2+1)})  d^{abc} J^b R^c 
 +   \frac{2(N^2-4)}{N^2+1} (\frac{k_1(N^2-3)}{N}+2) f^{abc}
  J^b \pa K^c \nonu \\
&& +   \frac{8(N^2-4)}{N(N^2+1)} (2k_1 + \frac{N^2-3}{N}) J^b K^a K^b
 +   \frac{8(N^2-4)}{N(N^2+1)}(k_1 + \frac{3}{N} ) J^a K^b K^b
\nonu \\
& & -   
2 (k_2 + \frac{(N^2-4)(N^2-3)}{N(N^2+1)}) d^{abc} Q^b K^c 
 - \frac{2(N^2-4)}{N^2+1} (\frac{k_2(N^2-3)}{N} +2) f^{abc} \pa J^b K^c
\nonu \\ 
&& +  \frac{8(N^2-4)}{N(N^2+1)} (2k_2 + \frac{N^2-3}{N}) J^a J^b K^b
 +  \frac{8(N^2-4)}{N(N^2+1)} (k_2 + \frac{3}{N}) J^b J^b K^a
\nonu \\
&& - \left.  4(k_2 + \frac{2(2N^2-3)}{N(N^2+1)})d^{ace} d^{bde} J^c
  J^d K^b 
\right] +\cdots,
\label{JJKK}
\\
&& J'^a(z) d^{bcde} J^b K^c K^d K^e(w)  = 
J'^a(z) d^{bcde} J^b J^c J^d K^e(w)|_{k_1 \leftrightarrow k_2,J^a
  \leftrightarrow K^a},
\label{JKKK}
\\
&& J'^a(z) d^{bcde} K^b K^c K^d K^e(w)  = 
J'^a(z) d^{bcde} J^b J^c J^d J^e(w)|_{k_1 \leftrightarrow k_2,J^a
  \leftrightarrow K^a},
\label{KKKK}
\\
&& J'^a(z) d^{bcf} d^{def} J^b J^c J^d J^e(w)  = 
\frac{1}{(z-w)^4} 2 (N^2-4)(N+2k_1) J^a 
\nonu \\
&& - \frac{1}{(z-w)^3}
2(N^2-4)(N+2k_1)
\pa J^a 
+ \frac{1}{(z-w)^2} \left[ -4(N+k_1) d^{abc} J^b Q^c \right.
\nonu \\
&& -  \left.
\frac{(N^2-4)(N+2k_1)}{N}  f^{abc} (\pa J^b J^c - J^b \pa J^c ) \right] +\cdots,
\label{ddJJJJ}
\\
&& J'^a(z) d^{bcf} d^{def} J^b J^c J^d K^e(w)  = 
\frac{1}{(z-w)^4} 2k_1(N^2-4) K^a +\frac{1}{(z-w)^3}
(1-\frac{2k_1}{N})(N^2-4)f^{abc} 
J^b K^c 
\nonu \\
&& +  \frac{1}{(z-w)^2}\left[ (-(N+k_1) +\frac{2}{N}(N^2-4)) d^{abc} Q^b K^c 
\right.
\nonu \\
&& +   (-(N+2k_1)-\frac{2}{N}(N^2-4)) d^{ace} d^{bde} J^c J^d
  K^b 
 -   (N^2-4) f^{abc} \pa J^b K^c -k_2 d^{abc} J^b Q^c
\nonu \\
&& -  \left. \frac{k_2(N^2-4)}{N} f^{abc} (\pa J^b J^c -  J^b \pa J^c) 
-\frac{8(N^2-4)}{N^2} (J^a J^b K^b -J^b J^b K^a ) \right]
 + \cdots,
\label{ddJJJK}
\\
&& J'^a(z) d^{bcf} d^{def} J^b J^c K^d K^e(w)  = 
-\frac{1}{(z-w)^2} \left[ (N+2k_1) d^{abc} J^b R^c+(N+2k_2) d^{abc}
  Q^b K^c \right] \nonu \\
&& +\cdots, 
\label{ddJJKK}
\\
&& J'^a(z) d^{bcf} d^{def} J^b K^c K^d K^e(w)  = 
-\frac{1}{(z-w)^3} \frac{1}{N}(N^2-4)(N+2k_2) f^{abc} J^b K^c +
\frac{1}{(z-w)^2} \left[
\right. \nonu \\
&&    -k_1 d^{abc} K^b R^c -(N+k_2) d^{abc} J^b R^c -(N+2k_2) 
d^{ace} d^{bde} J^b K^c K^d \nonu \\
&& +  \left. \frac{1}{N}(N^2-4)(N+2k_2) f^{abc} J^b \pa K^c \right] +
\cdots,
\label{ddJKKK}
\\
&& J'^a(z) d^{bcf} d^{def} K^b K^c K^d K^e(w)  = 
 J'^a(z) d^{bcf} d^{def} J^b J^c J^d J^e(w)|_{k_1 \leftrightarrow
  k_2,J^a \leftrightarrow K^a},
\\
&& J'^a(z)  J^b J^b J^c J^c(w)  = 
\frac{1}{(z-w)^4} 4N(N+k_1) J^a -\frac{1}{(z-w)^3} 4N(N+k_1)\pa J^a
\nonu \\
&& +\frac{1}{(z-w)^2} \left[ -4(N+k_1) J^a J^b J^b -2(N+k_1) f^{abc} \pa J^b
J^c  +  2(N+k_1) f^{abc} J^b \pa J^c \right] +\cdots,
\label{newjjjj}
\\
&& J'^a(z)  J^b J^b J^c K^c(w)  = 
\frac{1}{(z-w)^4} 2N k_1 K^a + \frac{1}{(z-w)^3} 2(N-k_1) f^{abc} J^b K^c 
\nonu \\
&& +  \frac{1}{(z-w)^2} \left[ -k_2 J^a J^b J^b -k_2 f^{abc} \pa J^b J^c +k_2
f^{abc} J^b \pa J^c  +   (-k_1 +\frac{8}{N} ) J^b J^b K^a \right. 
\nonu \\
&& -  \left. (
 2(N+k_1)+\frac{8}{N} ) 
J^a J^b K^b 
- 
2N f^{abc} \pa J^b K^c -2 d^{ace} d^{bde} J^c J^d K^b + 2 d^{abc} Q^b K^c \right] 
+\cdots,
\label{newjjjknew}
\\
&& J'^a(z)  J^b J^b K^c K^c(w)  = 
-\frac{1}{(z-w)^2} 2\left[ (N+k_1) J^a K^b K^b +(N+k_2) J^b J^b K^a \right]
+\cdots,
\label{jjkknnew}
\\
&& J'^a(z)  J^b K^b K^c K^c(w)  =  -\frac{1}{(z-w)^3}
2(N+k_2) f^{abc} J^b K^c 
 + \frac{1}{(z-w)^2} \left[ -k_1 K^a K^b K^b \right. \nonu \\
&&- \left. k_2 J^a K^b K^b 
-2(N+k_2) J^b K^a K^b  +  2(N+k_2)f^{abc} J^b \pa K^c \right] +\cdots,
\label{jkkkjkkk}
\\
&& J'^a(z)  K^b K^b K^c K^c(w)  = 
 J'^a(z)  K^b K^b K^c K^c(w)|_{k_1 \leftrightarrow k_2,J^a
   \leftrightarrow K^a},
\\
&& J'^a(z)  \pa^2 J^b J^b(w)  = 
-\frac{1}{(z-w)^4} 2(2N+3k_1) J^a -\frac{1}{(z-w)^3}
2N \pa J^a \nonu \\
&& +  \frac{1}{(z-w)^2}\left[-\frac{k_1}{N} f^{abc} J^b \pa J^c -( 2
    +\frac{k_1}{N} ) f^{abc} \pa J^b J^c \right] +\cdots,
\label{jjjexp}
\\
&& J'^a(z)  \pa^2 J^b K^b(w)  =  -\frac{1}{(z-w)^4} 6 k_1 K^a -
\frac{1}{(z-w)^3} f^{abc} J^b K^c \nonu \\
&& +
 \frac{1}{(z-w)^2} \left[ -\frac{k_2}{N} f^{abc} (\pa J^b J^c 
 +  J^b \pa J^c)  - 2f^{abc} \pa J^b  K^c  \right]
 +  \cdots,
\label{ddjk}
\\
&& J'^a(z)  \pa^2 K^b K^b(w)  = 
J'^a(z)  \pa^2 J^b J^b(w)|_{k_1 \leftrightarrow k_2,J^a
  \leftrightarrow K^a},
\\
&& J'^a(z)  \pa J^b \pa J^b(w)  = 
-\frac{1}{(z-w)^4} 2N J^a -\frac{1}{(z-w)^3} 2(N+2k_1)\pa J^a 
\nonu \\
&& +  \frac{1}{(z-w)^2} f^{abc} (\pa J^b J^c - J^b \pa J^c)
+\cdots,
\label{newdjdj}
\\
&& J'^a(z)  \pa J^b \pa K^b(w)  =  -\frac{1}{(z-w)^3} 2\left[ k_1 \pa K^a
+k_2 \pa J^a \right]  +  \frac{1}{(z-w)^2} f^{abc} (\pa J^b K^c
  -J^b \pa K^c) \nonu \\
&& + \cdots,
\label{anotherrel}
\\
&& J'^a(z)  \pa K^b \pa K^c(w)  = 
J'^a(z)  \pa J^b \pa J^c(w)|_{k_1 \leftrightarrow k_2,J^a
  \leftrightarrow K^a},
\\
&& J'^a(z)  J^b \pa^2 K^b(w)  =  J'^a(z)  \pa^2 J^b K^b(w)|_{k_1
  \leftrightarrow k_2,J^a \leftrightarrow K^a}.
\label{JJJJend}
\eea
In the operator product expansion (\ref{JJJJ}), since the current
$K^a(z)$ commutes 
with $J^b(z)$'s, 
one gets this from the operator product expansion (\ref{JT40detail})
exactly
\footnote{Sometimes it is better to describe the operator product
expansions for arbitrary $k_1$ and $k_2$ instead of putting $k_1=1,
k_2=k$ due to the fact that  by using the symmetry between these
levels and the symmetry of the two currents $J^a(z), K^a(z)$, 
some of the operator product expansions can be determined by
known operator product expansions. For example, in (\ref{JKKK}), 
we do not have to compute the operator product expansion newly.
Once the operator product expansion $J'^a(z) d^{bcde} J^b 
J^c J^d K^e(w)$ is known, then the operator product expansion 
$J'^a(z) d^{bcde} J^b 
K^c K^d K^e(w)$ is automatically determined by taking $J^a
\leftrightarrow K^a$ and $k_1 \leftrightarrow k_2$ on the above known
operator product expansion. However,  once $k_1=1$ and $k_2=k$ are
fixed from the beginning, 
one should compute the operator product expansion $J'^a(z) d^{bcde} J^b 
K^c K^d K^e(w)$ independently because one cannot find $k_1$ dependence.}.
In the operator product expansion (\ref{JJJK}), one uses 
the operator product expansion (\ref{JT31detail}), the defining equation
(\ref{KKcoset}) and the relation (\ref{tripleJJJ}). In the operator
product expansion (\ref{JJKK}), one can use 
the operator product expansion (\ref{j22}) and its version for the
$K^a$ current,
that is, the operator product expansion between $K^a(z)$ and $d^{bcde}
K^d K^e(w)$. 
Then one should simplify further.
The following triple product $d J K K$ can be obtained from the
relation (\ref{dJJ}) by
replacing the current $J^a(w)$'s with $K^a(w)$'s and then we multiply $J^b$
current to the left with $d$ symbol as follows:  
\bea
d^{abcd} J^b K^c K^d(w) & = & 2 d^{ace} d^{bde} J^b K^c K^d(w) 
+ d^{abc} J^b R^c(w) -\frac{(N^2-4)(N^2-3)}{N(N^2+1)} f^{abc} J^b \pa K^c(w)
\nonu \\
& -& \frac{4(N^2-4)}{N(N^2+1)} \left[ 2 J^b K^a K^b(w) + J^a K^b K^b(w) \right],
\label{dabcdjkk}
\eea
where we introduce the spin $2$ field for the $K^a(w)$ current
corresponding to (\ref{spin2Q})
\bea
R^a(w) \equiv d^{abc} K^b K^c(w).
\label{R}
\eea
One also has the following relation 
\bea
f^{adf} f^{feg} d^{bcde} J^g K^b K^c(w) & = &
-\frac{2(N^2-4)(N^2-3)}{N(N^2+1)} d^{abc} J^b R^c(w)
+ \frac{4(N^2-4)}{(N^2+1)} f^{abc} J^b \pa K^c(w)
\nonu \\
&+ & \frac{24(N^2-4)}{N^2(N^2+1)} J^a K^b K^b(w)
+\frac{8(N^2-4)(N^2-3)}{N^2(N^2+1)} J^b K^a K^b(w)
\nonu \\
& - & \frac{8(2N^2-3)}{N(N^2+1)}
d^{ace} d^{bde} J^b K^c K^d(w),
\label{ffdjkk}
\eea
where the relation (\ref{dJJ}) is used, the $J^a$'s are exchanged with
the current $K^a$'s as before, the current $J^a$ is multiplied  and 
the spin $2$ field (\ref{R}) is used.
Then the identities (\ref{dff}), (\ref{ffdd}), (\ref{ddf}) and
(\ref{ffJacobi})
are used.
Similarly, one can write down 
\bea
d^{abcd} J^b J^c K^d(w) & = & 2 d^{ade} d^{bce} J^b J^c K^d(w) 
+ d^{abc} K^b Q^c(w) +\frac{(N^2-4)(N^2-3)}{N(N^2+1)} f^{abc} \pa J^b  K^c(w)
\nonu \\
& -& \frac{4(N^2-4)}{N(N^2+1)} \left[ 2 J^a J^b K^b(w) + J^b J^b K^a(w) \right],
\label{dabcdjjk}
\eea
which can be seen from the identity (\ref{dJJ})
and by multiplying two $f$ symbols into (\ref{dabcdjjk}) one has the relation
\bea
f^{adf} f^{feg} d^{bcde} J^b J^c K^g(w) & = &
-\frac{2(N^2-4)(N^2-3)}{N(N^2+1)} d^{abc} Q^b K^c(w)
- \frac{4(N^2-4)}{(N^2+1)} f^{abc} \pa J^b  K^c(w)
\nonu \\
&+ & \frac{24(N^2-4)}{N^2(N^2+1)} J^b J^b K^a(w)
+\frac{8(N^2-4)(N^2-3)}{N^2(N^2+1)} J^a J^b K^b(w)
\nonu \\
& - & \frac{8(2N^2-3)}{N(N^2+1)}
d^{ace} d^{bde} J^c J^d K^b(w).
\label{ffdjjk}
\eea

In the operator product expansion (\ref{JKKK}), one has 
the following relation
\bea
d^{abcd} K^b K^c K^d(w)  & = & 3 d^{abc} K^b R^c(w)
 -\frac{12(N^2-4)}{N(N^2+1)} K^a K^b K^b(w) +
 \frac{2(N^2-4)(N^2-3)}{N(N^2+1)}
f^{abc} \pa K^b K^c(w) \nonu \\
&- &\frac{(N^2-4)(N^2-3)}{N(N^2+1)} f^{abc} K^b \pa K^c(w).
\label{dabcdkkk}
\eea
This can be seen from the relation 
(\ref{tripleJJJ}) by replacing the currents $J^a$'s with the currents $K^a$'s.
In the operator product expansion 
(\ref{ddJJJJ}), one can use the relations 
(\ref{QQJJJJ}), (\ref{jqq}) and (\ref{Jjjexp}).
In the operator product expansion (\ref{ddJJJK}), one also need to have
\bea
d^{ade} d^{bce} J^b J^c J^d(w) = d^{abc} J^b Q^c(w) 
+ \frac{N^2-4}{N} f^{abc} \left[\pa J^b J^c(w)- J^b \pa J^c(w) \right]. 
\label{ddjjj}
\eea
One can check this by moving the current $J^d$ to the left together
with the identity (\ref{ddf}).
One also has the following relation
\bea
f^{ceg} f^{agh} J^h J^c K^e(w) & = &
d^{abc} Q^b K^c(w) -\frac{4}{N} \left[ J^a J^b K^b(w) - J^b J^b K^a(w) \right]
\nonu \\
& - & d^{ace} d^{bde} J^c J^d K^b(w) -N f^{abc} \pa J^b K^c(w),
\label{ffjjk}
\eea
where the identities (\ref{ffJacobi}) and (\ref{ddf}) are used.
In the operator product expansion 
(\ref{ddJJKK}), one uses the defining equation (\ref{JQ}).
In the operator product expansion 
(\ref{ddJKKK}), we also used the following relation
\bea
d^{bcf} d^{afg} J^b K^c K^g(w) = d^{ace} d^{bde} J^b K^c K^d(w) 
- \frac{N^2-4}{N} f^{abc} J^b \pa K^c(w). 
\label{ddjkk}
\eea
In the operator product expansion 
(\ref{newjjjj}), the properties (\ref{JJandJJ}), (\ref{jtt}) 
and (\ref{Jjjexp}) are used.
From the relation,
\bea
J^a J^a J^b(w) & = & -2(N+k_1) J^b T_{(1)}(w) + 2 f^{bcd} \pa J^c J^d(w) -N
\pa^2 J^b(w),
\label{jjjrel}
\eea
one can compute the operator product expansion 
(\ref{newjjjknew}) where the following operator
product expansions are used
\bea
J^a(z) J^b T_{(1)}(w) & = & \frac{1}{(z-w)^3} f^{abc} J^c +
\frac{1}{(z-w)^2} \left[ - k_1 \delta^{ab} T_{(1)} +
J^b J^a\right], 
\nonu \\ &+& 
\frac{1}{(z-w)} f^{abc} J^c T_{(1)}+\cdots, \nonu \\
J^a(z) f^{bcd} \pa J^c J^d(w) & = & -\frac{1}{(z-w)^4} 2 N k_1 \delta^{ab}
+
\frac{1}{(z-w)^3} (N+2k_1) f^{abc} J^c
\nonu \\
&+& \frac{1}{(z-w)^2} \left[ f^{ace} f^{bcd} J^e J^d -k_1 f^{abc} \pa
  J^c \right] \nonu \\
&+ & \frac{1}{(z-w)} \left[ f^{ace} f^{bcd} \pa J^e J^d -f^{ace} f^{bcd}
  \pa J^d J^e \right]+\cdots.
\label{near}
\eea
In the operator product expansion 
(\ref{jjkknnew}), the operator product expansion between 
$J^a(z)$ and $T_{(1)}(w)$ is used. 
In the operator product expansion (\ref{jkkkjkkk}), the first equation of (\ref{near})
(with the current $K^a$) can be used: that is, the operator product
expansion $K^a(z) K^b T_{(2)}(w) $.
In the operator product expansion 
(\ref{jjjexp}), the operator product expansion (\ref{Jjjexp}) is used.
In the operator product expansion 
(\ref{ddjk}), the defining equation (\ref{JJ}) is used.
Moreover, in the operator product expansion 
(\ref{newdjdj}), the relations (\ref{ddt}), (\ref{jd2t})
and (\ref{Jjjexp})  are used. 
In the operator product expansion 
(\ref{anotherrel}), the relation (\ref{JJ}) is used.

Therefore, the independent field contents from the equations (\ref{tripleJJJ}), 
(\ref{dabcdjkk}), (\ref{ffdjkk}), (\ref{dabcdjjk}), (\ref{ffdjjk}),
(\ref{dabcdkkk}),
(\ref{ddjjj}), (\ref{ffjjk}) and (\ref{ddjkk}) can be summarized by
eighteen fields in the $\frac{1}{(z-w)^2}$ terms in the operator
product expansions between the diagonal spin $1$ current and spin $4$
currents. 
See also the
expression (\ref{indep}) we will explain in next Appendix.

\section{The operator product expansions between the stress tensor and
various spin 4 currents  in subsection \ref{spin4section}}

Let us introduce the following simplified notation as in (\ref{That}) 
\bea
T_{(1)}(z) +T_{(2)}(z) \equiv \hat{T}(z).
\label{that}
\eea
Any spin 4 field  $\Phi(z)$ satisfies 
$\hat{T}(z) \Phi(w) = \cdots + \frac{1}{(z-w)^2} 4\Phi(w)
+\frac{1}{(z-w)} \pa \Phi(w) + \cdots$. 
For simplicity, we present the only higher order singular terms 
$\frac{1}{(z-w)^n}$ with $n >2 $ and we
assume $\frac{1}{(z-w)^2}$ and $\frac{1}{(z-w)}$ terms as above:
\bea
&& \hat{T}(z) d^{bcde} J^b J^c J^d J^e(w)  =  +\cdots,
\label{cosetJJJJfin}
\\
&& \hat{T}(z) d^{bcde} J^b J^c J^d K^e(w)  =  +\cdots,
\\
&& \hat{T}(z) d^{bcde} J^b J^c K^d K^e(w)  =  +\cdots,
\\
&& \hat{T}(z) d^{bcde} J^b K^c K^d K^e(w)  =  +\cdots,
\\
&& \hat{T}(z) d^{bcde} K^b K^c K^d K^e(w)  =  +\cdots,
\label{cosetKKKKfin}
\\
&& \hat{T}(z) d^{bcf} d^{def} J^b J^c J^d J^e(w)  =  
-\frac{1}{(z-w)^4} \frac{4}{N}(N^2-4)(N+2k_1) J^a J^a +\cdots,
\label{cosetddJJJJ}
\\
&& \hat{T}(z)  d^{bcf} d^{def} J^b J^c J^d K^e(w)   =  
-\frac{1}{(z-w)^4} \frac{4k_1}{N}(N^2-4) J^a K^a 
\nonu \\
&&- \frac{1}{(z-w)^3} 2(N^2-4) \pa J^a K^a + \cdots,
\label{cosetddJJJK}
\\
&& \hat{T}(z)  d^{bcf} d^{def} J^b J^c K^d K^e(w)   =
+\cdots,
\label{cosetddJJKK}
\\
&& \hat{T}(z)  d^{bcf} d^{def} J^b K^c K^d K^e(w)  =  
-\frac{1}{(z-w)^4} \frac{2}{N}(N^2-4)(N+2k_2) J^a K^a +\cdots, 
\label{cosetddJKKK}
\\
&& \hat{T}(z)  d^{bcf} d^{def} K^b K^c K^d K^e(w)  =  
\hat{T}(z) d^{bcf} d^{def} J^b J^c J^d J^e(w)|_{k_1 \leftrightarrow
  k_2, J^a \leftrightarrow K^a},
\\
&& \hat{T}(z)   J^b J^b J^c J^c(w)  =  
-\frac{1}{(z-w)^4} 2\left[2(N+k_1)+k_1(N^2-1) \right] J^a J^a +\cdots,
\label{jjjj}
\\
&& \hat{T}(z)   J^b J^b J^c K^c(w)   =  
-\frac{1}{(z-w)^4} k_1(N^2+1) J^a K^a-\frac{1}{(z-w)^3} 2N \pa J^a K^a
\nonu
\\
&&+  \cdots,
\label{jjjk}
\\
&& \hat{T}(z)   J^b J^b K^c K^c(w)  =  
-\frac{1}{(z-w)^4} (N^2-1)\left[ k_1 K^a K^a + k_2 J^a J^a
\right]
+ \cdots,
\label{jjkk}
\\
&& \hat{T}(z)    J^b K^b K^c K^c(w)   =  
-\frac{1}{(z-w)^4} \left[ 2(N+k_2) +(N^2-1)k_2 \right] J^a K^a +\cdots,
\label{jkkk}
\\
&& \hat{T}(z)   K^b K^b K^c K^c(w)   =  
\hat{T}(z)   J^b J^b J^c J^c(w)|_{k_1 \leftrightarrow k_2, J^a
 \leftrightarrow K^a}, 
\\
&& \hat{T}(z)   \pa^2 J^b J^b(w)    =  
-\frac{1}{(z-w)^6} 6k_1(N^2-1) +\frac{1}{(z-w)^4} 6 J^a J^a
\nonu
\\
&& +  \frac{1}{(z-w)^3} 6 \pa J^a J^a +\cdots,
\label{derjj}
\\
&& \hat{T}(z)   \pa^2 J^b K^b(w)   =  
\frac{1}{(z-w)^4} 6 J^a K^a(w) +\frac{1}{(z-w)^3} 6 \pa J^a K^a +\cdots,
\label{derjk}
\\
&& \hat{T}(z)   \pa^2 K^b K^b(w)   =  
\hat{T}(z)   \pa^2 J^b J^b(w)|_{k_1 \leftrightarrow k_2, J^a
  \leftrightarrow K^a}, 
\\
&& \hat{T}(z)    \pa J^b \pa J^b(w)   =  
-\frac{1}{(z-w)^6} 4 k_1(N^2-1) +\frac{1}{(z-w)^3} 2 \pa (J^a J^a)
+\cdots,
\label{derjjnew}
\\
&& \hat{T}(z)   \pa J^b \pa K^b(w)  = 
\frac{1}{(z-w)^3} 2\left[ J^a \pa K^a(w)+\pa J^a K^a \right]
+ \cdots,
\\
&& \hat{T}(z)   \pa K^b \pa K^c(w)  =  
\hat{T}(z)    \pa J^b \pa J^b(w)|_{k_1 \leftrightarrow k_2, J^a
  \leftrightarrow K^a},
\\
&& \hat{T}(z)   J^b \pa^2 K^b(w)  =  
\hat{T}(z)   \pa^2 J^b K^b(w)|_{k_1\leftrightarrow k_2, J^a
  \leftrightarrow K^a}.
\label{finfin}
\eea
According to the operator product expansions (\ref{T13}), (\ref{T22}),
(\ref{T31}) 
and (\ref{T40}) and those where the currents $J^a(z)$ are replaced by
the currents $K^a(z)$, 
one obtains the operator product expansions 
(\ref{cosetJJJJfin})-(\ref{cosetKKKKfin}) because the operator product
expansions of the stress energy
tensors $T_{(1)}(z)$ and $T_{(2)}(z)$ with spin-$4$ currents do not
have any higher order singular terms and the currents $J^a(z)$ and
$K^a(z)$ commute with each other.  
In the operator product expansion 
(\ref{cosetddJJJJ}), one uses the relations (\ref{QQJJJJ}) and 
(\ref{derJJ}) together with the operator product expansion (\ref{TQQ}).
By writing the spin 4 current, 
\bea
d^{abe} d^{cde} J^a J^b J^c K^d(w) & = & d^{abc} J^a Q^b K^c(w)
\nonu \\
& - & \frac{(N^2-4)}{N} 
f^{abc} \left[ J^b \pa J^c K^a(w) -\pa J^b J^c K^a(w) \right],
\label{otherrelation}
\eea
in order to compute the operator product expansion (\ref{cosetddJJJK}),
one need to compute the following operator product expansions
\bea
T_{(1)}(z) J^a Q^b(w) & = & -\frac{1}{(z-w)^4} (N+2k_1) d^{abc} J^c
+\frac{1}{(z-w)^3} f^{abc} Q^c \nonu \\
& +& \frac{1}{(z-w)^2} 3 J^a Q^b +\frac{1}{(z-w)} \pa (J^a Q^b)
+\cdots,
\label{T1JQ} 
\\
T_{(1)}(z) f^{abc} \pa J^b J^c(w) & = & \frac{1}{(z-w)^4} 4 N J^a
+\frac{1}{(z-w)^3}
2N \pa J^a \nonu \\
& + & \frac{1}{(z-w)^2} 3 f^{abc} \pa J^b J^c +\frac{1}{(z-w)} \pa
(f^{abc} 
\pa J^b J^c) +\cdots, 
\label{fderJJ}
\\
T_{(1)}(z) f^{abc} J^b \pa J^c(w) & = & \frac{1}{(z-w)^4} 2 N J^a
+\frac{1}{(z-w)^3}
4N \pa J^a \nonu \\
& + & \frac{1}{(z-w)^2} 3 f^{abc} J^b \pa J^c +\frac{1}{(z-w)} \pa
(f^{abc} 
J^b \pa J^c) +\cdots.
\label{fJderJ}
\eea

In the operator product expansion (\ref{cosetddJJKK}), 
one uses the defining equation (\ref{TQa}).
Due to the different index structures, one cannot obtain
the operator product expansion (\ref{cosetddJKKK}) directly
from the operator product expansion 
(\ref{cosetddJJJK}) via symmetry arguments($k_1\leftrightarrow
k_2$ and $J^a \leftrightarrow K^a$).
Instead, by using the relation (\ref{T1JQ}), one gets the operator
product expansion (\ref{cosetddJKKK}).
From the relations 
(\ref{JJandJJ}), (\ref{Tsquare}) and (\ref{derJJ}), it is easy to
see the operator product expansion (\ref{jjjj}).
With the relation 
(\ref{jjjrel})
and the following operator product expansions
\bea
T_{(1)}(z) J^b T_{(1)}(w) & = & \frac{1}{(z-w)^4} \left[ 1 + \frac{k_1(N^2-1)}{2(N+k_1)}
\right] J^b +\frac{1}{(z-w)^2} 3 J^b T_{(1)} +\frac{1}{(z-w)} \pa (J^b
T_{(1)}) \nonu \\
& + & \cdots,
\label{TJT}
\\
T_{(1)}(z) \pa^2 J^b(w) & = & \frac{1}{(z-w)^4} 6 J^b
+\frac{1}{(z-w)^3} 6 \pa J^b +\frac{1}{(z-w)^2} 3 \pa^2 J^b
+\frac{1}{(z-w)} \pa^3 J^b \nonu \\
& + & \cdots,
\label{secj}
\eea
together with the operator product expansion (\ref{fderJJ}),
one can check the relation (\ref{jjjk}).
The operator product expansion (\ref{secj}) can be obtained 
by adding the operator product expansions (\ref{fderJJ}) and (\ref{fJderJ}).
Of course, one also gets the operator product expansion
\bea
T_{(1)}(z) J^a J^a J^b(w) & = &
-\frac{1}{(z-w)^4} k_1(N^2+1) J^b -\frac{1}{(z-w)^3} 2N \pa J^b
\nonu \\
& + & \frac{1}{(z-w)^2}
3 J^a J^a J^b +\frac{1}{(z-w)} \pa (J^a J^a J^b) +\cdots.
\label{equ}
\eea
This operator product expansion 
(\ref{equ}) is also useful to check the operator product expansion (\ref{jjjk}).
For the operator product expansion (\ref{jjkk}), one uses the defining equation (\ref{TT}).
For the operator product expansion (\ref{jkkk}), the relation (\ref{TJT}) is used.
In order to compute the operator product expansion 
(\ref{derjj}), one can use the relation (\ref{derJJ}).
The relation (\ref{secj}) gives the result of (\ref{derjk}).
By using the identities (\ref{ddt}), (\ref{tddt}) and (\ref{derJJ}), one gets 
the operator product expansion 
(\ref{derjjnew}). The remaining operator product expansions can be
obtained similarly.

\section{The coefficient functions that satisfy the linear equations
 in subsection \ref{spin4section}}

By collecting the $\frac{1}{(z-w)^2}$ terms in the operator product
expansions
(\ref{JJJJ})-(\ref{JJJJend})
we have eighteen equations for twenty two unknown coefficient functions.
\bea
&& \left[-12 (N+k_1) c_1 - 3k_2 c_2 -4(N+k_1) c_6 -k _2 c_7\right]d^{abc} J^b Q^c =0,
\nonu \\
&& \left[ \frac{48(N^2-4)(N+k_1)}{N(N^2+1)} c_1 +\frac{12k_2(N^2-4)}{N(N^2+1)} 
c_2 -4(N+k_1) c_{11} -k_2 c_{12} \right] J^a J^b J^b =0,
\nonu \\
&& \left[ \frac{2(N^2-4)}{N(N^2+1)} ( 3N(N^2-5) +2k_1(N^2-3)) c_1 
+\frac{k_2(N^2-4)(N^2-3)}{N(N^2+1)} c_2 +\frac{(N^2-4)(N+2k_1)}{N} c_6 \right.
\nonu \\
&& \left. + 
\frac{(N^2-4)k_2}{N} c_7 + 2(N+k_1)
c_{11} +k_2 c_{12} -\frac{k_1}{N} c_{16} -\frac{k_2}{N} c_{17}-c_{19}
\right] f^{abc} J^b \pa J^c =0,
\nonu \\
&& \left[-\frac{2(N^2-4)}{N(N^2+1)} ( N(5N^2-21) +4k_1(N^2-3) 
) c_1 -\frac{2k_2(N^2-4)(N^2-3)}{N(N^2+1)} c_2 \right. \nonu \\
&& -\frac{(N^2-4)(N+2k_1)}{N} c_6
 \nonu \\
&& \left. -\frac{(N^2-4)k_2}{N} c_7 - 2(N+k_1)c_{11} -k_2 c_{12} -(2+\frac{k_1}{N}) c_{16}
-\frac{k_2}{N} c_{17} + c_{19}  \right] f^{abc} \pa J^b J^c= 0,
\nonu \\
&& \left[(-3k_1 -\frac{12(2N^2-3)}{N(N^2+1)}) c_2 +( -2k_2
  -\frac{2(N^2-4)(N^2-3)}{N(N^2+1)} ) c_3 +( -(N+k_1)
  +\frac{2(N^2-4)}{N} ) c_7  \right. \nonu \\
&& \left. -(N+2k_2) c_8 +2c_{12} \right] d^{abc} Q^b K^c =0,
\nonu \\
&& \left[ -\frac{(N^2-4)}{(N^2+1)} ( \frac{3k_1(N^2-3)}{N} +4(N^2-6) ) c_2
-\frac{2(N^2-4)}{(N^2+1)} ( \frac{k_2}{N} (N^2-3)  +2 ) c_3 
\right. \nonu \\
&& \left. -(N^2-4) c_7 -2N c_{12} -2 c_{17} +c_{20} \right]f^{abc} \pa J^b K^c =0,
\nonu \\
&& \left[ ( -6k_1 -\frac{6(N^4-3N^2+6)}{N(N^2+1)} ) c_2
+( -\frac{8(2N^2-3)}{N(N^2+1)} - 4k_2 ) c_3 +(
  -(N+2k_1) -\frac{2}{N}(N^2-4)) c_7
\right. \nonu \\
&& \left. -2 c_{12} \right]d^{ace} d^{bde} J^c J^d K^b  =0,
\nonu \\
&& \left[ \frac{12(N^2-4)}{N(N^2+1)} ( 2k_1 +\frac{N^2+3}{N} ) c_2
+ \frac{8(N^2-4)}{N(N^2+1)}( 2k_2 
  +\frac{N^2-3}{N}) c_3 \right. \nonu \\
&& \left. 
-\frac{8}{N^2}
(N^2-4) c_7 +( -2(N+k_1) -\frac{8}{N} ) c_{12} \right]J^a J^b K^b =0,
\nonu \\
&& \left[ \frac{12(N^2-4)}{N(N^2+1)} (k_1 + \frac{N^2-3}{N}) c_2 +
\frac{8(N^2-4)}{N(N^2+1)}
(k_2 + \frac{3}{N}) c_3 +\frac{8(N^2-4)}{N^2} c_7 +(-k_1 +
\frac{8}{N}) c_{12} \right. \nonu \\
&& \left. -2(N+k_2) c_{13} \right]J^b J^b K^a =0,
\nonu \\
&& \left[ ( -2k_1 -\frac{2(N^2-4)(N^2-3)}{N(N^2+1)} ) c_3 +(
  -3k_2 -\frac{12(2N^2-3)}{N(N^2+1)} ) c_4 -(N+2k_1) c_8
\right. \nonu \\
&& \left. -(N+k_2) c_9  \right] d^{abc} J^b  R^c=0,
\nonu \\
&& \left[ \frac{2(N^2-4)}{N^2+1} ( 2 + \frac{k_1}{N}(N^2-3) ) c_3
+\frac{(N^2-4)}{(N^2+1)} ( \frac{3k_2(N^2-3)}{N} +4(N^2-6)
) c_4 \right. \nonu \\
&& \left. 
+\frac{(N+2k_2)(N^2-4)}{N} c_9 + 2(N+k_2) c_{14} -c_{20} +2c_{22}
\right] f^{abc} J^b \pa K^c
=0,
\nonu \\
&&\left[( -4k_1 -\frac{8(2N^2-3)}{N(N^2+1)} 
) c_3 +( -6k_2 -\frac{6(N^4-3N^2+6)}{N(N^2+1)} ) c_4
-(N+2k_2) c_9 \right]d^{ace} d^{bde} J^b K^c K^d  =0,
\nonu \\
&& \left[( \frac{8(N^2-4)}{N(N^2+1)} (2k_1 +\frac{N^2-3}{N} ) c_3
+\frac{12(N^2-4)}{N(N^2+1)}
(2k_2 + \frac{N^2+3}{N}) c_4 -2(N+k_2) c_{14} \right]J^b K^a K^b =0,
\nonu \\
&& \left[ \frac{8(N^2-4)}{N(N^2+1)} ( k_1 + \frac{3}{N} ) c_3 
+\frac{12(N^2-4)}{N(N^2+1)} ( k_2 + \frac{N^2-3}{N} ) c_4 
-2(N+k_1) c_{13} -k_2 c_{14} \right]J^a K^b K^b =0,
\nonu \\
&& \left[ -3k_1 c_4 -12(N+k_2) c_5 -k_1 c_9 -4 (N+k_2) c_{10} \right] d^{abc} K^b R^c =0,
\nonu \\
&&\left[ \frac{12k_1(N^2-4)}{N(N^2+1)} c_4 +\frac{48(N^2-4)(N+k_2)}{N(N^2+1)}
c_5 -k_1 c_{14} -4(N+k_2) c_{15} \right]K^a K^b K^b =0,
\nonu \\
&& \left[ \frac{k_1(N^2-4)(N^2-3)}{N(N^2+1)} c_4 +
 \frac{2(N^2-4)}{N(N^2+1)} ( 3N(N^2-5) +2k_2 (N^2-3) ) c_5
\right. \nonu \\
&& \left. +\frac{(N^2-4)(N+2k_2)}{N} c_{10} +2(N+k_2) c_{15} 
 -\frac{k_2}{N} c_{18} -c_{21} -\frac{k_1}{N} c_{22} \right]f^{abc} K^b \pa K^c =0,
\nonu \\
&&\left[ -\frac{2k_1(N^2-4)(N^2-3)}{N(N^2+1)} c_4 
-\frac{2(N^2-4)}{N(N^2+1)} ( N(5N^2-21) +4k_2 (N^2-3) ) c_5 
\right. \nonu \\
&& -  \frac{(N^2-4)(N+2k_2)}{N} c_{10} 
 \left. -2 (N+k_2) c_{15} -( 2+\frac{k_2}{N} ) c_{18} +c_{21}
-\frac{k_1}{N} c_{22} \right] f^{abc} \pa K^b K^c  =0.
\label{rel1}
\eea
Therefore, there exist the following independent terms 
\bea
&& d^{abc} J^b Q^c(w), \,\, J^a J^b J^b(w), \,\, f^{abc} J^b \pa
J^c(w), \,\, 
f^{abc} \pa J^b J^c(w),
\,\, d^{abc} Q^b K^c(w), \,\, f^{abc} \pa J^b K^c(w), \nonu \\
&& d^{ace} d^{bde} J^c
J^d K^b(w), \,\, 
J^a J^b K^b(w), \,\, J^b J^b K^a(w), \,\, d^{abc} J^b R^c(w), \,\, f^{abc}
J^b \pa K^c(w), \nonu \\
&& d^{ace} d^{bde} J^b K^c K^d(w), \,\, J^b K^a K^b(w), \,\,
J^a K^b K^b(w), \,\, 
d^{abc} K^b R^c(w), \,\, K^a K^b K^b(w), \nonu \\
&& f^{abc} K^b \pa K^c(w), f^{abc} \pa
K^b K^c(w). 
\label{indep}
\eea
The $\frac{1}{(z-w)^3}$ terms in the operator product expansions 
(\ref{JJJJ})-(\ref{JJJJend})
provide three equations for the coefficient functions 
\bea
&& \left[ -\frac{4N(N^2-4)(N^2-9)}{(N^2+1)} c_1 -2(N^2-4)(N+2k_1) c_6 -4N(N+k_1)
c_{11} -2N c_{16} \right. \nonu \\
&& \left. -2(N+2k_1) c_{19} -2k_2 c_{20}\right] \pa J^a =0,
\nonu \\
&& \left[ -\frac{4N(N^2-4)(N^2-9)}{(N^2+1)} c_5 -2(N^2-4)(N+2k_2) c_{10}
-4N(N+k_2) c_{15} -2N c_{18} \right. \nonu \\
&& \left. -2k_1 c_{20} -2(N+2k_2) c_{21} \right] \pa K^a =0,
\nonu \\
&& \left[
\frac{2(N^2-4)(N^2-9)}{(N^2+1)} c_2 -\frac{2(N^2-4)(N^2-9)}{(N^2+1)} c_4
+(1-\frac{2k_1}{N}) (N^2-4) c_7 -\frac{N^2-4}{N} (N+2k_2)
c_9 \right.
\nonu \\
&& \left. 
-2(k_1-N) c_{12} -2(N+k_2) c_{14} -2 c_{17} +2 c_{22} \right]f^{abc} J^b K^c =0.
\label{rel2}
\eea
The $\frac{1}{(z-w)^4}$ terms in the operator product expansions 
(\ref{JJJJ})-(\ref{JJJJend})
provide two equations for the coefficient functions 
\bea
&& \left[\frac{4N(N^2-4)(N^2-9)}{(N^2+1)} c_1 + 2(N^2-4)(N+2k_1) c_6 +
4N(N+k_1) c_{11} - 2(2N+3k_1) c_{16}  \right. 
\nonu \\
& & \left. - 2N c_{19}-6k_2 c_{22} \right] J^a =0,
\nonu \\
&& \left[ \frac{4N(N^2-4)(N^2-9)}{(N^2+1)} c_5 + 2k_1 (N^2-4) c_7 +
2(N^2-4)(N+2k_2) c_{10} + 2N k_1 c_{12} \right. 
\nonu \\
&& + \left. 4N(N+k_2) c_{15} 
 - 6k_1 c_{17} -2(2N+3k_2) c_{18} -2N c_{21} \right] K^a =0. 
\label{rel3}
\eea

On the other hands,
the $\frac{1}{(z-w)^3}$ 
terms in the operator product expansions (\ref{cosetJJJJfin})-(\ref{finfin})
lead to the following equations:
\bea
&& \left[ -2(N^2-4) c_7 + 2c_{20} -2N c_{12} + 6 c_{17} \right]\pa J^a K^a  =0,
\nonu \\
&& \left[ 6 c_{16} + 4 c_{19} \right]\pa J^a J^a  =0,
\nonu \\
&& \left[ 6 c_{18} + 4 c_{21}  \right]\pa K^a K^a =0,
\nonu \\
&& \left[ 2c_{20} + 6 c_{22} \right]J^a \pa K^a =0.
\label{rel4}
\eea
The $\frac{1}{(z-w)^4}$ 
terms in the operator product expansions (\ref{cosetJJJJfin})-(\ref{finfin})
lead to the following equations:
\bea
&& \left[ -\frac{4(N^2-4)(N+2k_1)}{N} c_6 - 2\left[2(N+k_1) +k_1(N^2-1)\right] c_{11}
-k_2(N^2-1) c_{13} + 6 c_{16} \right] J^a J^a =0,
\nonu \\
&& \left[ -\frac{4k_1(N^2-4)}{N} c_7 -\frac{2(N^2-4)(N+2k_2)}{N} c_9 
-k_1(N^2+1) c_{12} \right. \nonu \\
&& \left. +( -2(N+k_2) -k_2(N^2-1)) c_{14}
+ 6c_{17} + 6c_{22} \right] J^a K^a=0,
\nonu \\
&& \left[ -\frac{4(N^2-4)(N+2k_2)}{N} c_{10} -k_1 (N^2-1) c_{13} -2(
  2(N+k_2) +
k_2(N^2-1) ) c_{15} \right.
\nonu \\
&& \left. + 6c_{18} \right]K^a K^a =0. 
\label{rel5}
\eea
Finally, the $\frac{1}{(z-w)^6}$ terms 
in the operator product expansions (\ref{cosetJJJJfin})-(\ref{finfin})
provide the relation
\bea
&& \left[ 
-6k_1(N^2-1) c_{16} -6k_2(N^2-1) c_{18} -4 k_1(N^2-1) c_{19}
-4k_2(N^2-1)c_{21} 
\right] =0.
\label{relfin}
\eea
Then it is easy to see that $c_{16}=0=c_{18}$ from the relation (\ref{rel4}) by
using the conditions $c_{19}=0=c_{21}$ which was shown in 
subsection \ref{spin4section}. Then the equation (\ref{relfin}) is
automatically satisfied.
Now we are left with 18 equations in the relations (\ref{rel1}), 3 equations
in the relations (\ref{rel2}), 2 equations in the relations 
(\ref{rel3}), 2 equations in the relations (\ref{rel4}) and 3
equations in the relations (\ref{rel5}). 
Totally, there are 28 equations we have to solve. 

The coefficients satisfying the equations (\ref{rel1})-(\ref{relfin})
are fixed and they can be written in terms of $c_1$
\bea
c_2  & = &  \frac{-4N(N+k_1)}{3k_2(N+2k_1)D(k_1,k_2,N)} 
\left[ 90k_1^3 +270 k_1^2 k_2 +180 k_1 k_2^2 + 255k_1^2 N +363 k_1 k_2
N \right. \nonu \\
&- &  72 k_2^2 N -36 k_1^2 k_2^2 N 
 + 48 k_1 N^2 -100k_1^3 N^2 -102 k_2 N^2-332 k_1^2 k_2 N^2 \nonu \\
&- &  202k_1 k_2^2 N^2 -60N^3 -266 k_1^2 N^3-478k_1 k_2 N^3 +52
  k_2^2 N^3 + 12 k_1^2 k_2^2 N^3
\nonu \\
& -&  106 k_1 N^4 + 10k_1^3 N^4 +84 k_2 N^4 +62 k_1^2 k_2 N^4 +
50 k_1 k_2^2 N^4 +72N^5 \nonu \\
&+ & \left. 35 k_1^2 N^5 + 115 k_1 k_2 N^5 + 8 k_2^2 N^5 + 30 k_1 N^6 + 18
k_2 N^6  \right] c_1,
\nonu \\
c_3 & = & \frac{2N(N+k_1)(2N+3k_1)(N^2-3)}{k_2 D(k_1,k_2,N)}
\left[-5k_1-5k_2-10N +2 k_1 k_2 N + 7k_1 N^2+ 7k_2 N^2 \right.
\nonu \\
& + &\left.  12 N^3 \right] c_1,
\nonu \\
c_4 & = & \frac{4N(N+k_1)(2N+3k_1)}{3k_2(2N+3k_2)D(k_1,k_2,N)} 
\left[ 45 k_1 k_2 + 45k_2^2 +30 k_1 N + 120k_2 N +18 k_1^2 k_2 N +60
  N^2
\right. 
\nonu \\
& + &  28k_1^2 N^2 -34 k_1 k_2 N^2 -50 k_2^2 N^2 -4k_1 N^3 -142
  k_2 N^3 -6k_1^2 k_2 N^3 
\nonu \\
&- & \left.  72 N^4 -16 k_1^2 N^4 -11k_1 k_2 N^4 + 5k_2^2 N^4 -26k_1
  N^5 +10 k_2 N^5 \right] c_1,
\nonu \\
c_5  & = & \frac{k_1(N+k_1)(2N+3k_1)}{k_2(N+k_2)(2N+3k_2)D(k_1,k_2,N)} 
\left[-18k_1^2 k_2 -36 k_1 k_2^2-18k_2^3 -12 k_1^2 N -75 k_1 k_2 N \right.
\nonu \\
& - & 63 k_2^2 N
-34 k_1 N^2 
-  64 k_2 N^2 + 14 k_1^2 k_2 N^2 + 40 k_1 k_2^2 N^2 + 20 k_2^3 N^2
-20 N^3 +4 k_1^2 N^3 \nonu \\
&+ & 78 k_1 k_2 N^3 + 70 k_2^2 N^3 +28 k_1 N^4 + 74 k_2 N^4 -4 k_1
k_2^2 N^4 -2k_2^3 N^4 \nonu \\
& +& \left.
24 N^5 +4k_1^2 N^5 -3k_1 k_2 N^5 -7k_2^2 N^5 + 6k_1 N^6-6k_2 N^6   
\right]c_1,
\nonu \\
c_6 & = & \frac{2(N^2-9)}{(N+2k_1)(N^2+1)D(k_1,k_2,N)} \left[ 
-6k_1^4 -12 k_1^3 k_2 -6k_1^2 k_2^2 -24k_1^3 N -48k_1^2 k_2 N \right.
\nonu \\
&- & \left. 24k_1
k_2^2 N -53k_1^2 N^2 -46 k_2 N^3 + 12 k_1^2 k_2 N^3 + 10 k_1 k_2^2 N^3
-20 N^4 + 22k_1^2 N^4 \right. \nonu \\
&+ & 6k_1^4 N^4 + 46 k_1 k_2 N^4 +12 k_1^3 k_2 N^4 + 20 k_2^2 N^4 +
6k_1^2 k_2^2 N^4 + 46 k_1 N^5 
\nonu \\
&+ &  24 k_1^3 N^5 + 44 k_2 N^5 + 36 k_1^2 k_2 N^5 + 10 k_1
  k_2^2 N^5  + 24 N^6  + 31 k_1^2 N^6 \nonu \\
&+ & \left. 31 k_1 k_2 N^6 + 14 k_1 N^7 + 2k_2 N^7 
-77 k_1 k_2 N^2 -24 k_2^2 N^2-56 k_1 N^3 \right] c_1,
\nonu \\
c_7 & = & -\frac{8N(N^2-9)(N+k_1)}{k_2(N+2k_1)(N^2+1)D(k_1,k_2,N)}
\left[ 5 k_1^3 -2k_1^2 k_2 -7 k_1 k_2^2 -7k_1^2 N 
 -  33 k_1 k_2 N \right. \nonu \\
&- &  26
  k_2^2 N -36 k_1 N^2 -46 k_2 N^2 + 12 k_1^2 k_2 N^2 + 
14 k_1 k_2^2 N^2 -20 N^3 + 22 k_1^2 N^3 + 46 k_1 k_2 N^3 \nonu \\
&+& 20 k_2^2 N^3 + 42 k_1 N^4 - 5k_1^3 N^4 + 44 k_2 N^4 -10 k_1^2 k_2
N^4 - 3k_1 k_2^2 N^4 +24 N^5-15 k_1^2 N^5 \nonu \\
&- & \left. 13 k_1 k_2 N^5 + 2k_2^2 N^5 -10 k_1 N^6 + 2k_2 N^6
\right] c_1,
\nonu \\
c_8 & = & -\frac{4N(N^2-9)(N+k_1)(2N+3k_1)(N^2-3)}
{k_2(N+2k_1)(N+2k_2)(N^2+1)D(k_1,k_2,N)} \left[-8 k_1^2 k_2 -8k_1
  k_2^2-13 k_1^2 N \right. \nonu \\
&- &  26 k_1 k_2 N  -13 k_2^2 N -23 k_1 N^2 -23 k_2 N^2 + 4k_1^2 k_2
N^2  + 4 k_1 k_2^2 N^2 -10N^3+9k_1^2 N^3 \nonu \\
& + & \left. 20 k_1 k_2 N^3 + 9k_2^2 N^3 + 21 k_1 N^4 + 21 k_2 N^4 +
  12 N^5 \right] c_1,
\nonu \\
c_9 & = & -\frac{8N(N^2-9)(N+k_1)(2N+3k_1)}
{k_2(N+2k_2)(2N+3k_2)(N^2+1)D(k_1,k_2,N)} \left[-17 k_1^2 k_2 -22 k_1
  k_2^2 - 5k_2^3 -22 k_1^2 N \right. 
\nonu \\
& - &  59 k_1 k_2 N -37 k_2^2 N -42 k_1 N^2 -52 k_2 N^2 + 10k_1^2 k_2
N^2 + 12k_1 k_2^2 N^2 -20 N^3 + 20 k_1^2 N^3  \nonu \\
&+ &  46 k_1 k_2 N^3 + 22 k_2^2 N^3 + 44 k_1 N^4 + 46 k_2 N^4 + 3 k_1^2
k_2 N^4 + 10 k_1 k_2^2 N^4 + 5k_2^3 N^4 \nonu \\
&+ & \left. 24 N^5 -2k_1^2 N^5 + 13 k_1 k_2 N^5 + 15 k_2^2 N^5 -2k_1 N^6 + 10
k_2 N^6 \right] c_1,
\nonu \\
c_{10} & = &
\frac{2k_1(N^2-9)(N+k_1)(2N+3k_1)}{k_2(N+k_2)(N+2k_2)(2N+3k_2)
(N^2+1)D(k_1,k_2,N)} \left[ -6k_1^2 k_2^2 -12 k_1 k_2^3 -6 k_2^4
\right.
\nonu \\
& - &  24
k_1^2 k_2 N -48 k_1 k_2^2 N -24 k_2^3 N-24 k_1^2 N^2 -77k_1 k_2 N^2
-53 k_2^2 N^2 -46 k_1 N^3 \nonu \\
&- & 56 k_2 N^3 + 10 k_1^2 k_2 N^3  + 12k_1 k_2^2 N^3 -20N^4 + 20
k_1^2 N^4 + 46 k_1 k_2 N^4 + 22 k_2^2 N^4 \nonu \\
& + &  
6k_1^2 k_2^2 N^4  + 12 k_1 k_2^3 N^4 + 6k_2^4 N^4 + 44 k_1 N^5
+ 46 k_2 N^5 + 10 k_1^2 k_2 N^5 + 36 k_1 k_2^2 N^5 \nonu \\
&+ & \left. 24 k_2^3 N^5 + 24 N^6 +31 k_1 k_2 N^6 + 31 k_2^2 N^6 +
  2k_1 N^7 + 14 k_2 N^7  \right] c_1,
\nonu \\
c_{11} & = & -\frac{4(N^2-4)(N^2-9)}{N(N^2+1)D(k_1,k_2,N)} 
\left[ -6k_1^3 -12 k_1^2 k_2 - 6k_1 k_2^2 -21 k_1^2 N -33 k_1 k_2 N
  \right. 
\nonu \\
&+ & 6k_1^3 N^2 -26 k_2 N^2 + 12 k_1^2 k_2 N^2 + 6k_1 k_2^2 N^2 -10
N^3 +21 k_1^2 N^3 + 33 k_1 k_2 N^3 \nonu \\
&+ & \left. 10k_2^2 N^3 + 26 k_1 N^4+ 26 k_2 N^4 + 12 N^5 
 -12 k_2^2 N -26 k_1 N^2 \right] c_1,
\nonu \\
c_{12} & = &
\frac{16(N^2-4)(N^2-9)(N+k_1)}{k_2(N+2k_1)(N^2+1)D(k_1,k_2,N)} \left[
10k_1^3 + 14 k_1^2 k_2 + 4k_1 k_2^2 + 19 k_1^2 N + 3 k_1 k_2 N \right.
\nonu \\
&- & 6 k_1 N^2 -10 k_1^3 N^2 -26 k_2 N^2-14 k_1^2 k_2 N^2 -10 N^3 -19
k_1^2 N^3 -3 k_1 k_2 N^3 \nonu \\
&+ & \left. 14 k_2^2 N^3 + 2k_1 N^4 + 26 k_2 N^4 + 12 N^5 
-16
k_2^2 N \right]c_1,
\nonu \\
c_{13} & = &
-\frac{8(k_1+k_2)(N^2-4)(N^2-9)(N+k_1)(2N+3k_1)(N^2-3)}{k_2(N^2+1)D(
k_1, k_2, N)} c_1,
\nonu \\
c_{14} & = & \frac{16(N^2-4)(N^2-9)(N+k_1)(2N+3k_1)}
{k_2(2N+3k_2)(N^2+1)D(k_1,k_2,N)} \left[ -8k_1^2 -13 k_1 k_2 -5k_2^2
  -18 k_1 N - 18 k_2 N \right. 
\nonu \\
& - & \left. 10 N^2 + 6k_1^2 N^2 
+   13 k_1 k_2 N^2 + 5k_2^2 N^2 + 18 k_1 N^3 + 18 k_2 N^3 +
  12N^4 
\right] c_1,
\nonu \\
c_{15} & = & -\frac{4k_1(N^2-4)(N^2-9)(N+k_1)(2N+3k_1)}{k_2
  N(N+k_2)(2N+3k_2)(N^2+1)D(k_1,k_2,N)} 
\left[ -6k_1^2 k_2 -12 k_1 k_2^2 -6k_2^3 -12 k_1^2 N \right.
\nonu \\
&- & 33 k_1 k_2 N -21
k_2^2 N -26 k_1 N^2 -26 k_2 N^2 + 6k_1^2 k_2 N^2 + 12 k_1 k_2^2 N^2 +
6k_2^3 N^2 \nonu \\
&- & \left.
10N^3 + 10 k_1^2 N^3 + 33 k_1 k_2 N^3 + 21 k_2^2 N^3 + 26 k_1 N^4
+ 26 k_2 N^4 + 12 N^5 \right] c_1,
\nonu \\
c_{16}& = & 0,
\nonu \\
c_{17} & = & -\frac{4(N^2-4)(N^2-9)(N+k_1)(N^2-3)}
{3k_2(N+2k_1)(N^2+1)D(k_1,k_2,N)}
\left[ 6k_1^4 +12 k_1^3 k_2 + 6k_1^2 k_2^2 + 29 k_1^3 N + 46k_1^2 k_2
  N \right.
\nonu \\ 
&+ & 17 k_1 k_2^2 N + 46 k_1^2 N^2 -6 k_1^4 N^2 + 44 k_1 k_2 N^2 -12
k_1^3 k_2 N^2 -2 k_2^2 N^2 -6k_1^2 k_2^2 N^2 \nonu \\
&+ &  20 k_1 N^3 - 29 k_1^3 N^3-46 k_1^2 k_2 N^3 - 13 k_1 k_2^2 N^3 -46
k_1^2 N^4 -44 k_1 k_2 N^4 + 2k_2^2 N^4 \nonu \\
& - & \left. 24 k_1 N^5 \right] c_1,
\nonu \\
c_{18} & = & 0,
\nonu \\
c_{19}& = & 0,
\nonu \\
c_{20} & = & -
\frac{4(k_1+k_2)(N^2-1)(N^2-4)(N^2-9)(N+k_1)(2N+3k_1)(2N+k_1+k_2)(N^2-3)}
{k_2(N^2+1)D(k_1,k_2,N)} c_1,
\nonu \\
c_{21} & = & 0,
\label{coeff}
\\
c_{22} & = &
\frac{4(k_1+k_2)(N^2-1)(N^2-4)(N^2-9)(N+k_1)(2N+3k_1)(2N+k_1+k_2)(N^2-3)}
{3k_2(N^2+1)D(k_1,k_2,N)} c_1,
\nonu
\eea
where we introduce the function $D(k_1,k_2,N)$ which depends on the
levels $k_1, k_2$ and $N$ \footnote{Since the `solve' command with 28
  equations simultaneously in the
  mathematica does not give us the final answer, we consider several
  simple equations and find the coefficient functions in terms of
  $c_1$ and then solve the remaining linear equations completely.   }
\bea
D(k_1,k_2,N) & \equiv & -18k_1^3 -36k_1^2 k_2 -18k_1 k_2^2-63k_1^2 N-75k_1
k_2 N-12 k_2^2 N-64k_1 N^2 \nonu \\
&+& 20 k_1^3 N^2-34k_2 N^2+ 40k_1^2 k_2 N^2+ 14 k_1 k_2^2 N^2-20N^3
+ 70 k_1^2 N^3 \nonu \\
&+ & 78 k_1 k_2 N^3 + 4k_2^2 N^3 + 74 k_1 N^4 - 2k_1^3 N^4 + 28 k_2 N^4
- 4k_1^2 k_2 N^4 + 24 N^5 \nonu \\
& - & 7k_1^2 N^5 - 3k_1 k_2 N^5 + 4 k_2^2 N^5 - 6k_1 N^6 + 6k_2 N^6.
\label{dcoeff}
\eea
Note that for $N=3$, the coefficients $c_6, \cdots, c_{22}$, that have
the factor $(N^2-9)$, vanish.
Also we present the large $N$ limit (\ref{limit}) for these coefficients as follows:
\bea
c_2 & = & \left[\frac{4\lambda(4+5\lambda)}{3(-1+\lambda)(2+\la)} \right]c_1, \qquad
c_3 = \left[\frac{2\la^2(7+5\la)}{(-1+\la)^2(2+\la)} \right]c_1, \nonu \\
c_4  & = & \left[\frac{20\la^2(1+\la)}{3(-3+\la)(-1+\la)(2+\la)}\right]c_1, \qquad
c_5 = -\left[\frac{\la^2(1+\la)}{N(-3+\la)(-1+\la)}\right] c_1, \nonu \\
c_6 & = & \left[\frac{2\la}{2+\la}\right] c_1, \qquad 
c_7 = \left[\frac{8\la}{(-1+\la)(2+\la)}\right] c_1, \nonu \\ 
c_8 & = & \left[\frac{12\la^2(3+\la)}{(-2+\la)(-1+\la)^2(2+\la)}\right] c_1, \qquad
c_9 = \left[\frac{40\la^2(1+\la)}{(-3+\la)(-2+\la)(-1+\la)(2+\la)}\right] c_1,
\nonu \\  
c_{10} & = &
-\left[\frac{2\la^2(1+\la)(6+\la^2)}{N(-3+\la)(-2+\la)(-1+\la)(2+\la)}\right] c_1,
\qquad 
c_{11} = 0, \qquad
c_{12}  =  0, \nonu \\
c_{13} & = &  \left[\frac{8\la^2}{N(-1+\la)(2+\la)}\right] c_1, \qquad
c_{14}  = \left[ \frac{16\la^2(-5-8\la+\la^2)}{N(-3+\la)(-1+\la)^2(2+\la)}\right]
c_1,
\nonu \\
c_{15} & = & \left[\frac{4\la^2(3+\la^2)}{N^2(-3+\la)(-1+\la)^2}\right] c_1, 
\qquad
c_{17} = -\left[\frac{4N^2 \la}{3(2+\la)}\right] c_1, \nonu \\
c_{20} & = & \left[\frac{4N^2\la(1+\la)}{(-1+\la)(2+\la)}\right] c_1, \qquad
c_{22} = -\left[\frac{4N^2\la(1+\la)}{3(-1+\la)(2+\la)}\right] c_1.
\label{limitc}
\eea
Note that there are the $\frac{1}{N}$ dependence in $c_5, c_{10},
c_{13}, c_{14}$ and
$\frac{1}{N^2}$ dependence in $c_{15}$ and $N^2$ dependence in
$c_{17}, c_{20}, c_{22}$ \footnote{Here we substitute $N =
\frac{\la}{1-\la} k$ with $k_1=1, k_2=k$ into the relations (\ref{coeff}),
look at the power of $k$ in the numerator and denominator of the
coefficient functions and take $k \rightarrow \infty$ limit. By
reading off the power of $k$ and resubstituting $k =\frac{1-\la}{\la}
N$ into the leading term, 
one gets the final results (\ref{limitc}).}. 
Therefore, in the large $N$ limit, the only 11 independent terms in (\ref{WWWW})
survive.

\section{The independent terms of $\widetilde{W}(z)$  
in subsection \ref{spin4section}}

Let us check how many independent terms in the coset spin $4$ current
$\widetilde{W}(z)$
in (\ref{cosetspin4}) occur.
It is convenient to write down the terms of $c_2, c_3, c_4$  in the
equation
(\ref{cosetspin4})
explicitly from the relation (\ref{dabcd})
\bea
&& d^{abcd} J^a J^b J^c K^d(z) = 
3 d^{abc} J^a Q^b K^c(z)
 -\frac{12(N^2-4)}{N(N^2+1)} J^a J^b J^b K^a(z) \nonu \\
&& +
 \frac{2(N^2-4)(N^2-3)}{N(N^2+1)}
f^{abc} \pa J^a J^b K^c(z) 
- \frac{(N^2-4)(N^2-3)}{N(N^2+1)} f^{abc} J^a \pa J^b K^c(z),
\nonu \\
&& d^{abcd} J^a J^b K^c K^d(z)  =  
 2 d^{ace} d^{bde} J^a J^b K^c K^d(z) 
+ d^{abc} J^a J^b R^c(z) \nonu \\
&& -
\frac{(N^2-4)(N^2-3)}{N(N^2+1)} f^{abc} J^a J^b \pa K^c(z)
 - \frac{4(N^2-4)}{N(N^2+1)} \left[ 2 J^a J^b K^a K^b(z) + J^a J^a K^b
  K^b(z) \right], 
\nonu \\
&& d^{abcd} J^a K^b K^c K^d(z)  =  
 3 d^{abc} J^a K^b R^c(z)
-\frac{12(N^2-4)}{N(N^2+1)} J^a K^a K^b K^b(z) 
\nonu \\
&& +
 \frac{2(N^2-4)(N^2-3)}{N(N^2+1)}
f^{abc} J^a \pa K^b K^c(z) 
 - \frac{(N^2-4)(N^2-3)}{N(N^2+1)} f^{abc} J^a K^b \pa K^c(z).
\label{independent}
\eea
See also the relevant equations (\ref{tripleJJJ}), 
(\ref{dabcdjjk}), (\ref{dabcdjkk}) and (\ref{dabcdkkk}).
The $f^{abc} \pa J^a J^b K^c(z)$ can be written as a linear combination of
$f^{abc} J^a \pa J^b K^c(z)$ and $ \pa^2 J^a K^a(z)$.
For given independent terms $c_1$-$c_{22}$ in (\ref{cosetspin4}), one 
regards $d^{abcd} J^a J^b J^c K^d(z) $  as $ f^{abc} J^a \pa J^b K^c(z)$ and
other terms,  
$d^{abcd} J^a J^b K^c K^d(z)$ as both $ d^{ace} d^{bde} J^a J^b K^c K^d(z)$ and
$J^a J^b K^a K^b(z)$ (and other terms), 
and 
$d^{abcd} J^a K^b K^c K^d(z) $ as $ f^{abc} J^a K^b \pa K^c(z)$ (and
other terms). 
That is,
\bea
d^{abcd} J^a J^b J^c K^d(z) & \sim & f^{abc} J^a \pa J^b K^c(z) +\cdots,
\nonu \\
d^{abcd} J^a J^b K^c K^d(z) & \sim & 
d^{ace} d^{bde} J^a J^b K^c K^d(z), J^a J^b K^a K^b(z) + \cdots, 
\nonu \\
d^{abcd} J^a K^b K^c K^d(z) & \sim & f^{abc} J^a K^b \pa K^c(z) + \cdots.
\label{indepen1}
\eea
This implies that one may further generalize the $c_3$ term by
introducing two independent terms with two arbitrary coefficients. 
The $c_1$ term can be rewritten as the equation (\ref{dddd}) and $c_5$ term can be
expressed similarly by changing $J^a(z)$ into $K^a(z)$ from (\ref{dddd}).
We will use the explicit relations (\ref{independent}) to find the
exact relation between the spin-$4$ current (\ref{generalW}) 
and the one from \cite{BBSS1} where we do not see any quartic Casimir
operator because in this case, the spin-$4$ current is obtained from
the operator product expansions of two Casimir operators of rank $3$. 
Therefore, the relations (\ref{independent}) are very important to
identify these two spin-$4$ currents. 

Moreover, the other contractions of $d$ symbol (\ref{dabcd}) 
with various spin 4 fields
lead to
\bea
d^{ace} d^{bde} J^a J^b J^c J^d(z) &=& d^{abe} d^{cde} J^a J^b J^c J^d(z)
-(N^2-4)
\left[ \pa J^a \pa J^a(z) -\pa^2 J^a J^a(z) \right], \nonu \\
d^{ade} d^{bce} J^a J^b J^c J^d(z) & = &  d^{abe} d^{cde} J^a J^b J^c J^d(z)
-2(N^2-4) \pa J^a \pa J^a(z) +(N^2-4) \pa^2 J^a J^a(z),
\nonu \\
d^{ace} d^{bde} J^a J^b J^c K^d(z) &=& d^{abe} d^{cde} J^a J^b J^c K^d(z) 
+\frac{N^2-4}{N} f^{abc} J^a \pa J^b K^c(z),
\nonu \\
d^{ade} d^{bce} J^a J^b J^c K^d(z) & = & d^{abe} d^{cde} J^a J^b J^c K^d(z) 
+\frac{N^2-4}{N} f^{abc} \left[ J^a \pa J^b K^c(z) - \pa J^a J^b K^c(z) \right],
\nonu \\
d^{ade} d^{bce} J^a J^b K^c K^d(z) & = & 
d^{ace} d^{bde} J^a J^b K^c K^d(z) -(N^2-4) \pa J^a \pa K^a(z),
\nonu \\
d^{ace} d^{bde} J^a K^b K^c K^d(z) &=&  d^{abe} d^{cde} J^a K^b K^c K^d(z) 
+\frac{N^2-4}{N} f^{abc} J^a \pa K^b K^c(z), \nonu \\
d^{ade} d^{bce} J^a K^b K^c K^d(z) & = & 
d^{abe} d^{cde} J^a K^b K^c K^d(z) 
+\frac{N^2-4}{N} f^{abc} [J^a \pa K^b K^c(z)-J^a K^b \pa K^c(z)],
\nonu \\
d^{ace} d^{bde} K^a K^b K^c K^d(z) &=& d^{abe} d^{cde} K^a K^b K^c K^d(z)
-(N^2-4)
\left[ \pa K^a \pa K^a(z) -\pa^2 K^a K^a(z) \right], \nonu \\
d^{ade} d^{bce} K^a K^b K^c K^d(z) & = &  d^{abe} d^{cde} K^a K^b K^c K^d(z)
-2(N^2-4) \pa K^a \pa K^a(z) \nonu \\
&+ & (N^2-4) \pa^2 K^a K^a(z).
\label{g3}
\eea
From the relations (\ref{cosetspin4}), (\ref{independent}) and (\ref{indepen1}),
there is no additional independent term. 
The contractions with delta symbol in the relation (\ref{dabcd}) can be written as
\bea
J^a J^b J^a J^b(z) & = &   J^a J^a J^b J^b(z) + f^{abc} J^a \pa J^b J^c(z), \nonu \\
J^a J^b J^b J^a(z) & = &  J^a J^a J^b J^b(z) + f^{abc} J^a \pa J^b J^c(z)
-f^{abc} J^a J^b \pa J^c(z),
\nonu \\
J^a J^b J^a K^b(z) & = & J^a J^a J^b K^b(z) +f^{abc} J^a \pa J^b K^c(z), \nonu \\
J^a J^b J^b K^a(z) & = & J^a J^a J^b K^b(z) +f^{abc} J^a \pa J^b K^c(z) -
f^{abc} \pa J^a J^b K^c(z), \nonu \\
J^a J^b K^b K^a(z) & = &  J^a J^b K^a K^b(z) - f^{abc} J^a J^b \pa K^c(z), \nonu \\
J^a K^b K^a K^b(z) & = & J^a K^a K^b K^b(z) + f^{abc} J^a \pa K^b K^c(z), \nonu \\
J^a K^b K^b K^a(z) & = & J^a K^a K^b K^b(z) + f^{abc} J^a \pa K^b K^c(z)
-f^{abc}
J^a K^b \pa K^c(z), \nonu \\
K^a K^b K^a K^b(z) & = &  K^a K^a K^b K^b(z) + f^{abc} K^a \pa K^b K^c(z), \nonu \\
K^a K^b K^b K^a(z) & = & K^a K^a K^b K^b(z) + f^{abc} K^a \pa K^b K^c(z)
-f^{abc} K^a K^b \pa K^c(z). 
\label{g4}
\eea
Note that one sees that the field $J^a J^b K^a K^b(z)$ occurs in (\ref{g4}).
We do not see any further independent terms.
Therefore, if we impose two independent terms in $c_3$ term, then we
should include the field $J^a J^b K^a K^b(z)$ in the candidate for the
spin $4$ field (\ref{cosetspin4}). This can be also seen from the normal ordered
field product $(\widetilde{T} \widetilde{T})(z)$. 

In order to check the independent fields in 
$\widetilde{T}^2(z)$, one should write down the following 
normal ordered products  
\bea
(J^a K^a)(J^b K^b)(z) & = & J^a K^a J^b K^b(z) + N \pa^2 J^a K^a(z)-\frac{k_1}{2}
\pa^2 K^a K^a(z) + f^{abc} J^a \pa K^b K^c(z) \nonu \\
& - & \frac{k_2}{2} J^a \pa^2 J^a(z)
-  f^{abc}
J^a \pa J^b K^c(z),
\nonu \\
(J^a K^a)(K^b K^b)(z) & = &
J^a K^a K^b K^b(z) -(N+k_2) \pa^2 J^a K^a(z),
\nonu \\
(J^a J^a)(J^b K^b)(z) & = &
J^a J^a J^b K^b(z) -(2N+k_1) \pa^2 J^a K^a(z) + 2 f^{abc} J^a \pa J^b K^c(z),
\nonu \\
(K^aK^a)(J^b K^b)(z) & = & 
J^b K^a K^a K^b(z) -(2N+k_2) J^a \pa^2 K^a(z) + 2 f^{abc} J^a K^b \pa K^c(z),
\nonu \\
(J^a K^a)(J^b J^b)(z) & = &
J^a K^a J^b J^b(z) -(N+k_1) J^a \pa^2 K^a(z).
\label{prod}
\eea
The spin 4 current can be written, by using the above relations (\ref{prod}),  as
\bea
\widetilde{T}^2(z) & = & \frac{k_2^2}{4(N+k_1)^2(N+k_1+k_2)^2}\left[- 2 (k_1+N)
\pa^2 J^a J^a(z) + 
J^a J^a J^b J^b(z)\right] \nonu \\
&+&   \frac{k_1^2}{4(N+k_2)^2(N+k_1+k_2)^2}\left[- 2 (k_2+N)
\pa^2 K^a K^a(z) + 
K^a K^a K^b K^b(z)\right] \nonu \\
& +&  \frac{k_1 k_2}{2(N+k_1)(N+k_2)(N+k_1+k_2)^2} J^a J^a K^b K^b(z)
\nonu \\
& + & \frac{1}{(N+k_1+k_2)^2} \left[
J^a K^a J^b K^b(z) + N \pa^2 J^a K^a-\frac{k_1}{2}
\pa^2 K^a K^a(z) + f^{abc} J^a \pa K^b K^c(z) \right. \nonu \\
& - & \left. \frac{k_2}{2} J^a \pa^2 J^a(z)
-  f^{abc}
J^a \pa J^b K^c(z)\right]
\nonu \\
& - & \frac{k_1}{2(N+k_2)(N+k_1+k_2)^2} \left[J^a K^a K^b K^b(z) -(N+k_2) \pa^2 J^a K^a(z)
\right. \nonu \\
&+ & \left. J^b K^a K^a K^b(z) -(2N+k_2) J^a
\pa^2 K^a(z) + 
2 f^{abc} J^a K^b \pa K^c(z) \right]
\nonu \\
& - & \frac{k_2}{2(N+k_1)(N+k_1+k_2)^2} \left[J^a K^a J^b J^b(z) -(N+k_1) J^a \pa^2 K^a(z) 
\right. \nonu \\
& + & \left. J^a J^a J^b K^b(z) -(2N+k_1)
\pa^2 J^a K^a(z) + 
2 f^{abc} J^a \pa J^b K^c(z)\right].
\label{tt}
\eea
Therefore, one can think of $\widetilde{T}^2(z)$ 
as an independent term $J^a J^b K^a K^b(z)$ plus others:
\bea
\widetilde{T}^2(z) \sim J^a J^b K^a K^b(z) +\cdots.
\label{tt1}
\eea
The terms $f^{abc} J^a \pa J^b K^c(z)$ and $f^{abc} J^a K^b \pa
K^c(z)$
can be understood as $c_2$ term and $c_4$ term respectively 
from the relations (\ref{indepen1}).
Although we introduce the extra terms $d^{ace} d^{bde}$ multiplied by 
the fields appearing in $c_6$-$c_{10}$, 
the extra terms  $d^{ade} d^{bce}$ multiplied by 
the fields appearing in $c_6$-$c_{10}$, 
the extra terms $\delta^{ac} \delta^{bd}$ multiplied by 
the fields appearing in $c_{11}$-$c_{15}$, and 
the extra terms $\delta^{ad} \delta^{bc}$ multiplied by 
the fields appearing in $c_{11}$-$c_{15}$,
there exists only one further independent term which is given by
$J^a J^b K^a K^b(z)$.
We will use the explicit form of (\ref{tt}) in order to extract the
spin-$4$ field next Appendix $I$ later and the large $N$ limit for
(\ref{tt})
can be obtained.

Or one can add $\widetilde{T}^2(z)$(or $\widetilde{\Lambda}(z) \equiv \widetilde{T}^2(z)
-\frac{3}{10} \pa^2 \widetilde{T}(z)$) to the $\widetilde{W}(z)$ (\ref{cosetspin4}) in order
to see any further generalized spin $4$ field.
We can use the operator product expansion
\bea
\widetilde{T}(z) \widetilde{\Lambda}(w) & = & 
\frac{1}{(z-w)^4} \left[ \frac{22}{5}
  +\widetilde{c} 
\right] \widetilde{T}(w) +\frac{1}{(z-w)^2} 4 \widetilde{\Lambda}(w)
 +  \frac{1}{(z-w)} \pa \widetilde{\Lambda}(w)
+\cdots
\label{tlambda}
\eea
and $J'^a(z) \widetilde{\Lambda}(w) =\mbox{regular}$ and the central
charge $\widetilde{c}$ is given by the equation (\ref{ctilde}).
This particular combination eliminates the singular terms with
$\frac{1}{(z-w)^6}$ and $\frac{1}{(z-w)^3}$ that appear in (\ref{Tsquare}).
The fields $\widetilde{T}^2(z)$ and $ \widetilde{\Lambda}(z)$ are
quasiprimary \cite{BS} because there are higher order singular terms
(\ref{tlambda}) and (\ref{Tsquare}) where $T(z)$ and $c$ are 
replaced with $\widetilde{T}(z)$ and $\widetilde{c}$ respectively.
In the subsection \ref{other}, we include the field $J^a J^b K^a
K^b(z)$ rather than $\widetilde{T}^2(z)$.
Contrary to the spin-$3$ case, one should, in general, take into
account of the normal ordered field product of stress energy tensor
for higher spin greater than $3$. 

\section{The coefficient functions in the subsection \ref{other} }

The operator product expansion between the diagonal spin $1$ current
$J'^a(z)$ (\ref{diag}) and the spin $4$ field $J^b J^c K^b K^c(w)$ is given by
\bea
&& J'^a(z) J^b J^c K^b K^c(w) = \frac{1}{(z-w)^3} (-N) \left[k_1 \pa
  K^a+k_2 \pa J^a \right] 
+\frac{1}{(z-w)^2} \left[-\frac{4}{N} J^a
  K^b K^b \right. \nonu \\ 
&&  +(\frac{4}{N} -2k_1) J^b K^a K^b - d^{abc} J^b R^c - d^{abc}
  Q^b K^c  + d^{ace} d^{bde} J^b K^c K^d 
 + d^{ace} d^{bde} J^c J^d K^b \nonu \\
&& \left.  -(N-k_1) f^{abc} J^b \pa K^c +(N-k_2) f^{abc} \pa J^b K^c
 -\frac{4}{N} J^b J^b K^a +(\frac{4}{N}-2k_2) J^a J^b K^b \right] 
\nonu \\
&& + \cdots.
\label{jprimeeq}
\eea
By realizing the independent field contents described in (\ref{indep}),
one can arrange each field in the $\frac{1}{(z-w)^2}$ term in
the relation (\ref{jprimeeq})
and combine them with the linear equations (\ref{rel1}), (\ref{rel2})
and (\ref{rel4}).
The operator product expansion between the stress energy tensor (\ref{that}) and 
$J^b J^c K^b K^c(z)$ is 
\bea
\hat{T}(z)  J^b J^c K^b K^c(w) = -\frac{1}{(z-w)^4}  \left[ k_2 J^a
J^a + k_1 K^a K^a \right] +\frac{1}{(z-w)^3} N \pa (J^a K^a) 
+\cdots,
\label{tprimeeq}
\eea
where the $\frac{1}{(z-w)^2}$ term and $\frac{1}{(z-w)}$ term are ignored. 
One should include this equation into (\ref{rel4}) and (\ref{rel5}).

As in the subsection \ref{other},
let us introduce the coefficient $c_{23}$ in front of the extra field
$J^b J^c K^b K^c(z)$. 
Among 18 equations (\ref{rel1}), the first four and the last four
equations are unchanged and the rest  should be modified because of
the extra terms in $\frac{1}{(z-w)^2}$ terms of the operator product
expansion (\ref{jprimeeq}).
The first two equations of (\ref{rel2}) have to be modified due to the 
$\frac{1}{(z-w)^3}$
  term in the operator product expansion (\ref{jprimeeq}).
The first and fourth equations in (\ref{rel4}) should be changed due
to the $\frac{1}{(z-w)^3}$ in the operator product expansion (\ref{tprimeeq}).
Finally, the first and third equations in (\ref{rel5})
have to be changed from the $\frac{1}{(z-w)^4}$ terms in the operator
product expansion (\ref{tprimeeq}).
Then the coefficients $c_i$ should be modified as 
$(c_i + b_i)$, when we add the extra field $J^a J^b K^a K^b(z)$,  
where the coefficients $b_i$ depend on $c_{23}$ (there is no
constraint on $c_{23}$) and 
they are given by 
\bea
b_2 & = &
\frac{(k_2^2-1)N^2(N^2+1)^2 }{3(N^2-4)(N+2k_1)(N^2-3)D(k_1,k_2,N)}
\left[ -6k_1^3-6k_1^2 k_2 -17 k_1^2 N -5k_1 k_2 N \right. \nonu \\
&- & \left. 12 k_1 N^2 + 2k_1^3 N^2 + 2k_2 N^2 + 2k_1^2 k_2 N^2 + 7
  k_1^2 N^3 + 3 k_1 k_2 N^3 + 6k_1 N^4 \right] c_{23},
\nonu \\
b_3 & = & -\frac{N(N^2+1)}{4(N^2-4)D(k_1,k_2,N)} \left[ 12 k_1^3 +
  24k_1^2 k_2 + 12 k_1 k_2^2 + 44 k_1^2 N + 52 k_1 k_2 N + 2k_1^3 k_2
  N \right. \nonu \\
&+ & \left. 8 k_2^2 N + 2k_1^2 k_2^2 N  + 48k_1 N^2 -8k_1^3 N^2 + 24
  k_2 N^2 - 9k_1^2 k_2 N^2 -k_1 k_2^2 N^2 + 16N^3 \right.
\nonu \\
& - &  26k_1^2 N^3 -20 k_1 k_2 N^3 
+ 2k_1^3 k_2 N^3 + 2k_2^2 N^3 + 2k_1^2 k_2^2 N^3 -24 k_1 N^4 - 4k_2
N^4 + 7k_1^2 k_2 N^4 
\nonu \\
&+ & \left.
3k_1 k_2^2 N^4 - 6N^5 + 6k_1 k_2 N^5 \right] c_{23},
\nonu \\
b_4 & = & -\frac{N^2 (N^2+1)^2 }
{6(N^2-4)(2N+3k_2)(N^2-3)D(k_1,k_2,N)}
\left[ -18 k_1^2 k_2 + 18 k_1^4 k_2 -18 k_1 k_2^2 + 18 k_1^3 k_2^2  \right.
\nonu \\
&+ & 12 k_1^4 N -60 k_1 k_2 N + 75 k_1^3 k_2 N -12 k_2^2 N + 12
k_1^2 k_2^2 N -15 k_1 k_2^3 N -32 k_1 N^2 + 42 k_1^3 N^2  \nonu
\\
& - &  32 k_2 N^2 +
84 k_1^2 k_2 N^2  - 6 k_1^4 k_2 N^2 -40 k_1 k_2^2 N^2 - 6k_1^3 k_2^2 N^2
-10 k_2^3 N^2 - 16 N^3 + 52 k_1^2 N^3 \nonu \\
& + &   16 k_1 k_2 N^3 
 - 21 k_1^3 k_2 N^3 - 32 k_2^2 N^3 - 4k_1^2 k_2^2 N^3 + 5 k_1 k_2^3
N^3 + 26 k_1 N^4 - 18 k_2 N^4 \nonu \\
&- & \left. 18 k_1^2 k_2 N^4 + 10 k_1 k_2^2 N^4 - 12
  k_1^2 N  \right] c_{23},
\nonu \\
b_5 & = & -\frac{k_1(k_1-k_2)N^2(3N+2k_1+2k_2)(N^2+1)^2 }
{8(N^2-4)(2N+3k_2)(N+k_2)(N^2-3)D(k_1,k_2,N)} \left[ 
- 3k_1^2 k_2 - 3k_1 k_2^2 \right. \nonu \\
& - &  \left. 2k_1^2 N - 8k_1 k_2 N 
-   2k_2^2 N - 4k_1 N^2 - 4k_2 N^2 + k_1^2 k_2 N^2 + k_1 
k_2^2 N^2 - 2N^3 + 2k_1 k_2 N^3 \right] c_{23}, 
\nonu \\
b_6 & = & -\frac{(k_2^2-1)k_2N^2(N^2+1) }
{2(N^2-4)(N+2k_1)D(k_1,k_2,N)} \left[ k_1^2 + k_1 k_2 +
2 k_1 N + 2k_2 N + 2N^2 + k_1^2 N^2 \right.
\nonu \\
&+ & \left. k_1 k_2 N^2 + 2k_1 N^3  \right] c_{23},
\nonu \\
b_7 & = & -\frac{(k_2^2-1)N^3(N^2+1)}
{(N^2-4)(N+2k_1)(N^2-3)} \left[ k_1^2 +
13 k_1 k_2 + 12 k_1 N + 14 k_2 N + 12 N^2 
+2 k_1^2 N^2 \right.
\nonu \\
& - & \left. 2 k_1 k_2 N^2 - 2k_1 N^3 -2k_2 N^3 -4N^4+
k_1^2 N^4 + k_1 k_2 N^4 + 2k_1 N^5 \right] c_{23},
\nonu \\
b_ 8 & = & \frac{N(N^2+1)}
{2(N^2-4)(N+2k_1)(N+2k_2)D(k_1,k_2,N)} 
\left[ 72 k_1^4 k_2 + 144 k_1^3 k_2^2 + 72 k_1^2 k_2^3 +
36 k_1^4 N \right. \nonu \\
& + &  360 k_1^3 k_2 N 
+  408 k_1^2 k_2^2 N + 84 k_1 k_2^3 N + 144 
k_1^3 N^2 + 592 k_1^2 k_2 N^2 - 44 k_1^4
k_2 N^2 + 80 k_2^2 N^3 \nonu \\
& + & 
 328 k_1 k_2^2 N^2 -88 k_1^3 k_2^2 N^2+24 
k_2^3 N^2 -44 k_1^2 k_2^3 N^2 + 
212 k_1^2 N^3 - 4 k_1^4 N^3 + 388  k_1 k_2 N^3 
\nonu \\
& - & 
184 k_1^3 k_2 N^3 -219 k_1^2 k_2^2 N^3 -39 k_1 k_2^3 
N^3 + 136 k_1 N^4 - 16 k_1^3 N^4 + 88 k_2 N^4 - 
251 k_1^2 k_2 N^4 \nonu \\
&+ & 4k_1^4 k_2 N^4 - 137 k_1 
k_2^2 N^4 + 8k_1^3 k_2^2 N^4-
10 k_2^3 N^4 + 4k_1^2 k_2^3 N^4 + 
32 N^5 -26 k_1^2 N^5 \nonu \\
&- &  128 k_1 k_2 N^5 + 16 k_1^3 k_2 N^5 - 
26 k_2^2 N^5 + 17 k_1^2 k_2^2 N^5 + 
k_1 k_2^3 N^5 - 20 k_1 N^6 - 22 k_2 N^6 \nonu \\
& + & \left. 19 k_1^2 k_2 N^6 
+  5 k_1 k_2^2 N^6 - 6N^7 + 
6 k_1 k_2 N^7    \right] c_{23},
\nonu \\ 
b_9 & = & \frac{N^3(N^2+1)}{(N^2-4)(N+2k_2)(2N+3k_2)
(N^2-3)D(k_1,k_2,N)} \left[33 k_1^2 k_2 + 18 k_1^4 k_2 -
3k_1 k_2^2 + 18 k_1^3 k_2^2 \right.
\nonu \\
& - & 51 k_1^2 k_2^3 -15 k_1 k_2^4
+   30 k_1^2 N +36 k_1^4 N+ 46 k_1 k_2 N + 99 k_1^3 k_2 N
-2 k_2^2 N - 54 k_1^2 k_2^2 N  \nonu \\
& - &   10 k_2^4 N + 44 k_1 N^2 + 126 k_1^3 N^2
+ 16 k_2 N^2 + 76 k_1^2 k_2 N^2 + 16 k_1^4 k_2 N^2 -212 k_1 k_2^2
N^2 \nonu \\
& + &   16 k_1^3 k_2^2 N^2 
-74 k_2^3 N^2 + 2k_1^2 k_2^3 N^2 - 10 k_1 k_2^4 N^2 +16N^3
+ 116k_1^2 N^3 - 4 k_1^4 N^3 -82 k_1 k_2 N^3 \nonu \\
& + & 
52 k_1^3 k_2 N^3 - 114 k_2^2 N^3 + 44 k_1^2 k_2^2 N^3
-18 k_1 k_2^3 N^3 - 10 k_2^4 N^3 + 10k_1 N^4 - 14 k_1^3 N^4-
66 k_2 N^4 \nonu \\
& + &   69 k_1^2 k_2 N^4 
-  2k_1^4 k_2 N^4 + 25 k_1 k_2^2 N^4 - 2k_1^3 k_2^2 N^4 - 30 
k_2^3 N^4 + 5k_1^2 k_2 ^3 N^4 + 5k_1 k_2^4 N^4 \nonu \\
& - &   16 N^5
-10 k_1^2 N^5  + 40 k_1 k_2 N^5 - 7k_1^3 k_2 N^5
- 24 k_2^2 N^5 + 2k_1^2 k_2^2 N^5 + 15 k_1 k_2^3 N^5
\nonu \\
&+ &    \left. 2k_1 N^6 - 6k_2 N^6 - 6k_1^2 k_2 N^6 + 
10 k_1 k_2^2 N^6 - 145 k_1 k_2^3 N  \right] c_{23},
\nonu \\
b_{10} & = & \frac{k_1 N^2(N^2+1)}
{4(N^2-4)(N+2k_2)(N+k_2)(2N+3k_2)(N^2-3)D(k_1,k_2,N)} 
\left[-18k_1^2 k_2^2 - 18 k_1 k_2^3\right.
\nonu \\
&+ &  18 k_1 k_2^5- 54 k_1^2 k_2 N -18 k_1^4 k_2 N- 66 k_1 k_2^2 N -18
k_1^3 k_2^2 N -12 k_2^3 N +72 k_1^2 k_2^3 N \nonu \\
&+ &  84 k_1 k_2^4 N + 12 k_2^5 N -36 k_1^2 N^2 - 36 k_1^4 N^2
-108 k_1 k_2 N^2 - 99 k_1^3 k_2 N^2 -36 k_2^2 N^2 + 48 k_1^2 k_2^2 N^2 
 \nonu \\
&+ & 195 k_1 k_2^3 N^2 + 48 k_2^4 N^2 + 12 k_1^2 k_2^4 N^2 +
12 k_1 k_2^5 N^2 -60 k_1 N^3 - 126 k_1^3 N^2 - 48 k_2 N^3
\nonu \\
&- & 90 k_1^2 k_2 N^3 - 16 k_1^4 k_2 N^3 + 178 k_1 k_2^2 N^3 -16
k_1^3 k_2^2 N^3 + 94 k_2^3 N^3 + 12 k_1^2 k_2^3 N^3+60 k_1 k_2^4 N^3 \nonu \\
&+ &  12 k_2^5 N^3 - 24 N^4 -120 k_1^2 N^4 + 4k_1^4 N^4 + 36
k_1 k_2 N^4 -52 k_1^3 k_2 N^4 + 84 k_2^2 N^4 - 34 k_1^2 k_2^2 N^4  
\nonu \\
&+ &  70 k_1 k_2^3 N^4 + 48 k_2^4 N^4 - 6k_1^2 k_2^4 N^4 -
6k_1 k_2^5 N^4 - 22 k_1 N^5 + 14 k_1^3 N^5 + 34 k_2 N^5 
\nonu \\
&- & 62 k_1^2 k_2 N^5 + 2k_1^4 k_2 N^5 + 6k_1 k_2^2 N^5 +
2k_1^3 k_2^2 N^5 + 62 k_2^3 N^5 -12 k_1^2 k_2^3 N^5 -24 k_1 k_2^4 N^5 
\nonu \\
& +&  8N^6 + 12 k_1^2 N^6 -24 k_1 k_2 N^6 + 7k_1^3 k_2 N^6
+ 32 k_2^2 N^6 - 4k_1^2 k_2^2 N^6 - 31 k_1 k_2^3 N^6 + 2k_1 N^7
\nonu \\
&+ & \left. 
6k_2 N^7 + 6k_1^2 k_2 N^7 - 14 k_1 k_2^2 N^7  + 18 k_1^2 k_2^4   \right] c_{23},
\nonu \\
b_{11} & = & \frac{(k_2^2-1)k_2 N (N^2+1)
(3N+2k_1 +2k_2)}{2D(k_1,k_2,N)} c_{23},
\nonu \\
b_{12} & = & 
\frac{2(k_2^2-1)N^2(N^2+1)}{(N+2k_1)
(N^2-3)D(k_1,k_2,N)}
\left[ 2k_1^2 + 8k_1 k_2 + 9k_1 N + 10 k_2 N \right.
\nonu \\
& +& \left. 9N^2 + 2k_1^2 N^2 + k_1 N^3 -2k_2 N^3 -3N^4 
  \right] c_{23},
\nonu \\
b_{13} & = & \frac{1}{D(k_1,k_2,N)} \left[-18k_1^3 -36 k_1^2 k_2
-18 k_1 k_2^2 -63 k_1^2 N -75 k_1 k_2 N -12 k_2^2 N -67 k_1 N^2
\right.
\nonu \\
&+ &  2k_1^3 N^2-34 k_2 N^2 + 4k_1^2 k_2 N^2 -k_1 k_2^2 N^2
-22 N^3 + 7 k_1^2 N^3 \nonu \\
&+ &  \left. 3k_1 k_2 N^3 - 6k_2^2 N^3 + 5k_1 N^4 - 
6k_2 N^4 + k_1 k_2^2 N^4 \right] c_{23}, \nonu \\
b_{14} & = & -\frac{2N^2(N^2+1)}
{(2N+3k_2)(N^2-3)D(k_1,k_2,N)} 
\left[ 6k_1^2 + 18 k_1^4 - 3k_1 k_2 + 
18 k_1^3 k_2 - 24 k_1^2 k_2^2 -15 k_1 k_2^3 
\right.
\nonu \\
& + & 7k_1 N + 63 k_1^3 N - 2k_2 N + 
12 k_1^2 k_2 N -70 k_1 k_2^2 N - 10 k_2^3 N + 2N^2  +
64 k_1^2 N^2 \nonu \\
& -& 2k_1^4 N^2 -39 k_1 k_2 N^2 - 2k_1^3 k_2 N^2 - 
36 k_2^2 N^2 + 8k_1^2 k_2^2 N^2 + 5k_1 k_2^3 N^2 + 
13 k_1 N^3 \nonu \\
&- & \left. 7 k_1^3 N^3 -
28 k_2 N^3 + 4k_1^2 k_2 N^3 + 18 k_1 k_2^2 N^3-
6N^4 - 6k_1^2 N^4 + 12 k_1 k_2 N^4  
\right] c_{23}, 
\nonu \\
b_{15} & = & -\frac{N k_1(3N+2k_1+2k_2)(N^2+1)}
{2(N+k_2)(2N+3k_2)(N^2-3)D(k_1,k_2,N)} \left[
-9k_1 k_2 + 9k_1 k_2^3  -9k_1 N -
9k_1^3 N \right.
\nonu \\
&- &  6k_2 N + 18 k_1 k_2^2 N + 6k_2^3 N -
6N^2 -18 k_1^2 N^2 + 15 k_1 k_2 N^2 +
12 k_2^2 N^2 - 3k_1 k_2^3 N^2 \nonu \\
&- & \left. 3k_1 N^3 + k_1^3 N^3 + 8 k_2 N^3 -
6k_1 k_2^2 N^3 + 2N^4 + 2k_1^2 N^4 -4k_1 k_2 
N^4   \right] c_{23}, \nonu \\
b_{16} & = & 0, 
\nonu \\
b_{17} & = & - \frac{(k_2^2-1)N^2(N^2+1)}
{6(N+2k_1)D(k_1,k_2,N)} \left[ -6k_1^3 - 6k_1^2 k_2 - 
17 k_1^2 N -5k_1 k_2 N \right.
\nonu \\
&- & \left.
12 k_1 N^2 + 2k_1 ^3 N^2 + 2k_2 N^2 + 2k_1^2 k_2 N^2 +
7k_1^2 N^3 + 3k_1 k_2 N^3 + 6k_1 N^4  \right] c_{23}, \nonu \\
b_{18} & = & 0,
\nonu \\
b_{19} & = & 0,
\nonu \\
b_{20} & = & \frac{N}{2D(k_1,k_2,N)} 
\left[ 18k_1^3 + 36 k_1^2 k_2 + 18 k_1 k_2^2 +
66 k_1^2 N + 78 k_1 k_2 N + 12k_2^2 N -
3 k_1^2 k_2^2 N \right.
\nonu \\
&- & 3k_1 k_2^3 N+ 72 k_1 N^2 - 20 k_1^3 N^2 + 
36 k_2 N^2 - 40 k_1^2 k_2 N^2 - 22 k_1 k_2^2 N^2 
-2 k_2^3 N^2 \nonu \\
&+ & 24 N^3 - 68 k_1^2  N^3 - 76 k_1 k_2 N^3 -
8 k_2^2 N^3 - 2k_1^2 k_2^2 N^3 - 2k_1 k_2^3 
N^3  -68 k_1 N^4 \nonu \\
&+ & 2k_1^3 N^4 - 26 k_2 N^4 + 4k_1^2 k_2 N^4 -
6k_1 k_2^2 N^4 - 2k_2^3 N^4 - 20N^5 + 6k_1^2 N^5
+ 2k_1 k_2 N^5 
\nonu \\
&- & \left. 8 k_2^2 N^5 +k_1^2 k_2^2 N^5 + k_1 k_2^3 N^5 + 
4k_1 N^6 - 6k_2 N^6 + 2k_1 k_2^2 N^6  
\right] c_{23},
\nonu \\
b_{21} & = & 0,
\nonu \\
b_{22} & = & -\frac{(k_2^2-1)N^2(2N+k_1+k_2)(N^2+1)
(-3k_1-2N+k_1 N^2) }{6D(k_1,k_2,N)} c_{23}. 
\label{b}
\eea
Since the second and third equations in (\ref{rel4}) and the equation
(\ref{relfin})
remain unchanged, one sees that the coefficients $b_{16}, b_{18},
b_{19}$ and $b_{21}$ should vanish.
The $D(k_1,k_2,N)$ is defined as (\ref{dcoeff}).
Furthermore, the large $N$ limit (\ref{limit}) on these coefficients leads to 
\bea
b_2 & = & -\left[\frac{(-1+\la)(1+\la)}{2\la(2+\la)}\right] c_{23}, \qquad
b_3 = -\left[\frac{3(1+\la)}{8(2+\la)}\right] c_{23}, \nonu \\
b_4  & = & - \left[\frac{5(-1+\la)(1+\la)}{12(-3+\la)(2+\la)}\right] c_{23}, \qquad  
b_5 = \left[\frac{(-1+\la)(1+\la)}{16N(-3+\la)}\right] c_{23}, \nonu \\ 
b_6 & = & -\left[\frac{(-1+\la)^2(1+\la)}{4\la^2(2+\la)}\right] c_{23}, \qquad
b_7  =  \left[\frac{(-1+\la)(1+\la)}{2\la(2+\la)}\right] c_{23}, \nonu \\  
b_8 & = & -\left[\frac{(1+\la)(1+2\la)}{4(-2+\la)(2+\la)}\right] c_{23}, \qquad
b_9 = -\left[\frac{5(-1+\la)(1+\la)}{2(-3+\la)(-2+\la)(2+\la)}\right] c_{23}, \nonu \\
b_{10} & = & \left[\frac{(-1+\la)(1+\la)(6+\la^2)}{8N(-3+\la)(-2+\la)(2+\la)}
\right] c_{23},
\qquad 
b_{11} = \left[\frac{(-1+\la)^2}{4\la^2}\right] c_{23}, \nonu \\
b_{12} & = & \left[\frac{(-1+\la)}{\la}\right] c_{23}, \qquad
b_{13} = -\left[\frac{(-1+\la)}{2N(2+\la)}\right] c_{23}, \nonu \\
b_{14} & = &  -\left[\frac{(-5-8\la+\la^2)}{N(-3+\la)(2+\la)}\right] c_{23}, 
\qquad
b_{15} = -\left[\frac{(3+\la^2)}{4N^2(-3+\la)}\right] c_{23}, \nonu \\
b_{17} & = & \left[\frac{N^2(-1+\la)(1+\la)}{4\la(2+\la)} \right] c_{23}, \qquad
b_{20} = -\left[\frac{N^2(-1+\la)(1+\la)}{4\la(2+\la)}\right] c_{23}, \nonu \\
b_{22}  & = & \left[\frac{N^2(-1+\la)(1+\la)}{12\la(2+\la)}\right] c_{23}.
\label{limitb}
\eea
There are the $\frac{1}{N}$ dependence in $b_5, b_{10}, b_{13}, b_{14}$ and
$\frac{1}{N^2}$ dependence in $b_{15}$ and $N^2$ dependence in
$b_{17}, b_{20}, b_{22}$. 
Then, the more general spin $4$ field, in the large $N$ limit, can be written 
as (\ref{limitW}).
The extra terms proportional to $c_{23}$ in (\ref{limitW}) appear
also in (\ref{WWWW}). The contributions from $J^a J^a J^b J^b(z), J^a J^a
J^b K^b(z)$ and $J^a J^b K^a K^b(z)$ in the large $N$ limit become zero.

\section{The coset spin $4$ primary field from the operator product  
expansion between the third-order Casimirs}

The primary field of dimension $4$ under the stress energy tensor 
(\ref{Ttilde}) and (\ref{stressetal})
has been constructed in \cite{BBSS1} from the operator product expansion between 
the primary fields of dimension $3$.  
That is, the coset spin $3$ primary field is given by 
\bea
\widetilde{T}^{(3)}(z) = B d^{abc} \left[ A_1 J^a J^b J^c(z) + A_2 J^a J^b K^c(z) + A_3
  J^a K^b K^c(z) + A_4 K^a K^b K^c(z) \right], 
\label{t3tilde}
\eea
where the coefficient functions that depend on $k_1, k_2$ or $N$ 
are determined by \cite{BBSS1}
\bea
B^2 & = & -\frac{N}{18(N+k_1)^2(N+k_2)^2(N+k_1+k_2)^2(N+2k_1)
(N+2k_2)(3N+2k_1+2k_2)(N^2-4)}, \nonu \\
A_1 & = & k_2 (N+k_2)(N+2k_2), \nonu \\
A_2 & = & -3 (N+k_1)(N+k_2)(N+2k_2), \nonu \\
A_3 & = & 3(N+k_1)(N+k_2)(N+2k_1), \nonu \\
A_4 & = & -k_1 (N+k_1)(N+2k_1).
\label{BA}
\eea

By focusing on the $\frac{1}{(z-w)^2}$ term in the operator product
expansion between $\widetilde{T}^{(3)}(z)$ and
$\widetilde{T}^{(3)}(w)$
which can be constructed from the relation (\ref{t3tilde}),
one obtains
\bea
&& \widetilde{R}^{(4)}(w) + \frac{32}{22 +5\widetilde{c}} 
  \widetilde{\Lambda}(w) + \frac{3}{10}  \pa^2 \widetilde{T}(w)  =  
\nonu \\
&& d^{abcd} \left( 
d_2  J^a J^b J^c K^d(w) +
d_3  J^a J^b K^c K^d(w)
+
d_4  J^a K^b K^c K^d(w)
 \right) \nonu \\
&&+     d^{abe} d^{cde} \left( d_6 J^a J^b J^c J^d(w)+
d_8   J^a J^b K^c K^d(w)
+   
d_{10}  K^a K^b K^c K^d(w) \right)
\nonu \\
&&+  
d_{12} J^a J^a J^b K^b(w) 
+
d_{13} J^a J^a K^b K^b(w)  +
d_{14} J^a K^a K^b K^b(w)  
+    d_{16} \pa^2 J^a J^a(w)
\nonu \\
&&+ 
d_{17} \pa^2 J^a K^a(w) 
+
d_{18} \pa^2 K^a K^a(w) + d_{20} \pa J^a \pa K^a(w) 
+   
d_{22} J^a \pa^2 K^a(w) 
\nonu \\
&&+  d_{23} J^a J^b K^a K^b(w) +d_{24} f^{abc} J^a \pa
J^b K^c(w)  
+d_{25} f^{abc} J^a K^b \pa K^c(w).
\label{r4}
\eea
Then a new spin-$4$ field $\widetilde{R}^{(4)}(w)$ can be read off
from (\ref{r4}). While there are new terms characterized by 
$d_{23}, d_{24}$ and $d_{25}$, but there are missing terms
corresponding to $c_1, c_5, c_7, c_9, c_{11}$, and $c_{15}$. 
Of course, it is not so simple to see the first three terms on the
right hand side of (\ref{r4}). 
Here the coefficient functions on the right hand side of (\ref{r4})
depend on $k_1, k_2$ or $N$ and they are determined by, together with
the relation (\ref{BA}), 
\bea
d_2 & = & \left[-4 A_1 A_2 (N+k_1) -\frac{2}{3} A_2 A_3 (N+2k_2)
+\frac{2}{3} 
A_2^2 N \right] B^2  \nonu \\
& = & 12(N+k_1)^2(N+k_2)^2(N+k_1+k_2)(N+2k_2)^2 B^2, \nonu \\
d_3 & = & \left[-A_2^2 (N+2k_1) -A_3^2(N+2k_2) \right]B^2 \nonu \\
& = & -18(N+k_1)^2(N+2k_1)(N+k_2)^2(N+k_1+k_2)(N+2k_2) B^2, \nonu \\
d_4 & = & \left[-\frac{2}{3} A_2 A_3 (N+2k_1) - 4A_3 A_4(N+k_2) +\frac{2}{3}
A_3^2 N \right]B^2 \nonu \\
& = & 12(N+k_1)^2(N+2k_1)^2(N+k_2)^2(N+k_1+k_2) B^2, \nonu \\
d_6 & = & \left[ - 9 A_1^2 (N+k_1) - A_2^2 k_2  \right]B^2=
-9k_2(N+k_1)(N+k_2)^2(N+k_1+k_2)(N+2k_2)^2 B^2, \nonu \\
d_8 & = & \left[ -6 A_1 A_3 (N+k_1) + 4 A_2 A_3 N-6 A_2 A_4 (N+k_2)
+ 2 A_2^2 k_1 + 2A_3^2 k_2  \right]B^2 \nonu \\
& = & -36N(N+k_1)^2(N+k_2)^2(N+k_1+k_2)^2 B^2, \nonu \\
d_{10} & = & \left[ -9 A_4^2 (N+k_2) -A_3^2 k_1\right] B^2
=-9k_1(N+k_1)^2(N+2k_1)^2
(N+k_2)(N+k_1+k_2) B^2, \nonu \\
d_{12} & = & \left[ -\frac{48(N^2-4)(N+k_1)}{N(N^2+1)} A_1 A_2 -
\frac{8(N^2-4)(N+2k_2)}{N(N^2+1)} A_2 A_3 +\frac{8(N^2-4)}{N^2+1} A_2^2  \right]B^2 \nonu \\
& = & \frac{144}{N(N^2+1)} (N^2-4)(N+k_1)^2(N+k_2)^2(N+k_1+k_2)(N+2k_2)^2 B^2, \nonu \\
d_{13} & = & \left[-\frac{4(N^2-4)(N+2k_1)}{N(N^2+1)} A_2^2 -
\frac{4(N^2-4)(N+2k_2)}{N(N^2+1)} A_3^2  \right] B^2 \nonu \\
& = & -\frac{72}{N(N^2+1)}
(N^2-4)(N+k_1)^2(N+2k_1)(N+k_2)^2(N+k_1+k_2)(N+2k_2) B^2, \nonu \\
d_{14} & = & \left[ -\frac{8(N+2k_1)(N^2-4)}{N(N^2+1)} A_2 A_3
  -\frac{48(N+k_2)(N^2-4)}{N(N^2+1)}
 A_3 A_4 + \frac{8(N^2-4)}{N^2+1} A_3^2  \right] B^2 \nonu \\
& = & \frac{144}{N(N^2+1)}
(N^2-4)(N+k_1)^2(N+2k_1)^2(N+k_2)^2(N+k_1+k_2) B^2,\nonu \\
d_{16} & = & \left[ \frac{9}{N} A_1^2 (N+k_1)(N+2k_1)(N^2-4) +
\frac{2k_2}{N} (N^2-4)(N+2k_1) A_2^2 \right.
\nonu \\
&+ & \left. \frac{1}{N} (N^2-4) k_2 (N+2k_2) 
A_3^2  \right] B^2 \nonu \\
& = & \frac{1}{N}
9k_2(N^2-4)(N+k_1)(N+2k_1)(N+k_2)^2(N+k_1+k_2)(N+2k_2)(3N+2k_1+2k_2) B^2,\nonu \\
d_{17} & = & \left[\frac{8(N+k_1)(N^2-4)(N^2+3)}{N^2+1} A_1 A_2
  +\frac{3}{N} A_1 A_2 (N+k_1)(N+2k_1)(N^2-4)
  \right. \nonu \\
&+ &  \frac{4(N+2k_2)(N^2-4)(N^2+3)}{3(N^2+1)} A_2 A_3 +
  \frac{1}{N} (N+2k_1)(N+2k_2)(N^2-4) A_2 A_3 \nonu \\
&+ &  \frac{3}{N}
  (N+k_2)(N+2k_2)(N^2-4) 
A_3 A_4  + (-2(N+2k_1)(N^2-4) \nonu \\
& - & \left. \frac{4N(N^2-4)(N^2+3)}{3(N^2+1)})A_2^2
 - 2 A_3^2 (N+2k_2)(N^2-4)
\right] B^2 \nonu \\
& = & \frac{1}{N(N^2+1)}
3(N^2-4)(N+k_1)^2(N+k_2)^2(N+k_1+k_2)(N+2k_2)\left[
12k_1^2 +12k_1 k_2 + 36 k_1 N \right.
\nonu \\
& + & \left. 54k_2 N + 39N^2 + 12 k_1 k_2 N^2 +
36 k_1 N^3 + 22 k_2 N^3 + 23 N^4
\right] B^2, \nonu \\
d_{18} & = & \left[ \frac{9}{N} (N+k_2)(N+2k_2)(N^2-4) A_4^2
  +\frac{1}{N}
(N^2-4)k_1(N+2k_1) A_2^2  \right.
\nonu \\
&+ & \left. \frac{2k_1}{N} (N^2-4)(N+2k_2) A_3^2 \right] B^2 \nonu \\
& = & \frac{9}{N}
k_1(N^2-4)(N+k_1)^2(N+2k_1)(N+k_2)(N+k_1+k_2)(N+2k_2)(3N+2k_1+2k_2) B^2, \nonu \\
d_{20} & = & \left[ -\frac{(N+2k_1)(N^2-4)(N^2-3)}{N^2+1}  A_2^2
  -\frac{(N+2k_2)(N^2-4)(N^2-3)}{N^2+1} A_3^2 \right] B^2 \nonu \\
& = & -\frac{18}{(N^2+1)}
(N^2-4)(N+k_1)^2(N+2k_1)(N+k_2^2)(N+k_1+k_2)(N+2k_2)(N^2-3) B^2, \nonu \\
d_{22} & = & \left[\frac{3}{N}(N+k_1)(N+2k_1)(N^2-4) A_1 A_2
  +\frac{4(N+2k_1)
(N^2-4)(N^2-3)}{3(N^2+1)} A_2 A_3 \right. \nonu \\
&+&   \frac{1}{N}(N+2k_1)(N+2k_2)(N^2-4) A_2 A_3 +
\frac{8(N^2-4)(N^2-3)(N+k_2)}{N^2+1}
A_3 A_4 \nonu \\
&+&   \frac{3}{N} (N+2k_2)(N^2-4)(N+k_2) A_3 A_4 -
  2(N+2k_1)(N^2-4) A_2^2 
 +\frac{8N(N^2-4)}{N^2+1} A_3^2  \nonu \\
&+ & \left. (-2(N+2k_2)(N^2-4) - \frac{4N(N^2-4)(N^2+3)}
{3(N^2+1)} )A_3^2  \right] B^2 \nonu \\
& = & -\frac{1}{N(N^2+1)}
3(N^2-4)(N+k_1)^2(N+2k_1)(N+k_2)^2(N+k_1+k_2)\left[ 12 k_1 k_2 +
  12k_2^2 \right. \nonu \\
& -& \left. 42 k_1 N + 36 k_2 N - 9N^2 + 12 k_1 k_2 N^2 + 12 k_2^2 N^2 + 22
k_1 N^3 + 36 k_2 N^3 + 23 N^4  \right] B^2, \nonu \\
d_{23} & = & \left[ -\frac{8(N^2-4)(N+2k_1)}{N(N^2+1)} A_2^2-
\frac{8(N^2-4)(N+2k_2)}{N(N^2+1)} A_3^2  \right] B^2 \nonu \\
& = & -\frac{144}{N(N^2+1)}
(N^2-4)(N+k_1)^2(N+2k_1)(N+k_2)^2(N+k_1+k_2)(N+2k_2) B^2, \nonu \\
d_{24} & = & \left[ - \frac{12(N+k_1)(N^2-4)(N^2+5)}{N(N^2+1)} A_1 A_2
-\frac{2(N+2k_2)(N^2-4)(N^2+5)}{N(N^2+1)} A_2 A_3 \right. \nonu \\
&+ & \left. (\frac{2}{N} (N+2k_1)(N^2-4) +\frac{2(N^2-4)(N^2+5)}{N^2+1})
  A_2^2
+\frac{2}{N} (N^2-4)(N+2k_2) A_3^2  \right] B^2 \nonu \\
& = & \frac{1}{N(N^2+1)}
72(N^2-4)(N+k_1)^2(N+k_2)^2 B^2, \nonu \\
d_{25} & = & \left[ -\frac{2(N+2k_1)(N^2-4)(N^2-3)}{N(N^2+1)} A_2 A_3
  +
\frac{12(N+k_2)(N^2-4)(N^2+5)}{N(N^2+1)} A_3 A_4 \right.  \nonu \\
&+ &  (\frac{2}{N} (N+2k_2)(N^2-4) +\frac{2(N^2-4)(N^2-3)}{N^2+1}
  )A_3^2 
+ \frac{2(N^2-4)(N+2k_1)}{N} A_2^2  \nonu \\
 &-& \left.  \frac{16(N^2-4)}{N^2+1} A_3^2 \right] B^2
\nonu \\
& = & \frac{72}{N(N^2+1)}
(N^2-4)(N+k_1)^2(N+2k_1)(N+k_2)^2(N+k_1+k_2) \left[-3k_1+k_2-
N \right. \nonu \\
& + & \left.  k_1 N^2 + k_2 N^2 +N^3 \right] B^2,
\label{dcoeff1}
\eea
where $B(k_1,k_2,N)$ is given by (\ref{BA}).
The large $N$ limit (\ref{limit}) can be obtained as before 
\bea
d_2 & = & \frac{2(-2+\la)\la}{3N^3(2+\la)}, \qquad 
d_3 = \frac{\la^2}{N^3(2+\la)}, \qquad
d_4  =  \frac{2\la^2}{3N^3(-2+\la)(2+\la)}, \nonu \\
d_6 & = & \frac{(-2+\la)(-1+\la)}{2N^3(2+\la)}, \qquad
d_8  =  -\frac{2\la^2}{N^3(-2+\la)(2+\la)}, \qquad
d_{10} = -\frac{\la^4}{2N^4(-2+\la)(2+\la)}, \nonu \\
d_{12} & = & \frac{8(-2+\la)\la}{N^4(2+\la)}, \qquad
d_{13} =  \frac{4\la^2}{N^4(2+\la)}, \qquad
d_{14}  =  \frac{8\la^3}{N^4(-2+\la)(2+\la)}, \nonu \\
d_{16} & = &  \frac{-1+\la}{2N}, \qquad
d_{17} = \frac{\la(22+\la)}{6N(2+\la)}, \qquad
d_{18}  =  -\frac{\la^2}{2N^2}, \nonu \\
d_{20}  & = &  \frac{\la^2}{N(2+\la)}, \qquad
d_{22} = \frac{\la(-12-12\la +\la^2)}{6N(-2+\la)(2+\la)}, \qquad
d_{23} =  \frac{8\la^2}{N^4(2+\la)}, \nonu \\
d_{24} & = & -\frac{4\la}{N^2(2+\la)}, \qquad
d_{25}  =  \frac{4\la^2}{N^2(-2+\la)(2+\la)}.
\label{dlimit}
\eea
Note that there are the $\frac{1}{N^4}$ dependence in $d_{10}, d_{12},
d_{13}, d_{14}, d_{23}$ and
$\frac{1}{N}$ dependence in $d_{16}, d_{17}, d_{20}, d_{22}$ and 
$\frac{1}{N^2}$ dependence in $d_{18}, d_{24}, d_{25}$. 
The $\widetilde{\Lambda}(w)$ is defined in previous Appendix $G$ when we
discuss the operator product expansion (\ref{tlambda}). 
The field $J^a J^b K^a K^b(w)$ appears on the right hand side of
(\ref{r4}) and also on the left side of (\ref{r4}) through
$\widetilde{\Lambda}$ together with the relation (\ref{tt1}). 

We would like to see the explicit form for the $\widetilde{R}^{(4)}(z)$.
Let us present the relevant $\frac{1}{(z-w)^2}$ terms in the operator
product expansions in details and explain how we get (\ref{r4}):
\bea
Q(z) Q(w)|_{\frac{1}{(z-w)^2}} 
& = & -9(N+k_1) Q^a Q^a 
 -  \frac{9}{N}(N+k_1)(N+2k_1)(N^2-4)J^a \pa^2
J^a, \nonu \\
Q(z)  
Q^a K^a(w)|_{\frac{1}{(z-w)^2}} & = &
-3(N+k_1)(2 d^{abc} J^a Q^b K^c 
 -\frac{1}{N} (N+2k_1)(N^2-4) \pa^2 J^a
K^a), \nonu \\
Q(z) J^a R^a(w)|_{\frac{1}{(z-w)^2}} & = &
-3(N+k_1) Q^a R^a, \nonu \\
Q^a K^a(z)
Q(w)|_{\frac{1}{(z-w)^2}} & = &  
-3(N+k_1)(2 d^{abc} J^a Q^b K^c 
-\frac{1}{N} (N+2k_1)
(N^2-4) J^a \pa^2 K^a), \nonu \\
Q^a K^a(z)
Q^b K^b(w) |_{\frac{1}{(z-w)^2}} & = &
  d^{abcd} \left[ 
\frac{2}{3}  N  J^a J^b J^c K^d 
- (N+2k_1)  J^a J^b K^c K^d
 \right] \nonu \\
&+&      d^{abe} d^{cde} \left[ -  k_2  J^a J^b J^c J^d+
2  k_1    J^a J^b K^c K^d \right] 
\nonu \\
&+&  
\frac{8(N^2-4)}{N^2+1}  J^a J^a J^b K^b 
-\frac{4(N^2-4)(N+2k_1)}{N(N^2+1)} J^a J^a K^b K^b  
\nonu \\
&+ & \frac{2k_2}{N} (N^2-4)(N+2k_1)   \pa^2 J^a J^a
\nonu \\
&+& 
\left[-2(N+2k_1)(N^2-4)  -  \frac{4N(N^2-4)(N^2+3)}{3(N^2+1)}\right]
 \pa^2 J^a K^a 
\nonu \\
&+ & \frac{1}{N}
(N^2-4)k_1(N+2k_1)
 \pa^2 K^a K^a \nonu \\
&-&   \frac{(N+2k_1)(N^2-4)(N^2-3)}{N^2+1}
 \pa J^a \pa K^a 
\nonu \\
&-&
  2(N+2k_1)(N^2-4)
 J^a \pa^2 K^a 
\nonu \\
&-& 
 \frac{8(N^2-4)(N+2k_1)}{N(N^2+1)}
 J^a J^b K^a K^b  \nonu \\
& + &  \frac{2(N^2-4)(N+2k_1)}{N} 
f^{abc} J^a K^b \pa K^c
\nonu \\
&+&   
\left[\frac{2}{N} (N+2k_1)(N^2-4) +\frac{2(N^2-4)(N^2+5)}{N^2+1}\right]
f^{abc} J^a \pa
J^b K^c,
\nonu 
\\
Q^a K^a(z)  
J^b R^b(w) |_{\frac{1}{(z-w)^2}} & = & 
-(N+2k_2) d^{abc} J^a Q^b K^c 
 + 2N Q^a R^a \nonu \\
&-& (N+2k_1) d^{abc} J^a K^b
R^c  \nonu \\
& + &
\frac{1}{N} (N+2k_1)(N+2k_2)(N^2-4) \pa^2 J^a  K^a,\nonu \\
Q^a K^a(z) R(w)  |_{\frac{1}{(z-w)^2}} & = & -3(N+k_2) Q^a R^a,
\nonu \\
J^a R^a(z)  
Q(w) |_{\frac{1}{(z-w)^2}} & = & \left[ Q^a K^a(z)  
R(w) |_{\frac{1}{(z-w)^2}}\right]_{J^c \leftrightarrow K^c, 
k_1 \leftrightarrow k_2}, \nonu \\
J^a R^a(z)  Q^b K^b(w) |_{\frac{1}{(z-w)^2}} & = &  \left[ Q^a K^a(z)  
J^b R^b(w) |_{\frac{1}{(z-w)^2}}\right]_{J^c \leftrightarrow K^c, 
k_1 \leftrightarrow k_2}, \nonu \\
J^a R^a(z)  J^a R^a(w) |_{\frac{1}{(z-w)^2}} & = &  \left[Q^a K^a(z)  
Q^b K^b(w) |_{\frac{1}{(z-w)^2}}\right]_{ J^c \leftrightarrow K^c, 
k_1 \leftrightarrow k_2}, \nonu \\
J^a R^a(z)  R(w) |_{\frac{1}{(z-w)^2}} & = &  \left[Q^a K^a(z)  
Q(w) |_{\frac{1}{(z-w)^2}}\right]_{J^c \leftrightarrow K^c, 
k_1 \leftrightarrow k_2}, \nonu \\
R(z)  Q^a K^a(w) |_{\frac{1}{(z-w)^2}} & = &  \left[Q(z)  
J^a R^a(w) |_{\frac{1}{(z-w)^2}}\right]_{J^c \leftrightarrow K^c, 
k_1 \leftrightarrow k_2}, \nonu \\
R(z)  J^a R^a(w) |_{\frac{1}{(z-w)^2}} & = & \left[Q(z)  
Q^a K^a(w) |_{\frac{1}{(z-w)^2}}\right]_{J^c \leftrightarrow K^c, 
k_1 \leftrightarrow k_2}, \nonu \\
R(z)  R(w) |_{\frac{1}{(z-w)^2}} & = &  \left[Q(z)  
Q(w) |_{\frac{1}{(z-w)^2}}\right]_{J^c \leftrightarrow K^c, 
k_1 \leftrightarrow k_2}, 
\label{squareterms} 
\eea
where $R(w) \equiv d^{abc} K^a K^b K^c(w)$ corresponding to (\ref{Q}).
The right hand side of the first equation of (\ref{squareterms})
corresponds to the spin 4 field $R_4(w)$ in $(3.2)$ 
of \cite{BBSS} up to an overall constant.  
Here the last seven operator product
expansions (\ref{squareterms})
can be obtained from the first seven operator product expansions 
using the properties
$J^a \leftrightarrow K^a$ and $k_1 \leftrightarrow k_2$.

At first sight, the spin $4$ field (\ref{r4}) looks different from (\ref{cosetspin4}),
but the independent field contents are the same except 
the one field $J^a J^b K^a K^b(z)$.
Let us look at the structures in details.
First of all, one has the following identities
\bea
f^{abc} J^a \pa J^b K^c(w)  & = & \frac{N(N^2+1)}{(N^2-4)(N^2-3)} \left[ 
-  d^{abe} d^{cde} J^a J^b J^c K^d(w) + \frac{1}{3} d^{abcd} J^a J^b J^c
K^d(w)  \right. \nonu \\
& + & \left.
\frac{4(N^2-4)}{N(N^2+1)} J^a J^a J^b K^b(w)
 +  \frac{(N^2-4)(N^2-3)}{3(N^2+1)} \pa^2 J^a K^a(w)  \right],
\nonu \\
f^{abc} J^a  K^b  \pa K^c(w)  & = & \frac{N(N^2+1)}{(N^2-4)(N^2-3)} \left[ 
d^{abe} d^{cde} J^a K^b K^c K^d(w) - \frac{1}{3} d^{abcd} J^a K^b K^c
K^d(w)  \right. \nonu \\
& - & \left.
\frac{4(N^2-4)}{N(N^2+1)} J^a K^a K^b K^b(w)
 +   \frac{2(N^2-4)(N^2-3)}{3(N^2+1)} J^a \pa^2 K^a(w)  \right],
\nonu \\ 
\pa J^a \pa J^a(w) & = & \frac{(N^2+1)}{3(N^2-4)(N^2-3)} \left[ 
- d^{abcd} J^a J^b J^c J^d(w) + 3 d^{abe} d^{cde}  J^a J^b J^c J^d(w)
\right. 
\nonu \\ 
& - &  \left.
\frac{12(N^2-4)}{N(N^2+1)} J^a J^a J^b J^b(w) 
 +  \frac{2(N^2-4)(N^2-3)}{(N^2+1)} \pa^2 J^a J^a(w) \right],
\nonu \\   
\pa K^a \pa K^a(w) & = & \frac{(N^2+1)}{3(N^2-4)(N^2-3)} \left[ 
- d^{abcd} K^a K^b K^c K^d(w) + 3 d^{abe} d^{cde}  K^a K^b K^c K^d(w)
\right. \nonu \\
& - & \left.
\frac{12(N^2-4)}{N(N^2+1)} K^a K^a K^b K^b(w) 
 +  \frac{2(N^2-4)(N^2-3)}{(N^2+1)} \pa^2 K^a K^a(w) \right].
\label{rrela}
\eea
The first and second relations of (\ref{rrela}) 
can be obtained from the first and third relations of
(\ref{independent}), respectively. These two fields appear in (\ref{squareterms}).
We want to reexpress (\ref{r4}) in terms of the fields in (\ref{cosetspin4}).
Then, the $d_{24}$ term in (\ref{r4}) can be absorbed into $d_2,
d_{12}, d_{17}$
and a new term $ d^{abe} d^{cde} J^a J^b J^c K^d(w)$ corresponding to
$c_{7}$ term in (\ref{cosetspin4}).
Similarly, 
 the $d_{25}$ term in (\ref{r4}) can be absorbed into $d_4,
d_{14}, d_{22}$
and a new term $ d^{abe} d^{cde} J^a K^b K^c K^d(w)$ corresponding to
$c_{9}$ term in (\ref{cosetspin4}).
The third and fourth relations of (\ref{rrela})
can be obtained from the relation (\ref{dddd}).
The reason why we use these equations comes from the discussion in the
subsection \ref{spin4section} where we keep $c_{19}=0=c_{21}$.
Since the left hand side of (\ref{r4}) contains $\widetilde{\Lambda}(w)$ and
$\pa^2 \widetilde{T}(w)$, one sees $\pa J^a \pa J^a(w)$ term and $\pa K^a
\pa K^a(w)$ term in the field $\widetilde{R}^{(4)}(w)$. 
So we have to use the third and fourth equations of (\ref{rrela})
and reexpress them in terms of other independent fields.

Moreover, note that there are the following identities
\bea
d^{abc} J^a Q^b K^c(w)  & = & \frac{1}{3} d^{abcd} J^a J^b J^c K^d(w) + 
\frac{4(N^2-4)}{N(N^2+1)} J^a J^a J^b K^b(w) \nonu \\
& + &
\frac{(N^2-4)(N^2+5)}{N(N^2+1)} f^{abc} J^a \pa J^b K^c(w)
 -  \frac{2(N^2-4)(N^2+3)}
{3(N^2+1)} \pa^2 J^a K^a(w), \nonu \\
d^{abc} J^a K^b R^c(w)  & = & \frac{1}{3} d^{abcd} J^a K^b K^c K^d(w) + 
\frac{4(N^2-4)}{N(N^2+1)} J^a K^b K^b K^a(w) 
\label{jqkjkr}
\\
& + &
\frac{(N^2-4)(N^2+5)}{N(N^2+1)} f^{abc} J^a K^b \pa K^c(w)
 -  \frac{2(N^2-4)(N^2+3)}
{3(N^2+1)} J^a \pa^2 K^a(w).
\nonu
\eea
Using the first equation of (\ref{jqkjkr}), 
for example, one simplifies 
the $\frac{1}{(z-w)^2}$ term in the operator product
expansion 
$Q(z) Q^a K^a(w)$ in
(\ref{squareterms}). 
See also the relation
(\ref{otherrelation}).
Similarly, by the second equation of
(\ref{jqkjkr}), 
one can simplify the $\frac{1}{(z-w)^2}$ term in the 
operator product expansion $Q^a K^a(z) J^b K^b(w)$ in (\ref{squareterms}).
Of course, the above relations (\ref{jqkjkr}) 
can be further simplified using the equations (\ref{rrela}).
By plugging the expressions (\ref{rrela}) and (\ref{jqkjkr}) into
(\ref{squareterms}), one has the
independent terms as in the relation (\ref{cosetspin4})  as well as the term $J^a
J^b K^a K^b(z)$. 

All the coefficients turn out to be (\ref{dcoeff1}). 
It turns out that the coset spin $4$ field 
$\widetilde{R}^{(4)}(w)$ is given by the previous one
$\widetilde{W}^{(4)}(w)$ 
(\ref{generalW}) by 
realizing that the undetermined two coefficients 
$c_{1}$ and $c_{23}$ become 
\bea
c_1 & = &  -\frac{k_2(N^2+1)}{2(N^2-4)(N^2-3)(N+k_1)(N+k_1+k_2) d(k_1,k_2,N)}
\left[ -3k_1^2 k_2 -2k_1^2 N - 8k_1 k_2 N  \right. \nonu \\
& - & \left. 2k_2^2 N 
-  4 k_1 N^2 -4k_2 N^2
+ k_1^2 k_2 N^2 + k_1 k_2^2 N^2 - 2N^3 + 2k_1 k_2 N^3 -3 k_1 k_2^2 \right],
\nonu \\
c_{23} & = & \frac{-8}{(N+k_1+k_2)(3N+2k_1+2k_2)(N^2+1) 
d(k_1,k_2,N)} \left[ -9 k_1^2 k_2 - 9 k_1 k_2^2 - 14 k_1^2 N \right. \nonu \\
& - &  28 k_1
k_2 N
-   14 k_2^2 N - 24 k_1 N^2 - 24 k_2 N^2 + 3k_1^2 k_2 N^2 +
  3k_1 k_2^2 N^2 - 10 N^3 + 8k_1^2 N^3 \nonu \\
& + & \left.  18 k_1 k_2 N^3 + 8k_2^2 N^3 +
20 k_1 N^4 + 20 k_2 N^4 + 12 N^5 \right],
\label{c1c23}
\eea
where the function $d(k_1,k_2,N)$ is introduced as follows:
\bea
d(k_1,k_2,N) & \equiv & 17k_1^2 k_2 + 17k_1 k_2^2 + 22k_1^2 N + 56 k_1 k_2 N +
22 k_2^2 N + 44k_1 N^2 + 44 k_2 N^2 \nonu \\
&+ & 5k_1^2 k_2 N^2 + 5 k_1 k_2^2 N^2 +
22 N^3 + 10 k_1 k_2 N^3.
\label{ddefinition}
\eea
That is, the spin 4 field $\widetilde{R}^{(4)}(w)$ is nothing but the
field $\widetilde{W}^{(4)}(w)$,
\bea
\widetilde{R}^{(4)}(w) =\widetilde{W}^{(4)}(w), \qquad \mbox{with}
\,\,\,
(\ref{c1c23}) \,\,\, \mbox{and} \,\,\, (\ref{ddefinition}).
\label{r4w4}
\eea
The $c_1$ term comes from the $\pa^2 \widetilde{T}(w)$ in (\ref{r4}) 
together with (\ref{rrela}).
The $c_{23}$ term comes from  $\widetilde{T}^2(w)$ term and 
$J^a J^b K^a K^b(w)$ term characterized by $d_{23}$ in (\ref{r4}). 
One can easily obtain the large $N$ 't Hooft limit (\ref{limit}) for these
coefficients
and they are given by
\bea
c_1(N,\la) = \frac{-1+\lambda}{10N^3}, \qquad c_{23}(N,\la) =
\frac{32\lambda^2}{5N^3(-1+\lambda)
(1+\lambda)},
\label{c1c23limit}
\eea
where one uses the relations (\ref{dlimit}).
It is easy to check that the operator product expansion between
$J'^a(z)$ and $\widetilde{R}^{(4)}(w)$ does not have any singular
terms. That is, $J'^a(z) \widetilde{R}^{(4)}(w)=0$.
This implies that the operator product expansion between $J'^a(z)$ and
the right hand side of (\ref{r4}) gives no singular terms (the
operator product expansion $J'^a(z)
\widetilde{T}(w)$ vanishes). 
Similarly, one can check that the operator product expansion between 
$\hat{T}(z)$ and $\widetilde{R}^{(4)}(w)$ does not have any higher
order singular
terms $\frac{1}{(z-w)^n}, n> 2$ 
by using the fact that $T'(z) \widetilde{R}^{(4)}(w)=0$.  

The large $N$ behavior for the zero mode of $\widetilde{R}^{(4)}(z)$
is described at the end of the subsection \ref{other} and due to the
fact that this spin-$4$ field is the field
$\widetilde{W}^{(4)}(z)$, one has also the relation (\ref{EEigen}). 


\end{document}